\documentclass{stavanger}

\usepackage[utf8]{inputenc}
\usepackage{amssymb,amsmath,amsfonts,amsthm}
\usepackage{hyperref}
\usepackage{natbib}
\usepackage{graphicx}
\usepackage{tikz}
\usepackage{cleveref}
\usepackage{mathtools}
\usepackage{placeins}
\usepackage{xcolor}
\usepackage{bm}
\usepackage[export]{adjustbox} 

\startlocaldefs

\newcommand{\lt}{\tau_{\rm L}} 
\newcommand{\f}{\phi}    

\endlocaldefs

\begin{document}


\begin{frontmatter}

\begin{fmbox}


\dochead{Research Article - Preprint}{FP}


\title{Towards learning optimized kernels for complex Langevin}


\author[
   addressref={aff1},                   	  
   corref={aff1},                     		  
   email={daniel.alvestad@uis.no}   		  
]{\inits{DA}\fnm{Daniel} \snm{Alvestad}}
\author[
   addressref={aff1},                   	  
   email={rasmus.n.larsen@uis.no}   		  
]{\inits{RL}\fnm{Rasmus} \snm{Larsen}}
\author[
   addressref={aff1},                   	  
   email={alexander.rothkopf@uis.no}   		  
]{\inits{AR}\fnm{Alexander} \snm{Rothkopf}}


\address[id=aff1]{
  \orgname{Faculty of Science and Technology}, 	 
  \street{University of Stavanger},                     		 
  \postcode{4021}                               			 
  \city{Stavanger},                              				 
  \cny{Norway}                                   				 
}



\end{fmbox}


\begin{abstractbox}

\begin{abstract} 
We present a novel strategy aimed at restoring correct convergence in complex Langevin simulations. The central idea is to incorporate system-specific prior knowledge into the simulations, in order to circumvent the NP-hard sign problem. In order to do so, we modify complex Langevin using kernels and propose the use of modern auto-differentiation methods to learn optimal kernel values. The optimization process is guided by functionals encoding relevant prior information, such as symmetries or Euclidean correlator data. Our approach recovers correct convergence in the non-interacting theory on the Schwinger-Keldysh contour for any real-time extent. For the strongly coupled quantum anharmonic oscillator we achieve correct convergence up to three-times the real-time extent of the previous benchmark study. An appendix sheds light on the fact that for correct convergence not only the absence of boundary terms, but in addition the correct Fokker-Plank spectrum is crucial.
\end{abstract}


\begin{keyword}
\kwd{stochastic quantization, complex Langevin, sign problem, real-time, machine learning}
\end{keyword}

\end{abstractbox}


\end{frontmatter}


\section{Motivation}
\label{sec:introduction}

Strongly correlated quantum systems underlie some of the most pressing open questions in modern theoretical physics. Whether it is the transport of highly energetic partons through a liquid of deconfined quarks and gluons \cite{busza_heavy_2018}, created in heavy-ion collisions \cite{Foka:2016vta} or the transport of non-relativistic fermions \cite{chien_quantum_2015}, captured in the iconic Hubbard model \cite{qin_hubbard_2021} at low energies. When formulated in Minkowski time, quantum field theories so far have defied a treatment by conventional Monte-Carlo simulation techniques, due to the presence of the notorious sign problem \cite{gattringer_approaches_2016,Pan:2022fgf}. And while progress has been made in extracting real-time dynamics from Euclidean time simulations using e.g. Bayesian inference \cite{Rothkopf:2022ctl}, the sign problem prevails by rendering the extraction ill-posed and equally exponentially hard. 

The sign problem has been proven to be NP-hard \cite{Troyer:2004ge}, which entails that no generic solution method is likely to exist. In turn, if we wish to make inroads towards overcoming the sign problem, system-specific solutions are called for.

Over the past decade, several approaches to tackle the sign problem have been put forward \cite{gattringer_approaches_2016,berger_complex_2021}. They can be divided into system-specific and system-agnostic approaches. The reformulation strategies discussed e.g. in Refs. \cite{chandrasekharan_meron-cluster_1999,mercado_qcd_2011,Kloiber:2013rba} are an example of the former class, where the partition function of the original system is re-expressed in terms of new degrees of freedom, for which no sign problem exists. While highly successful in the systems for which a reformulation has been discovered, no systematic prescription exists to transfer the approach to other systems. The other approaches, among them reweighting, extrapolation from sign-problem free parameter ranges \cite{de_forcrand_constraining_2010,braun_imaginary_2013,braun_zero-temperature_2015,guenther_qcd_2017}, density of states \cite{wang_efficient_2001,langfeld_density_2012,gattringer_density_2015}, tensor networks \cite{orus_tensor_2019,hauschild_efficient_2018}, Lefschetz thimbles \cite{rom_shifted-contour_1997,cristoforetti_new_2012,Alexandru:2020wrj} and complex Langevin (CL) \cite{damgaard_stochastic_1987,namiki_stochastic_1992} all propose a generic recipe to estimate observables in systems with a sign problem. As the NP-hard sign problem however requires system-specific strategies, all of these methods are destined to fail in some form or the other. Be it that their costs scale excessively when deployed to realistic systems (e.g. reweighting, Lefschetz thimbles, tensor networks) or that they simply fail to converge to the correct solution (complex Langevin).

Both the Lefschetz Thimbles and complex Langevin belong to the class of complexification strategies \cite{berger_complex_2021}. They attempt to circumvent the sign problem by moving the integration of the Feynman integral into the complex plane. After complexifying the degrees of freedom, the former proposes to integrate over a specific subspace on which the imaginary part of the Feynman weight remains constant (thimble), while the latter proposes to carry out a diffusion process of the coupled real- and imaginary part of the complexified degrees of freedom.

In this paper our focus lies on the complex Langevin approach, as it has been shown to reproduce correctly the physics of several strongly correlated model systems, albeit in limited parameter ranges \cite{Seiler:2017wvd}. Most importantly in its naive implementation it scales only with the volume of the system, similar to conventional Monte-Carlo simulations. In the past, complex Langevin had suffered from two major drawbacks: the occurrence of unstable trajectories, called runaways and the convergence to incorrect solutions. In a previous publication \cite{Alvestad:2021hsi} we have shown how to avoid runaways by deploying inherently stable implicit solvers (c.f. the use of adaptive step size \cite{AartsJames2010}). In this study we propose a novel strategy to restore correct convergence in the complex Langevin approach.

One crucial step towards establishing complex Langevin as reliable tool to attack the sign problem is to identify when it converges to incorrect solutions. The authors of ref.~\cite{Aarts:2011ax} and later \cite{Nagata:2016vkn} discovered that in order for CL to reproduce the correct expectation values of the underlying theory, the histograms of the sampled degrees of freedom must fall off rapidly in the imaginary direction. Otherwise boundary terms spoil the proof of correct convergence. The absence of boundary terms has been established as necessary criterion and efforts are underway \cite{Sexty:2021nov} to compensate for their presence to restore correct convergence.

With QCD at the center of attention, the gauge cooling strategy \cite{Seiler:2012wz,berges_real-time_2008}, based on exploiting gauge freedom, has been proposed. It has recently been amended by the dynamic stabilization approach \cite{Attanasio:2018rtq,Aarts:2016qhx}, which modifies the CL stochastic dynamics with an additional drift term. Both are based on the idea that by pulling the complexified degrees of freedom closer to the real axis, boundary terms can be avoided. Their combination has led to impressive improvements in the correct convergence of complex Langevin in the context of QCD thermodynamics with finite Baryo-chemical potential \cite{Aarts:2022krz} and is currently explored in the simulation of real-time gauge theory \cite{Boguslavski:2022vjz}.

We focus here on scalar systems formulated in real-time on the Schwinger-Keldysh contour (for a Lefschetz thimble perspective see \cite{Alexandru:2016gsd,Alexandru:2017lqr}). For scalars, gauge freedom does not offer a rescue from the convergence problems. The fact that dynamical stabilization introduces a non-holomorphic modification of the drift term means that the original proof of convergence is not applicable, which is why we refrain from deploying it here. Furthermore the boundary term correction requires that the eigenvalues of the Fokker-Planck equation associated with the original system lie in the lower half of the complex plane, which is not necessarily the case in the scalar systems that we investigate. 

The convergence problem in real-time complex Langevin is intimately connected with the extent of the real-time contour \cite{BergesSexty2007}. In a previous publication \cite{Alvestad:2021hsi} we showed that for a common benchmark system, the strongly correlated quantum anharmonic oscillator, real-time simulations directly on the SK contour are feasible for times up to $m t_{\rm max}=0.5$. Convergence quickly breaks down when extending the contour beyond this point.

Within the complex Langevin community, coordinate transformations and redefinitions of the degrees of freedom have been used in the past to weaken the sign problem in a system specific manner (see e.g. discussion in \cite{Aarts:2012ft}). All of these reformulations can be captured mathematically by introducing a so called kernel for complex Langevin. It amounts to a simultaneous modification of the drift and noise contribution to the CL stochastic dynamics. In the past it has been used to improve the autocorrelation time in real-valued Langevin simulations \cite{namiki_stochastic_1992} and has been explored in simple model systems to restore the convergence of complex Langevin (see e.g. \cite{Okamoto:1988ru}). The construction of the kernels, as discussed in the literature applies to a specific system only and so far no systematic strategy exists to make kernels work in more realistic theories.

Our study takes inspiration from both conceptual and technical developments in the machine learning community. In machine learning, an optimization functional, based on prior knowledge and data is used to train an algorithm to perform a specific task. The algorithm depends on a set of parameters, e.g. the weights of a neural network, which need to be tuned to minimize the prescribed optimization functional. Highly efficient automatic differentiation programming techniques \cite{baydin2018automatic} have been developed to compute the dependence of the outcome of complex algorithms on their underlying parameters. Here we utilize them to put forward a systematic strategy to incorporate prior knowledge about the system into the CL evolution by learning optimal kernels.

In \cref{sec:kernelled_lang} we review the concept of kernelled Langevin, first in the context of Euclidean time simulations and subsequently for use in complex Langevin. In \cref{sec:const_kernel} we show how the concept of a kernel emerges in a simple model system and how it relates to the Lefschetz thimbles of the model. Subsequently we discuss that a constant kernel can be used to restore convergence of real-time complex Langevin for the quantum harmonic oscillator. The kernel found in this fashion will help us to improve the convergence of the interacting theory too. \Cref{sec:optimal_kernels} introduces the central concept of our study: a systematic strategy to learn optimal kernels for complex Langevin, based on system-specific prior information. Numerical results from deploying a constant kernel to the quantum anharmonic oscillator are presented in \cref{sec:num_res} (Source code for the kernel optimization and simulation is written in Julia and available at \cite{daniel_alvestad_2022_7373498}), leading to a significant extension of correct convergence. In the appendices, we discuss some of the limitations of constant kernels and show in the context of simple models that correct convergence requires not only the vanishing of boundary terms but in addition requires the spectrum of the associated Fokker-Plank equation to remain negative.


\section{Neutral and non-neutral modifications of Langevin dynamics}
\label{sec:kernelled_lang}

Stochastic quantization, the framework underlying Langevin simulations, sets out to construct a stochastic process for fields in an artificial additional time direction $\tau_{\rm L}$ with a noise structure, which correctly reproduces the quantum statistical fluctuations in the original theory. In the context of conventional Monte-Carlo simulations in Euclidean time, where expectation values of observables are given by the path integral
\begin{align}\label{eq:PathIntegral}
    \langle O \rangle = \frac{1}{Z} \int {\cal D}\phi\;O[\phi] e^{-S_E[\phi]}, \quad S_E[\phi]=\int d^dx L_E[\phi],
\end{align}
with Euclidean action $S_E$, the goal thus is to guarantee at late Langevin times a distribution of fields $\Phi[\phi]\propto{\rm exp}\big(-S_E[\phi]\big)$. The chain of configurations $\phi(\tau_L)$ underlying the distribution $\Phi[\phi]$, can then be used to evaluate the expectation values of observables $O$ from the mean of samples $\langle O \rangle = \lim_{\tau_{\rm L}\to\infty} \frac{1}{\tau_{\rm L}} \int_0^{\tau_{\rm L}} d\tau_{\rm L}^\prime O[\phi(\tau_{\rm L}^\prime)]$.  The simplest stochastic process, which realizes this goal and which is therefore commonly deployed is 
\begin{equation}\label{eq:LE}
\begin{aligned}
    & \frac{d\f}{d\lt} = -\frac{\delta S_E[\f]}{\delta \f(x)} + \eta(x,\lt) \quad \textrm{with}   \\
    &\langle \eta(x,\lt) \rangle = 0, \quad \langle \eta(x,\lt) \eta(x',\lt') \rangle = 2\delta(x-x')\delta(\lt-\lt').
\end{aligned}
\end{equation}

Its drift term is given by the derivative of the action $S_E$ and the noise terms $\eta$ are Gaussian. The associated Fokker-Planck equation reads
\begin{equation}\label{eq:FP_operator}
    F_{\textrm{FP}} = \int d^dx \; \frac{\partial}{\partial \f(x)}\left(\frac{\partial}{\partial \f(x)} + \frac{\delta S_E[\f]}{\delta \f(x)} \right), \quad \frac{\partial\Phi(\phi,\lt)}{\partial \lt} = F_{\textrm{FP}}\Phi(\phi,\lt).
\end{equation}
For an in-depth review of the approach see e.g. ref.~\cite{namiki_stochastic_1992}.

In the following we will discuss the fact that there exists the freedom to introduce a so called \textit{kernel} into \cref{eq:FP_operator}, which as a \textit{purely real} quantity allows us to modify the above Fokker-Planck equation without spoiling the convergence to the correct stationary solution $\Phi[\phi]=\lim_{\tau_L\to\infty}\Phi[\phi,\tau_L]\propto{\rm exp}\big(-S_E[\phi]\big)$. One may use this freedom to improve autocorrelation times of the simulation and for other problem-specific optimizations as has been explored in the literature.

Subsequently we will turn our attention to the case of \textit{complex Langevin}, where the simplest stochastic process proposed by stochastic quantization is not guaranteed to converge to the correct solution. In that case we will explore how a reparametrization of the associated Fokker Planck equations through in general \textit{complex kernels} can be used to not only change the convergence speed but actually to change the stationary distribution itself, allowing us to recover correct convergence where the naive process fails.

\subsection{Kernelled Real Langevin}

As alluded to above, there exists a freedom to reparametrize the Fokker-Planck \cref{eq:FP_operator} by introducing a real-valued kernel function $K_{ij}(x,x^\prime,\phi;\tau_{\rm L})$
\begin{equation}\label{eq:KFP_operator}
    F_{\textrm{FP}} = \sum_{i,j} \int d^dx \int d^dx' \; \frac{\partial}{\partial \f_i(x)} K_{ij}(x,x',\f;\tau_{\rm L}) \left(\frac{\partial}{\partial \f_j(x')} + \frac{\delta S_E[\f]}{\delta \f_j(x')} \right).
\end{equation}
Written in its most generic form, it may couple the different degrees of freedom of the system (according to the $ij$ indices), it may couple different space-time points (according to its $x$ and $x^\prime$ dependence) and may depend explicitly both on the Langevin time $\tau_{\rm L}$, as well as the field degrees of freedom $\phi$. 
The corresponding Langevin equation reads
\begin{equation}\label{eq:generalKLE}
\begin{aligned}
     \frac{d\f_i(x,\lt)}{d\lt} = & \sum_j \left\{  -\int d^dx' K_{ij}(x,x';\f) \frac{\delta S_E[\f]}{\delta \f_j(x',\lt)} + \int d^dx' \frac{\delta K_{ij}(x,x';\f)}{\delta \f_j(x',\lt)}  \right.
    \\& \left. + \int d^dx' H_{ij}(x,x';\f)\eta(x',\lt) \right\}    \quad \textrm{with} \\
      K(x,x';\f) =& \sum_k \int d^dx'' H_{ik}(x,x'';\f)H_{jk}(x',x'';\f),
\end{aligned}
\end{equation}
where in the last equation we assume that $K$ is factorizable. In practice we will either choose kernels, which can be factorized using the square root of their eigenvalues or will start directly by constructing the function $H$ that can be combined into an admissible $K$.

Let us gain a bit of intuition about the role of the kernel when considering it in its simplest form, a constant scalar kernel, which multiplies each d.o.f. with a real number $\gamma$. Inspecting \cref{eq:generalKLE} we find that, as it appears in front of the drift term and as square root in front of the noise term, $\gamma$ simply leads to a redefinition of the Langevin time coordinate $\tau_{\rm L}^\prime=\gamma \tau_{\rm L}$. While the stationary solution is left unchanged, the convergence time has been modified. 

Even for more general kernels, the fact that $K$ appears in the generalized Fokker-Planck \cref{eq:KFP_operator} on the outside of the parenthesis $\left(\frac{\partial}{\partial \f_i(x)} + \frac{\delta S_E[\f]}{\delta \f_i(x)} \right)$  tells us that the stationary distribution remains unchanged. It goes without saying that choosing $K_{ij}(x,x';\phi)=\delta_{ij}\delta(x' - x)$ we regain the standard Langevin \cref{eq:LE}. 

\subsection{Kernelled Complex Langevin}\label{sec:complexKernelLangevin}

Let us now consider the application of stochastic quantization to complex-valued path integrals, in particular to those describing real-time physics in Minkowski time. Here the observables are given by  Feynman's path integral
\begin{align}\label{eq:PathIntegralM}
    \langle O \rangle = \frac{1}{Z} \int {\cal D}\phi\;O[\phi] e^{iS_M[\phi]}, \quad S_M[\phi]=\int d^dx L_M[\phi],
\end{align}
which houses the Minkowski time action of the theory $S_M$. Stochastic quantization in this case \textit{proposes} to modify the real-valued stochastic process of \cref{eq:LE} via the substitution $-S_E\to iS_M$ such that
\begin{equation}\label{eq:CLE}
\begin{aligned}
    & \frac{d\f}{d\lt} = i\frac{\delta S_M[\f]}{\delta \f(x)} + \eta(x,\lt) \quad \textrm{with}   \\
    &\langle \eta(x,\lt) \rangle = 0, \quad \langle \eta(x,\lt) \eta(x',\lt') \rangle = 2\delta(x-x')\delta(\lt-\lt').
\end{aligned}
\end{equation}
It is obvious that even if one starts out with purely real degrees of freedom at $\tau_{\rm L}=0$, the presence of the complex drift term necessitates the complexification $\phi=\phi_{\rm R}+i\phi_{\rm I}$, each of which will obey a coupled stochastic evolution.

In the complexified scenario, the question of correct convergence is not as simple to answer as in the purely real case. The most stringent criterion refers to whether complex Langevin reproduces the correct expectation values
\begin{align}
\lim_{\tau_{\rm L}\to\infty} \frac{1}{\tau_{\rm L}} \int_0^{\tau_{\rm L}} d\tau_{\rm L}^\prime O[\phi_R+i\phi_I] \overset{?}{=} \frac{1}{Z} \int {\cal D}\phi\;O[\phi] e^{iS_M[\phi]}  \label{eq:corrconv}
\end{align}
of the theory, defined on the right. And indeed it has been found that the dynamics of \cref{eq:CLE} may violate the equal sign of \cref{eq:corrconv}. I.e. complex Langevin converges, but it does not converge to the correct solution. In this study we set out to recover correct convergence by introducing kernels into the complex Langevin dynamics.

To this end we consider a not-necessarily real kernel function $K(x,x^\prime;\phi)$ which enters the complexified dynamics as
\begin{equation}\label{eq:KCLE}
\begin{aligned}
    & \frac{d\f}{d\lt} = \int d^dx' \left\{ i K(x,x';\f) \frac{\delta S_M[\f]}{\delta \f(x',\lt)} + \frac{\partial K(x,x'; \f)}{\partial \f(x',\lt)} + H(x,x';\f)\eta(x,\lt) \right\}   \\
    & \textrm{with} \quad \langle \eta(x,\lt) \rangle = 0, \quad \langle \eta(x,\lt) \eta(x',\lt') \rangle = 2\delta(x-x')\delta(\lt-\lt') \\
    & \textrm{and} \quad K(x,x';\f) = \int d^dx'' H(x,x'';\f)H(x',x'';\f).
\end{aligned}
\end{equation}
Expressed as two separate but coupled stochastic processes for the real- and imaginary part of the complexified field we obtain
\begin{equation}\label{eq:KCLE_split}
\begin{aligned}
    & \frac{d\f_R}{d\lt} = \int d^dx'\left.\left\{ \textrm{Re}\left[ K[\f]  i\frac{\delta S_M[\f]}{\delta \f} + \frac{\delta K[\phi]}{\delta \phi} \right] + \textrm{Re}\left[ H[\phi] \right] \eta \right\}\right|_{\f = \f_R + i\f_I}, \\
    & \frac{d\f_I}{d\lt} = \int d^dx'\left.\left\{ \textrm{Im}\left[ K[\f]  i\frac{\delta S_M[\f]}{\delta \f} + \frac{\partial K[\phi]}{\partial \phi} \right] + \textrm{Im}\left[ H[\phi] \right] \eta \right\}\right|_{\f = \f_R + i\f_I}.
\end{aligned}
\end{equation}

Note that at this point we are dealing with two different concepts of Fokker-Planck equations. One describes how the probability distribution $\Phi[\phi_R,\phi_I]$ of the real- and imaginary part $\phi_R$, $\phi_I$ of the complexified field evolve under \cref{eq:KCLE}
\begin{equation}\label{eq:FP_KCLE}
\begin{aligned}
        \frac{\partial \Phi}{\partial \lt} =& \left[ \left(\frac{\partial}{\partial \phi^R} H_R + \frac{\partial}{\partial \phi^I} H_I \right)^2 - \frac{\partial}{\partial \phi^R} {\rm Re} \left\{iK\frac{\partial S_M}{\partial \f} + \frac{\partial K}{\partial \phi}\right\} \right. \\
        & \left. - \frac{\partial}{\partial \phi^I} {\rm Im} \left\{iK\frac{\partial S_M}{\partial \f} + \frac{\partial K}{\partial \phi}\right\} \right] \Phi = L_K\Phi.  
\end{aligned}
\end{equation}
We define the operator for the real Fokker Planck equation, which has been separated into real and imaginary part, as $L_K$, not to be confused with the original now complex Fokker Planck equation $F_{FP}$, which was only defined on the real part of $\phi$.
For the term quadratic in derivatives, we have split the kernel K into the product of H functions, as shown in \cref{eq:KCLE}, such that each derivative acts on either the real or the imaginary part of $H$ respectively. Since it is the noise term of the Langevin \cref{eq:KCLE} that translates into a term quadratic in derivatives in the Fokker-Planck language, it is there that $H$ appears in \cref{eq:FP_KCLE}.

It is important to recognize that the correct late Langevin-time distribution of this Fokker-Planck equation is purely real and therefore is not related in a trivial manner to the Feynman weight ${\rm exp[i S_M]}$ of the original path integral, as has been established in simple models in the literature as discussed e.g. in refs.~\cite{Klauder:1985kq,Giudice:2013eva,Abe:2016hpd,Seiler:2017vwj,Salcedo:2018uop}.

The other Fokker-Plank equation is not a genuine Fokker-Planck equation, in the statistical sense, as it does not describe the evolution of a real-valued probability density $P[\phi,\tau_L]$ but instead that of a complex-valued distribution $\rho(\phi,\tau_L)$
\begin{align}
     &\frac{\partial}{\partial \tau_L} \rho(\phi,t) = F_{FP} \rho(\phi,\tau_L),\label{eq:CFP}\\
     \nonumber &F_{FP} = \sum_{i,j} \int d^dx \int d^dx' \; \frac{\partial}{\partial \f_i(x)} K_{ij}(x,x',\f;\tau_{\rm L}) \left(\frac{\partial}{\partial \f_j(x')} - i\frac{\delta S_M[\f]}{\delta \f_j(x')} \right).
\end{align}
It is this equation whose late time limit we expect to reproduce the Feynman weight $\lim_{\tau_L\to\infty}\rho(\phi,\tau_L)={\rm exp[i S_M]}$ and we will refer to in the following as the complex Fokker-Planck equation.

Significant progress in the understanding of the convergence properties of complex Langevin had been made starting with ref.~\cite{Aarts:2011ax} in the form of so-called correctness criteria.

The criteria most often discussed in the literature are boundary terms (for a detailed exposition see Refs.~\cite{Aarts:2011ax,Nagata:2016vkn}). They tell us if the expectation value calculated from the real distribution $\Phi(\phi^R,\phi^I;\lt)$ (\cref{eq:FP_KCLE}), which we can sample using the CL, is the same as the expectation value obtained from the complex distribution $\rho(\phi;\lt)$. The latter one can only be obtained from solving the complex Fokker-Planck equation, \cref{eq:CFP}. The two expectation values, $\langle \mathcal O \rangle_\Phi(\lt) = \langle \mathcal O \rangle_\rho (\lt)$ only agree if $\Phi(\phi^R,\phi^I;\lt)$ falls off exponentially fast. If it does not fall of sufficiently fast, it will produce boundary terms and the equal sign in \cref{eq:corrconv} is not valid. This criterion is however not sufficient as it does not guarantee the equilibrium distribution of the complex Fokker-Planck equation to be ${\rm exp[i S_M]}$. These two criteria combined are however sufficient to claim convergence of the CL to the true solution. For a proof that the correctness criterion still holds after introducing a kernel into the CL, we revisit the proof in \cref{sec:correctnessCriterionWithAKernel}.

How can a kernel help to restore the correct convergence? Not only do we need to make sure that no boundary terms arise in sampling \cref{eq:KCLE_split} but also that the complex Fokker-Planck equation has a unique and correct complex stationary distribution. I.e. we need in general a non-neutral modification of the complex Langevin dynamics.

If we were to introduce a real-valued kernel, similarly to the case of conventional real-valued Langevin, we will be able to change the speed of convergence but not the stationary solution. On the other hand, since the drift term is complex, there is no reason not to consider also complex valued kernels, which will act differently on the stochastic process for $\phi_R$ and $\phi_I$, representing a genuine non-neutral modification in the corresponding Fokker-Planck \cref{eq:FP_KCLE}. Similarly in the complex Fokker-Planck \cref{eq:CFP} the presence of a complex $K_{ij}$ can change the stationary distribution through a reshuffling of the associated eigenvalues, as is discussed in more detail in \cref{sec:correctnessCriterionWithAKernel}. For a comprehensive discussion of different modifications to complex Langevin, including kernels, see also ref.~\cite{Aarts:2012ft}.

In the following sections we will start off with constructing an explicit example of a field-independent kernel that improves convergence in the free theory and find that it can restore correct convergence in the interacting theory to some degree. We will then continue to present our novel strategy to learn optimal kernels for the restoration of correct convergence and showcase their efficiency in a benchmark model system. Subsequently we discuss the limitations of field-independent kernels and shed light on how kernels connect to the correctness criteria.

\section{A field independent kernel for real-time complex Langevin}
\label{sec:const_kernel}

In this section, we will manually construct one specific field-independent kernel and demonstrate its use to improve convergence in real-time simulations of the quantum anharmonic oscillator. The form of the kernel is motivated by insight gained in a simple one d.o.f. model and reveals an interesting connection between kernelled Langevin and the thimble approach. Since in the following only low dimensional model systems are considered, we will refer to the dynamical degrees of freedom from now on as $x$.

\subsection{A kernel for the simplest real-time model}
\label{sec:simplestRTModel}

Following ref.~\cite{Okamoto:1988ru} let us investigate the simplest model of real-time physics, the integrals 
\begin{align}\label{eq:simpleModelObservableIntegral}
    \langle x^n\rangle =\frac{1}{Z}\int \; dx\, x^n\, {\rm exp}[-\frac{1}{2}ix^2], \quad Z= \int \; dx\, \, {\rm exp}[-\frac{1}{2}ix^2].
\end{align}
Attempting to solve this expression using the complex Langevin approach for $x(\tau_L)$, leads to a stochastic process 
\begin{align}\label{eq:simpleModelCLE_ix^2}
    \frac{dx}{d\tau_L} = -ix + \eta,
\end{align}
with Gaussian noise $\eta$. \Cref{eq:simpleModelCLE_ix^2} fails at reproducing correct values of $\langle x^n\rangle$. 

We can understand this failure by recognizing that without regularization the original integral in \cref{eq:simpleModelObservableIntegral} is not well defined and this lack of regularization is inherited by the Langevin \cref{eq:simpleModelCLE_ix^2}. 

One way to proceed is to explicitly modify the action by introducing a regulator term, such as $\epsilon x^2$. The integral becomes well-defined and its value is obtained when we let $\epsilon \rightarrow 0$ at the end of the computation. In a numerical setting this would require to explicitly include the regulator term, carry out the corresponding simulation for different values of $\epsilon$ and extrapolate $\epsilon \rightarrow 0$. There are two drawbacks to this strategy: first it requires several evaluations of the simulation, which can be expensive for large systems. The second reason is that the relaxation time for the simulation grows, the smaller $\epsilon$ becomes, and hence in practice we cannot make $\epsilon$ arbitrarily small. 

Let's consider an alternative strategy of solving the integral of \cref{eq:simpleModelObservableIntegral}, which  relies on contour deformations, the so-called Lefschetz thimble method. We will carry out a change of variables in the integral which moves the integration path into the complex plane and which in turn will weaken the oscillatory nature of the integrand. This method is based on a continuous change of variables according to the following gradient descent equation
\begin{equation}\label{eq:ThimbleFlow}
    \frac{d \tilde x}{d\tau} = \overline{\frac{d S_E[\tilde x]}{d\tilde x}},
\end{equation}
which complexifies the degree of freedom, $\tilde x = a+ib$. \Cref{eq:ThimbleFlow} evolves the formerly real-valued $x$ towards the so called Lefschetz thimble which is the optimal contour deformation where the imaginary part of the action stays constant.  

Following the steps outlined in \cite{Woodward:2022pet}, we solve the flow of \cref{eq:ThimbleFlow} analytically which gives $\tilde x(x,\tau) = x(\cosh(\tau) - i\sinh(\tau))$. For large values of $\tau$ it leads to
\begin{equation}
    \tilde x(x,\tau) \overset{\tau \gg1}{\approx} x(1-i)\frac{1}{2e^{-2\tau}} = \frac{x}{2e^{-2\tau}} e^{-i\frac{\pi}{4}}.
\end{equation}
The above equation tells us that the optimal thimble in this system lies on the downward $45^\circ$ diagonal in the complex plane $z(x)=xe^{-i\frac{\pi}{4}}$. On this contour the integrand of the original integral \cref{eq:simpleModelCLE_ix^2} reduces to a real Gaussian $e^{-x^2}$ for which no regularization is required.

If we flow for just a very small $\tau=\epsilon$, we obtain on the other hand $\cosh(\tau) - i \sinh(\tau) \approx 1 - i\epsilon$ and 
\begin{align}
    &\int \; dx\,  {\rm exp}[-\frac{1}{2}ix^2] 
        = \int \; d x\, \frac{\partial \tilde x}{\partial x} \;  {\rm exp}[-\frac{1}{2}i \tilde x^2] \\
        \nonumber&= (1 - i\epsilon) \int \; d x  \;  {\rm exp}[-\frac{1}{2}i x^2(1 - i\epsilon)^2] 
        \approx  (1 - i\epsilon) \int \; d x  \;  {\rm exp}[-\frac{1}{2}i x^2 - \epsilon x^2].
\end{align}
We see that the term $\epsilon$ here takes on the role of a regulator in the action but due to its presence also in the Jacobian, the value of the integral is not changed. This is different from introducing the regulator only in the action itself. 

Hence the obvious benefit of the deformation method is that we can introduce a regulator to tame oscillations without the need to extrapolate that regulator in the end. The closer we approach the optimal thimble, the easier the integral will be to solve numerically.

How can such a coordinate transformation be implemented in complex Langevin? Intuitively the action in the integral is what influences the drift in complex Langevin and the measure is related to the noise structure. The above tells us that the change we introduced will therefore affect the drift quadratically, while it occurs in the noise linearly. Thinking back to \cref{eq:KCLE}, we  see that this is just how a field-independent kernel modifies the complex Langevin equations.

For the optimal thimble with $z(x)=xe^{-i\frac{\pi}{4}}$ the modification in the drift therefore becomes $K=e^{-i\frac{\pi}{2}}=\frac{1}{i}$ and for the noise $H=\sqrt{K}=\sqrt{-i}$. This leads to the following stochastic process 
\begin{align}\label{eq:simpleModelCLE_ix^2_kernel}
    \frac{dx}{d\tau_L} = -x + \sqrt{-i}\eta,
\end{align}
which had been identified as optimal already in ref.~\cite{Okamoto:1988ru}. This stochastic process converges to the correct solution of the integral \cref{eq:simpleModelObservableIntegral}. Interestingly the imaginary unit has disappeared from the drift term since the kernel $K$ exactly canceled it there and instead moved it over into the noise term.

As the last step, let us show explicitly that the choice of kernel above indeed amounts to a coordinate transform. Following \cite{Aarts:2012ft} we have
\begin{align}
    \frac{dx}{d\tau_L} =& -HH^T\frac{\partial S_E(x)}{\partial x} + H\eta\\
    &\Rightarrow H^{-1}\frac{dx}{d\tau_L} = -H^T\frac{\partial S_E(x)}{\partial x} + \eta \\
    &\Rightarrow \frac{du}{d\tau_L} = -H^T (H^{T})^{-1} \frac{\partial T(u)}{\partial u} + \eta  = -\frac{\partial T(u)}{\partial u} + \eta.\label{eq:coordtransfCLE}
\end{align}
Here $x = Hu$ and $T(u) = S_E(H u) = S_E(x)$. We find that introducing a kernel $K=HH^T$ in the evolution equation for $x$ has the same effect as carrying out a coordinate transformation to $u=H^{-1}x$. 

\begin{figure}\centering
    \includegraphics[scale=0.28]{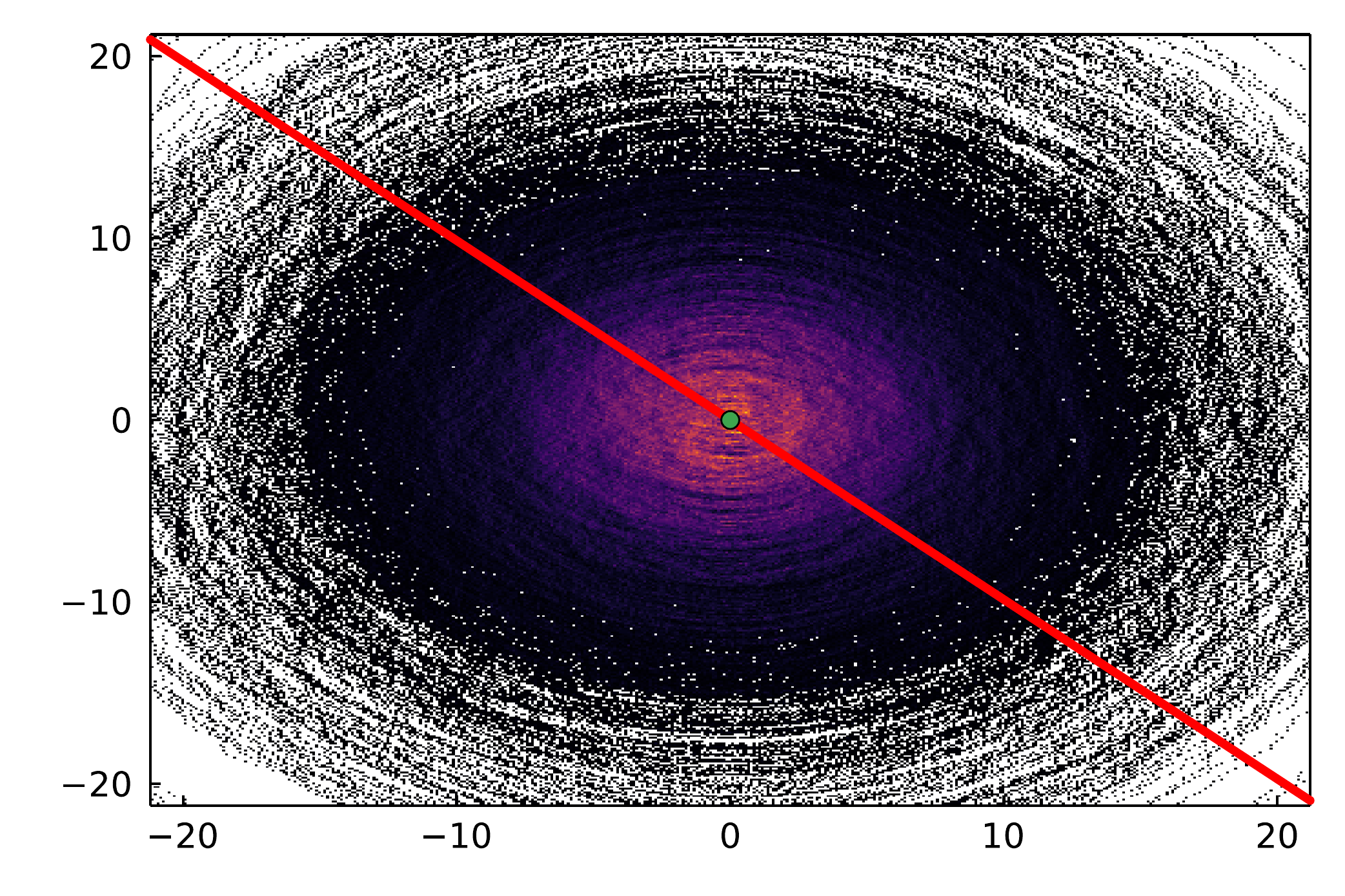}
    \includegraphics[scale=0.28]{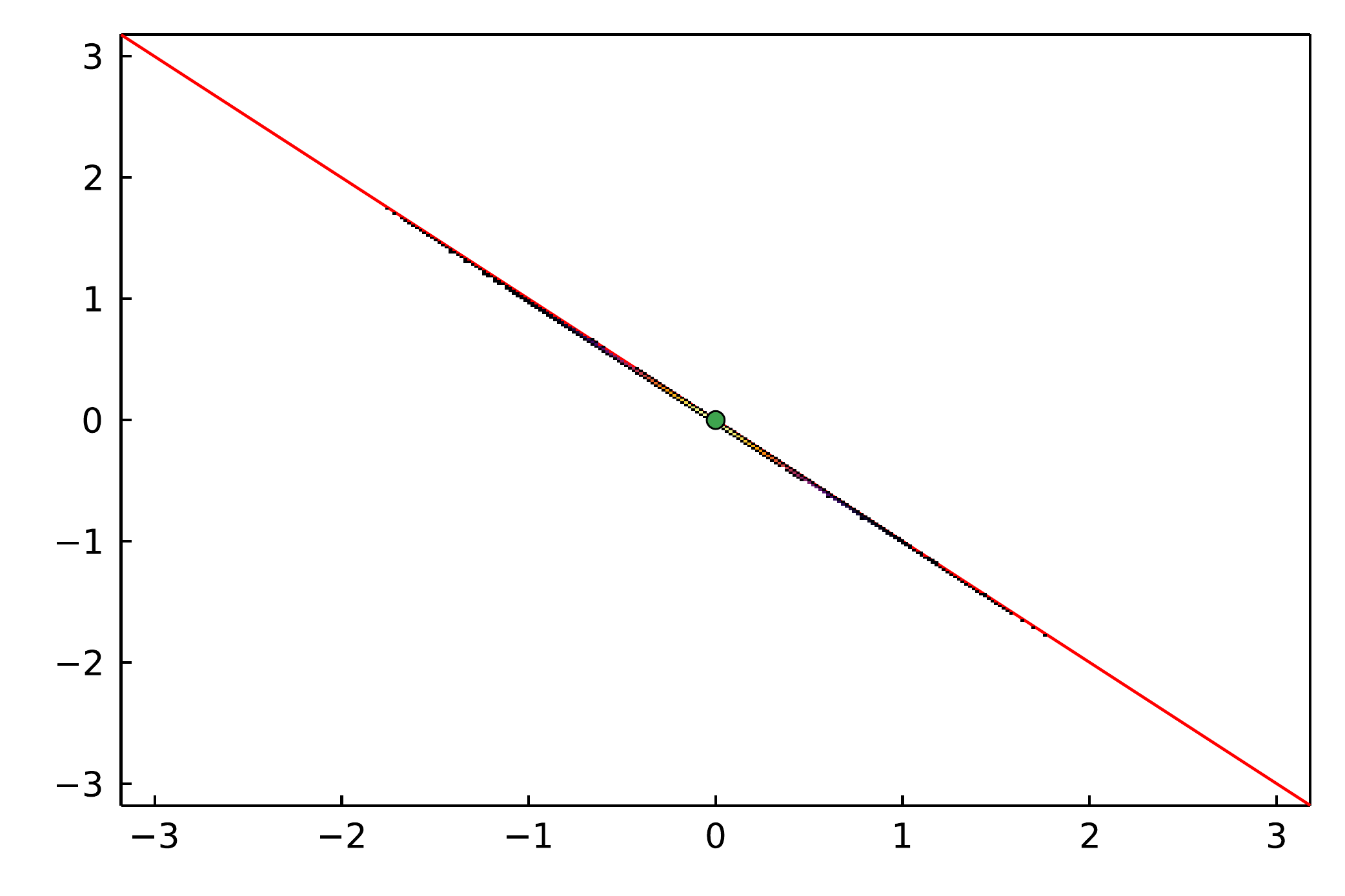}
    \caption{Distribution of a complex Langevin simulation (scatter points) and the Lefschetz thimble (red line) for the model \cref{eq:simpleModelObservableIntegral}. (left) simulation according to the naive CL \cref{eq:simpleModelCLE_ix^2} and (right) simulation after introducing the kernel in \cref{eq:simpleModelCLE_ix^2_kernel}. The color of the scatter points refers to the number of measurements recorded at the corresponding position, a lighter color indicates a larger number. Note that the optimal kernel has moved the sampling onto the single thimble present in this simple system.}
    \label{fig:SimpleModelCLEandKCLE}
\end{figure}

As an example of the complex Langevin dynamics in the absence (left panel) and presence (right panel) of the kernel discussed above, we show the corresponding scatter plots in \cref{fig:SimpleModelCLEandKCLE}. The kernel has indeed rotated the noise into the direction of the thimble, along which the system now samples. Note that while the naive CL dynamics have been implemented using the semi-implicit Euler-Maruyama scheme to avoid runaways, we are able to carry out the kernelled dynamics with a fully explicit solver without adaptive step size. The reason is that on the deformed contour the integral has already been regularized. 

This result shows that in the simple model discussed here we can find a kernel that both restores correct convergence of the complex Langevin dynamics and at the same time removes the need for a regulator. Both are related to the fact that the kernel effectively instituted a coordinate transform that amounts to a deformation of the integration contour into the complex plane. In the next section, we will consider a similar kernel for the harmonic oscillator. 

We investigate the relation between the Lefschetz thimbles and kernel controlled complex Langevin in \cref{sec:kernel_and_boundaryTerms}, where a similar analysis is performed in a system where  a $\frac{\lambda}{4}x^4$ term has been added to the action. As such a one-degree-of-freedom model has a non-linear drift term, a constant kernel in that case will not suffice to remove the imaginary unit from the drift term, and hence the complex Langevin will not sample directly on the thimble as was the case for the example above.

\subsection{A kernel for the harmonic oscillator}

When constructing a kernel for the harmonic oscillator, we encounter similar difficulties related to stability and convergence of complex Langevin process as in the previous section. In order to see how an optimal kernel can be chosen we revisit the discussion originally found in refs.~ \cite{Okamoto:1988ru,namiki_stochastic_1992}. 

The continuum action of this one-dimensional system is given by
\begin{equation}
    \label{eq:AHOAction}
    S_M = \int dt \left\{ \frac{1}{2}  \left( \frac{\partial x(t)}{\partial t} \right)^2 - \frac{1}{2}m^2x^2(t) \right\}=\int dt  \left\{ \frac{1}{2} x(t) \Big( -\partial_t^2 - \frac{1}{2}m^2 \Big) x(t)\right\},
\end{equation}
In quantum mechanics the fields $\phi$ are the position $x$ and the coordinates, previously called $x$, are the time $t$. The corresponding complex Langevin equation reads
\begin{align}
    \frac{dx(t,\tau_L)}{d\tau_L}=-i\Big(\partial_t^2+m^2\Big)x(t,\tau_L)+\eta(t,\tau_L)\label{eq:harmoscCLE}.
\end{align}
In the absence of a regularization, this stochastic process is unstable and does not show convergence to the correct result. In analogy with the results for the simple model system in the previous section we will argue analytically that correct convergence can be achieved in this system via a kernel with the property $-\Big(\partial_t^2+m^2\Big)K(t-t')=i\delta(t-t')$. This kernel will render the drift term trivial, proportional to $x$ itself and move all complex structure into the noise.

Following \cite{Okamoto:1988ru,namiki_stochastic_1992} we solve \cref{eq:harmoscCLE} analytically and obtain for the two-point correlator in Fourier space
\begin{equation}
    \langle x(\omega,\tau_L)x(\omega',\tau_L') \rangle = \delta(\omega+\omega') \frac{i}{\omega^2 - m^2} \left(e^{i(\omega^2 - m^2)|\tau_L-\tau_L'|} - e^{i(\omega^2-m^2)(\tau_L+\tau_L')} \right).
\end{equation}
Obviously this expression does not have a well defined value in the late Langevin-time limit. Introducing an explicit regulator of the form $i\epsilon x(t)^2$ yields
\begin{equation}\label{eq:ActionWithRegulator}
    S_M = \int dx \frac12 \left\{\partial_0 \f(x) \partial_0 \f(x) - (m^2 - i\epsilon)\f^2(x) \right\}
\end{equation}
and improves the situation, as now the stochastic process correctly converges to 
\begin{equation}
   \lim_{\tau_L\to\infty} \langle x(\omega,\tau_L)x(\omega',\tau_L) \rangle = \delta(\omega+\omega') \frac{i}{\omega^2 - m^2+i\epsilon}.
\end{equation}
A careful analysis of the associated Fokker-Planck equation in ref.~\cite{Nakazato:1986zy} however reveals that the relaxation time towards the correct solution scales with $1/\epsilon$. I.e. carrying out a CL simulation based on a small regulator $\epsilon$ will lead to slow convergence. In addition one also needs to take the limit $\epsilon\rightarrow 0$ as well as $\Delta \lt\rightarrow 0$, which may not commute \cite{Huffel:1985ma}.

Since the action of the harmonic oscillator in Fourier space decouples into a collection of non-interacting modes we may deploy a similar strategy for each mode as we considered in the simple model of the preceding section. I.e. we introduce a kernel, which moves the integration onto the single thimble for each mode. 
\begin{align}
    &\frac{\partial}{\partial \lt} x(\omega,\tau_L) = i \tilde K(\omega) \;  \frac{\delta S_M[x]}{\delta x(\omega)} + \sqrt{\tilde K(\omega)}\xi(\omega,\tau_L),\\
    &\langle \xi(\omega,\tau_L)\rangle =0,\quad \langle \xi(\omega,\tau_L)\xi(\omega',\lt')\rangle
=2\delta(\omega+\omega')\delta(\tau_L-\tau_L').
\end{align}

This train of thought leads us to choose the following field-independent kernel, which had been explored in ref.~\cite{Okamoto:1988ru} before
\begin{equation}\label{eq:Free_theory_momentum_kernel}
  \tilde K(\omega) = \frac{iA(\omega)}{\omega^2 - m^2 + i\epsilon}, \quad   K(t) = \int \frac{d\omega}{(2\pi)}\tilde K(\omega)e^{-i\omega t}, \quad 
\end{equation}
where $A(\omega)$ is a real, positive and even function of $\omega$. Thus for a constant $A(\omega)$, $\tilde K(\omega)$ is nothing but the propagator of the free theory in momentum space.

The corresponding correlation function is found to read 
\begin{equation}
   \lim_{\lt\to\infty} \langle \phi(\omega,\lt)\phi(\omega',\lt) \rangle = \delta(\omega+\omega')\frac{\tilde K(\omega)}{A(\omega)} = \delta(\omega+\omega')\frac{i}{\omega^2 - m^2 + i\epsilon}
\end{equation}
which is the correct result. The most important difference to simply introducing a regulator in the action however lies in the fact that now the relaxation time for each mode is proportional to $1/A(\omega)$ and not proportional to $1/\epsilon$ and no extrapolation in $\epsilon$ needs to be carried out. For completeness let us note the corresponding coordinate space complex Langevin process
\begin{align}
    &\frac{\partial}{\partial \lt} x(t,\tau_L) = i\int dt' \; K(t-t') \;  \frac{\delta S_M[x]}{\delta x(t')} + \chi(t,\tau_L),\\
    &\chi(t,\tau_L)=\int d\omega e^{i\omega t} \sqrt{\tilde K (\omega)}\xi(\omega,\tau_L).
\end{align}

\subsection{A kernel for real-time Langevin on the thermal SK contour}
\label{sec:freeTheoryKernel_FreeTheory}
The analytic study of the one d.o.f. model and the harmonic oscillator have provided us with insight into how a kernel can be used to both satisfy the need for regularization of the path integral and achieve convergence to the correct solution of the associated complex Langevin equation in practice.

We will now construct the corresponding kernel for the harmonic oscillator at finite temperature, discretized on the Schwinger-Keldysh contour. Numerical simulations will confirm the effectiveness of the kernel in the non-interacting theory.

\begin{figure}
\includegraphics[scale=0.4,trim=0 -2.5cm 0 0 ]{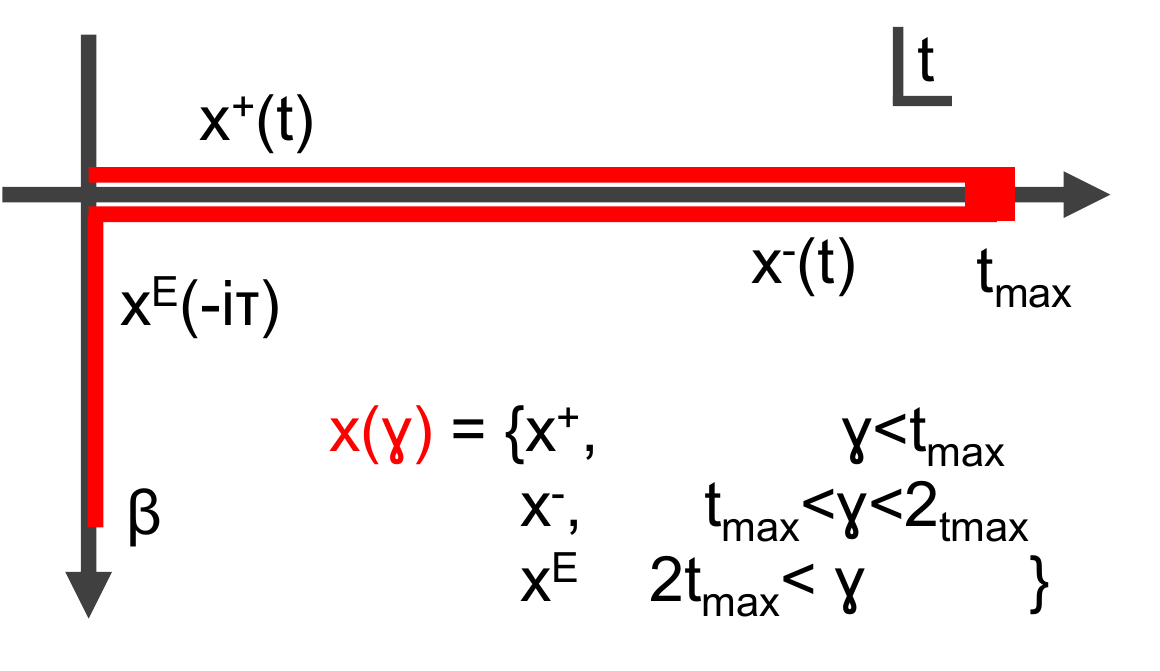}
\includegraphics[scale=0.35]{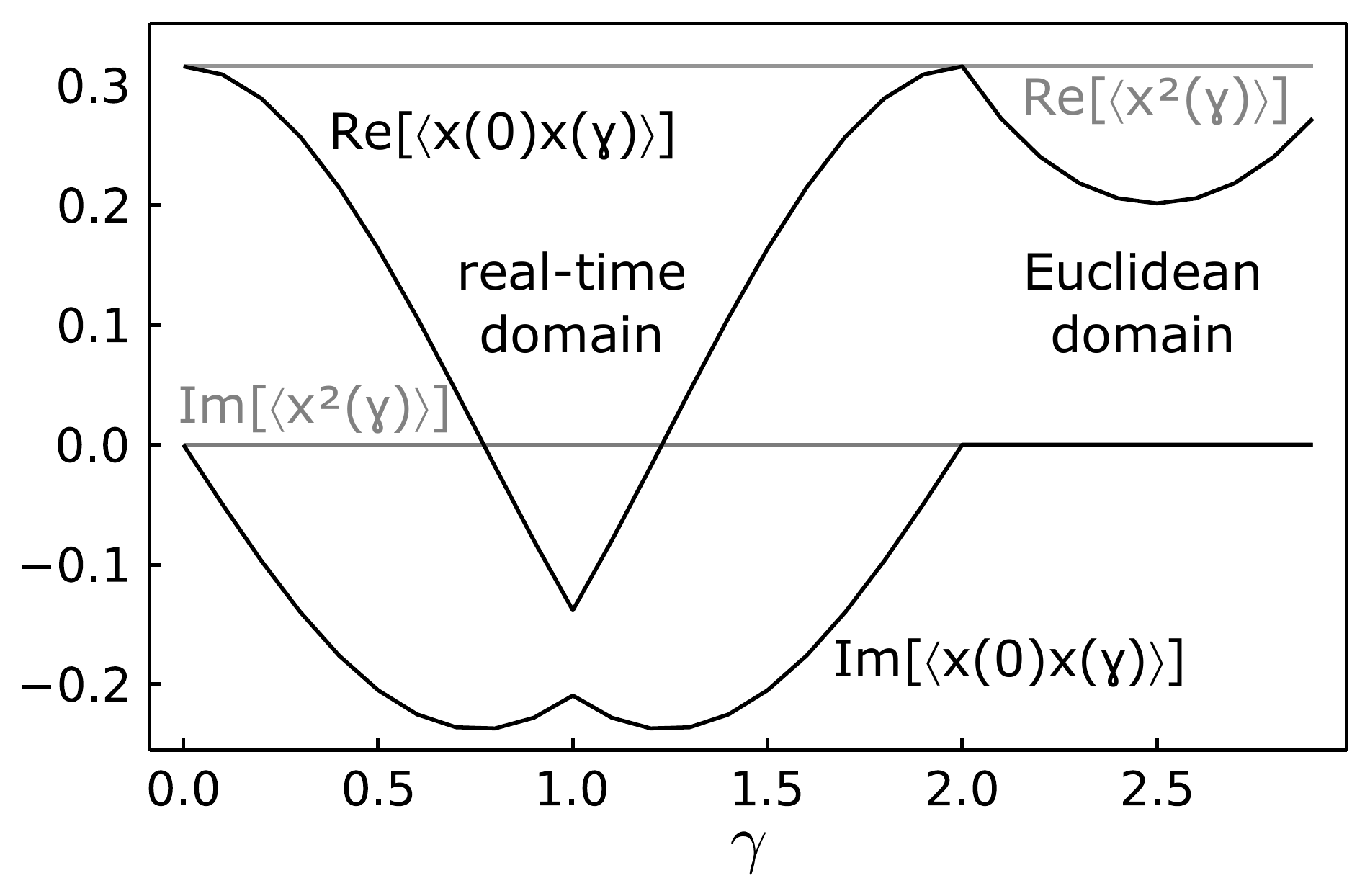}
\caption{(left) Sketch of the Schwinger-Keldysh contour deployed in our study with forward $x^+(t)$ and backward $x^-(t)$ branches on the real-time axis, connected to an imaginary time branch $x^E(-i\tau)$. The contour parameter $\gamma$ is used to address all branches in a unified manner. (right) Sketch of the visualization of our observables along the contour parameter $\gamma$ for the example of $m t_{\rm max}=1$. The analytic solution of the real- and imaginary part of the equal time correlator $\langle x^2(\gamma) \rangle$ and the unequal time correlator $\langle x(0)x(\gamma)\rangle$ will be plotted in the real-time $\gamma<2 m t_{\rm max}$ and subsequently Euclidean domain $\gamma> 2 m t_{\rm max}$.
}\label{fig:sketchSKObs}
\end{figure}

The Schwinger-Keldysh contour for a quantum system at finite temperature encompasses three branches. The forward branch along the conventional time axis reaches up to a real-time $t_{\rm max}$ and the degrees of freedom associated with it are labeled $x^+(t)$. The backward branch with $x^-(t)$ returns to the initial time $t_0$ in reverse and the Euclidean branch which houses $x^E(-i\tau)$ and extends along the negative imaginary time axis. The physical length of the imaginary time branch dictates the inverse temperature of the system. A sketch of our contour setup is shown in the left panel of \cref{fig:sketchSKObs}.

In the action of the system, the integration over time is rewritten into an integration over a common contour parameter $\gamma$. The d.o.f. on the different branches are then distinguished by the values of the contour parameter $x(\gamma)$ and we will drop the superscript in the remainder of the text.

As sketched in the right panel of \cref{fig:sketchSKObs}, we will refer to the equal- and unequal-time two-point correlation functions along the SK contour in the following, plotted against the contour parameter. The reader can identify the values along the forward and backward branch as being mirrored, connecting to the values on the Euclidean branch that show the expected periodicity of a thermal theory.

When discretizing the action for use in a numerical simulation the direction of each branch of the SK contour is encoded in a contour spacing $a_i\in\mathbb{C}$. Computing the drift term for an arbitrary contour yields
\begin{align}
        \label{eq:discretized_action_derivaitve}
    i\frac{\partial S_M[x]}{\partial x_j} =  \frac{i}{\frac{1}{2}\left(|a_{j}| + |a_{j-1}|\right)}\Big\{ &\frac{x_j - x_{j-1}}{a_{j-1}}- \frac{x_{j+1} - x_j}{a_j}
    - \frac12 \left[a_{j-1} + a_j\right] \frac{\partial V(x_j)}{\partial x_j} \Big\}.
\end{align}
This expression simplifies if we use a constant magnitude step-size $|a_i|=|a|$, such that the prefactor in the above equation can be reduced to $\frac{i}{|a|}$. In that case we can go over to a convenient matrix-vector notation
\begin{equation}
    i \nabla_x S_M[{\bm x}] = \frac{1}{|a|} \; iM {\bm x},
\end{equation}
where 
\begin{equation}\label{eq:MassMatrix}
    M_{jk} = 
\begin{cases}    
\frac{1}{a_{j-1}} + \frac{1}{a_j} - \frac12 \left[a_{j-1} + a_j\right]m^2,& j=k\\
 -\frac{1}{a_j},& j = k-1 \\
-\frac{1}{a_{j-1}},& j = k+1.
\end{cases}
\end{equation}

Based on the findings in the previous sections the form of the optimal discrete free theory kernel in coordinate space will be the inverse propagator
\begin{equation}\label{eq:freeTeoryPropKernel_discrete}
    K = H \; H^T = iM^{-1},
\end{equation}
where $H$ is the factorized kernel used in the noise term. The form of this kernel relies on the matrix $M$ to be invertible, and $M^{-1}$ to be factorizable, both of which holds. We obtain $H=\sqrt{iM^{-1}}$ by using the square root of the eigenvalues. 
Written in differential form with Wiener processes $d{\bm W}$, the corresponding Langevin equation reads
\begin{equation}
    d{\bm x} = \frac{1}{|a|}\left( \frac{i}{M}iM{\bm x} \right)d\tau_L + \sqrt{2 \frac{i}{M}}d{\bm W} = -\frac{1}{|a|}{\bm x} d\tau_L + \sqrt{2 \frac{i}{M}}d{\bm W},
\end{equation}
which leaves us with a complex non-diagonal noise coefficient $\sqrt{\frac{2i}{M}}$ and a drift term pointing in the direction of $-{\bm x}$.

\begin{figure}\centering
    \includegraphics[scale=0.4]{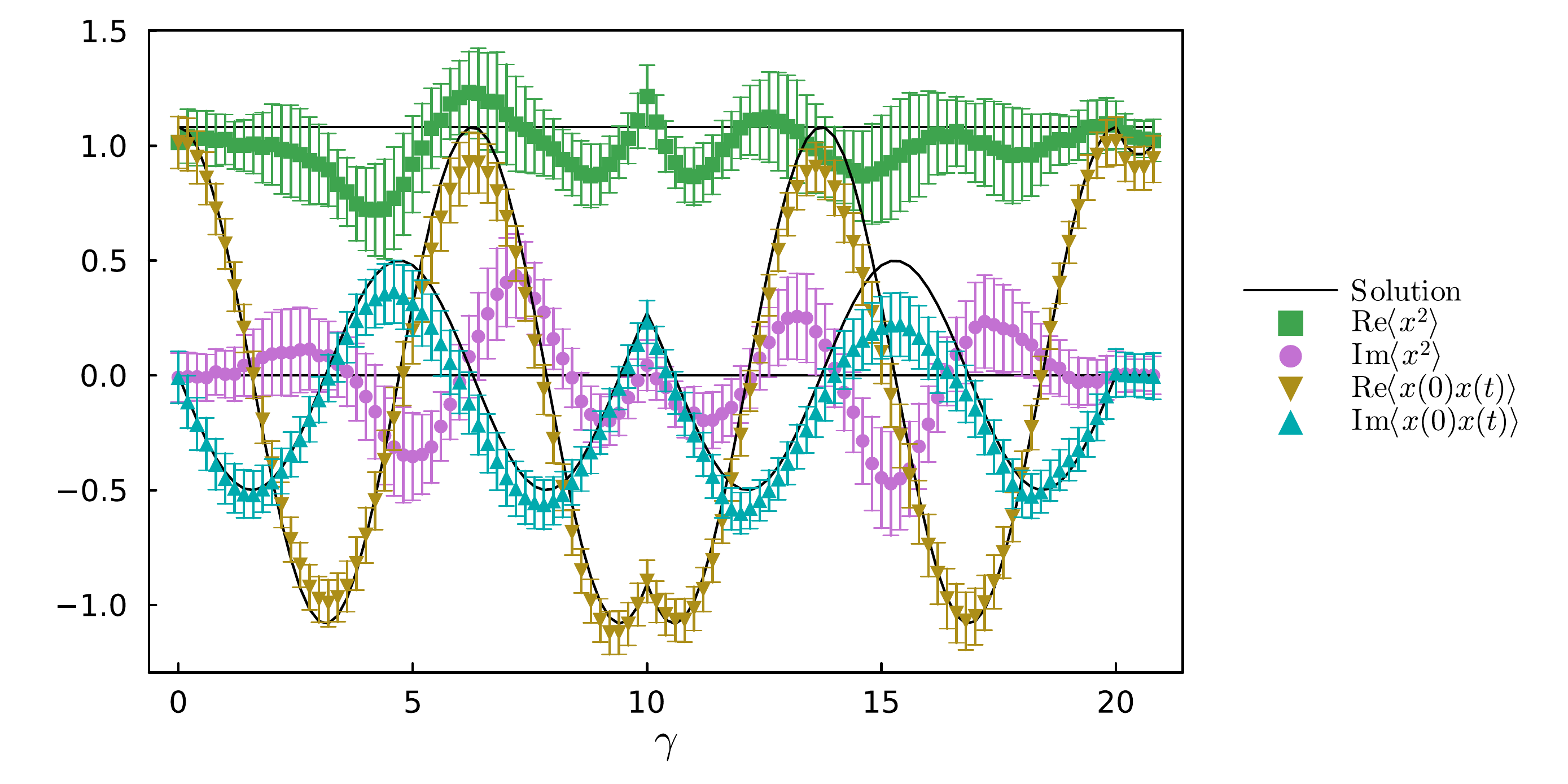}
    \includegraphics[scale=0.4]{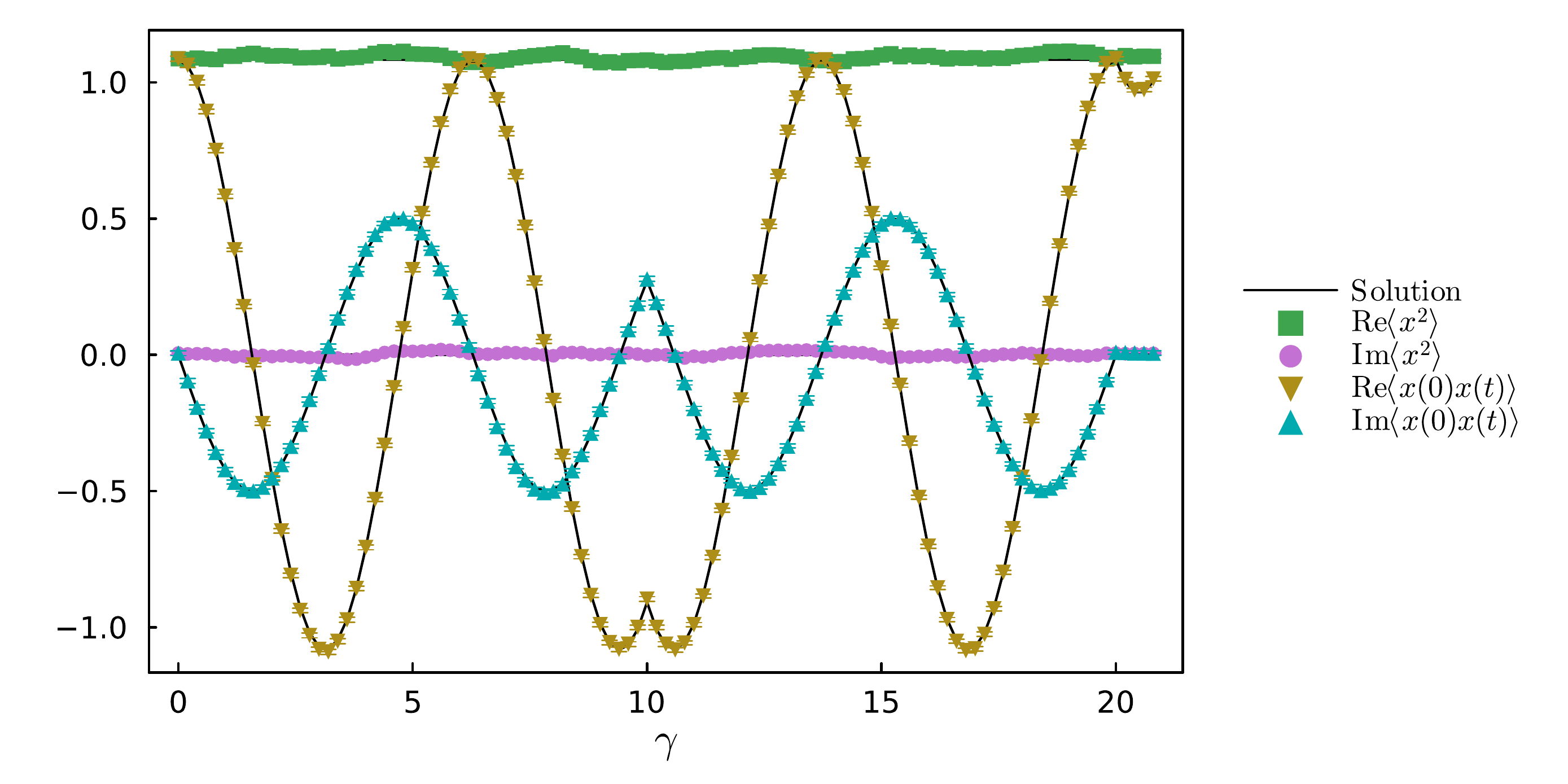}
    \caption{The result of a complex Langevin simulation of the real-time harmonic oscillator without a kernel (top) and with a kernel (bottom) for the observables $\langle x^2\rangle$, and the correlator $\langle x(0)x(t) \rangle$. Simulations are carried out with 100 trajectories and up to $m\tau_L=100$ in Langevin time, saving at every $m\Delta_{\tau_L}=0.1$ using the implicit scheme Euler-Maruyama with $\theta=1.0$ with adaptive step-size with tolerance $10^{-3}$ (top) and the explicit scheme Euler-Maruyama with $\theta=0$ with fixed step-size $\Delta \lt=10^{-3}$ (bottom). Values of the correlators from the solution of the Schr\"odinger equation are given as solid black lines.  }
    \label{fig:HO_freeKernel}
\end{figure}

Let us demonstrate the effect of this kernel by carrying out a simulation for the following parameters. We discretize the canonical SK contour with $N_t=50$ points on the forward and backward branch each and $N_\tau=5$ points on the imaginary branch. Note that we do not introduce any tilt here. Choosing a mass parameter $m=1$, the imaginary branch extends up to $m\tau_{\rm max}=1$. As real-time extent, we choose $mt_{\rm max}=10$. The value chosen here is arbitrary as the kernelled dynamics of the free theory are stable and converge for any real-time extent. The results of the simulation without a kernel are given in the top panel of \cref{fig:HO_freeKernel} and rely on the implicit Euler-Maruyama scheme to avoid the occurrence of runaway solutions. The results with our choice of kernel are shown in the bottom panel and were obtained using a simple forward-stepping Euler scheme at $\Delta \lt=10^{-3}$ without adaptive step size. In each case we generate $100$ different trajectories, saving configurations at every $m\Delta_{\tau_L}=0.1$ in Langevin time up to a total of $m\tau_L=100$. 
Each panel in \cref{fig:HO_freeKernel} showcases four quantities plotted against the contour parameter $\gamma$. Their values for $0<m\gamma<10$ are obtained on the forward branch, those for $10<m\gamma<20$ on the backward branch and the small piece  $20<m\gamma<21$ denotes the Euclidean time results. The real- and imaginary part of the equal time expectation value $\langle x^2(\gamma)\rangle$ are plotted as green and pink data points respectively. The real- and imaginary-part of the unequal time correlator $\langle x(0)x(\gamma)\rangle$ on the other hand are plotted as orange and blue data points. The analytically known values  from solving the Schr\"odinger equation are underlaid as black solid lines.

The results without a kernel show both deviations from the correct result and exhibit relatively large uncertainties. The reason lies in the slow relaxation rate to the correct result due to the presence of an explicit regulator. Here the regulator is provided by our use of an implicit numerical scheme (Euler-Maruyama with implicitness parameter $\theta=1.0$ and an adaptive step-size with a maximum step size of $10^{-3}$), but could equally well be introduced by adding a small term $i\epsilon x^2$ to the system action. Using a stronger regulator, e.g., tilting the contour, would yield a shorter relaxation time, but any such explicit regulator distorts the results away from the actual $\epsilon\rightarrow0$ physical solution. It is interesting to note that it is the equal time observable $\langle x^2 \rangle$ that is performing the worst. We will see later on that this is the hardest observable to accurately reproduce. 

For the bottom plot we use the free theory propagator kernel, \cref{eq:freeTeoryPropKernel_discrete}. This simulation now aligns excellently with the true solution for all the observables. 

After applying the kernel, the problem is regularized and thus less stiff, and we can revert to using a fixed step-size explicit Euler-Maruyama scheme. The step-size here is $d\lt=10^{-3}$. The fact that we do not need to impose an explicit regulator term is important as in this case we only need to take the limit $\Delta\lt\to0$ to obtain a physical result, and do not need to extrapolate the regulator term to zero ($\epsilon\rightarrow 0$). This might be important considering the recent work in ref.~\cite{Matsumoto:2022ccq}, which shows that one encounters subtleties in taking the limit of the regulator $\epsilon \rightarrow 0$. 

\subsection{A kernel for the quantum anharmonic oscillator}
\label{sec:FreeTheoryKernelInteractive}

Not only did the kernel in the free theory change the convergence behavior of the complex Langevin simulation, it also removed the need of a regulator in the action. The obvious next step is to explore the interacting theory where the problem of convergence to the wrong solution is more severe. The potential term in the action is now given by
\begin{equation}
    V(x) = \frac{1}{2}m x(t)^2 + \frac{\lambda}{4!}x(t)^4,
\end{equation}
where we use $m=1$ and $\lambda=24$. This choice of parameters has been deployed in the past as benchmark for strongly-interacting real-time complex Langevin in refs.~\cite{BergesSexty2007,Alvestad:2021hsi}. As  ref.~\cite{BergesSexty2007} formulated the real-time dynamics on a tilted contour they found correct convergence up to $t^{\textrm{max}}=0.8$, while ref.~\cite{Alvestad:2021hsi} worked with an untilted contour and observed onset of incorrect convergence already above $t^{\textrm{max}}=0.5$. In the following we will remain with an untilted contour.

We find that using the free kernel of \cref{eq:freeTeoryPropKernel_discrete} the convergence to the correct solution can be extended slightly to around $t^{\textrm{max}}=0.75$. If in addition we modify the free theory kernel by rescaling the contributions from the kinetic term with a common prefactor $g$ and modify the mass term away from the free theory value $m$
\begin{equation}\label{eq:MassMatrixMod}
    M_{jk}(g,m_g) = 
\begin{cases}    
\frac{g}{a_{j-1}} + \frac{g}{a_j} - \frac12 \left[a_{j-1} + a_j\right]m_g^2,& j=k\\
 -\frac{g}{a_j},& j = k-1 \\
-\frac{g}{a_{j-1}},& j = k+1.
\end{cases}
\end{equation}
convergence can be pushed up to $t_{\textrm{max}}=1.0$ by using the heuristically determined parameter values $g=0.8$ and $m_g=1.8$. The CL equation we simulate is given by
\begin{equation}\label{eq:freeTheoryPropagatorKernel_CLE}
    d{\bm x} = \frac{1}{|a|}\left[ \frac{i}{M(g,m_g)} i\left(M(1,m){\bm x} + \frac{\lambda}{3}x^2{\bm x} \right) \right]d\tau_L + \sqrt{\frac{2i}{M(g,m_g)}}d{\bm W}.
\end{equation}

We carry out simulations, assigning $N_t=10$ points to the forward and backward branches each and $N_\tau=10$ points to the imaginary branch of the contour. Here we use the implicit Euler-Maruyama scheme with implicitness parameter $\theta=0.5$. Even though we do not need a regulator in the presence of the kernel, the system retains some of its stiffness in contrast to the free theory. The use of an explicit scheme with e.g. adaptive step size is possible, however we find it more efficient to rely on an implicit scheme, as it allows the use of much larger Langevin step sizes.

\begin{figure}\centering
    \includegraphics[scale=0.4]{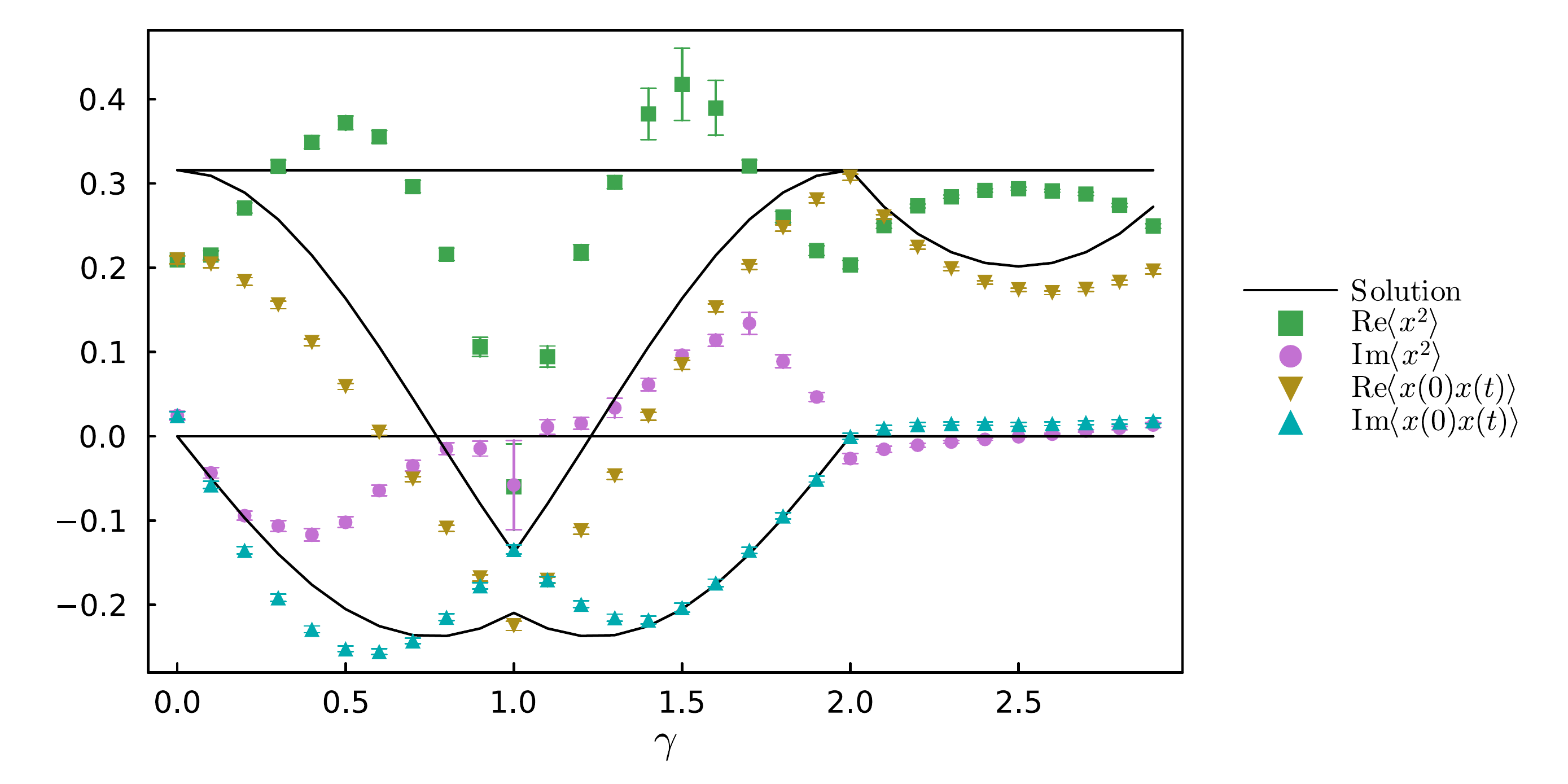}
    \includegraphics[scale=0.4]{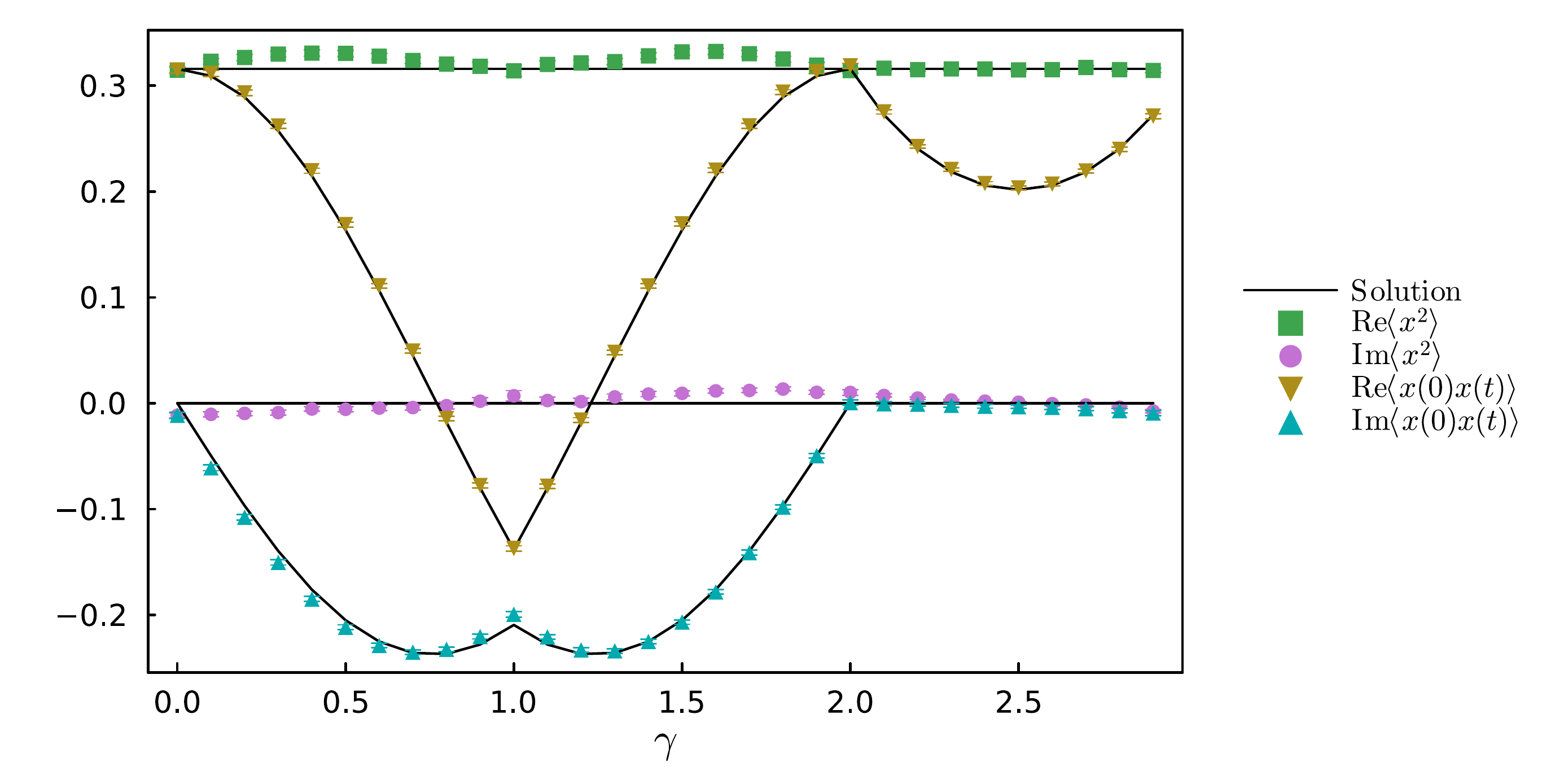}
    \caption{Anharmonic oscillator at $m=1$ and $\lambda=24$ up to $t_{\textrm{max}}=1$ in real-time without (top) and in the presence (bottom) of the heuristic free theory kernel with $g=0.8$ and $m_g=1.8$ from \cref{eq:freeTheoryPropagatorKernel_CLE}. Both simulations are carried out generating 100 trajectories simulated up to $m\tau_L=100$ in Langevin time, saving configurations at every $m\Delta \tau_L=0.01$. We deploy the Euler-Maruyama solver with $\theta=1.0$ without kernel (top), and $\theta=0.5$ with kernel (bottom). Values of the correlators from the solution of the Schr\"odinger equation are given as solid black lines.}
    \label{fig:AHO_freeKernel}
\end{figure}

The results of two simulations with a maximum real-time extent $t_{\textrm{max}} = 1.0$ are shown in \cref{fig:AHO_freeKernel}. One is carried out without a kernel and using an implicit scheme (top) and the other in the presence of a kernel based on the parameters $g=0.8$ and $m_g=1.8$ (bottom). The graphs show the real and imaginary part of the equal time $\langle x(t)x(t) \rangle$ (green and pink data points) and unequal time correlator $\langle x(0)x(t) \rangle$ (orange and blue datapoints) plotted against the contour parameter $\gamma$. For $0<m\gamma<1$ and $1<m\gamma<2$ it refers to the forward and backward branch of the contour and for $2<m\gamma<2.9$ denotes the imaginary time branch. 

The top panel indicates that naive complex Langevin fails to converge to the correct solution at this real-time extent of $mt_{\textrm{max}}=1$. It is interesting to point out the failure of CL at two specific points along the contour, the first one is the starting point at $\gamma=0$, which is connected by periodic boundary condition to the Euclidean path. Then at the turning point of the contour at maximum real-time extent, corresponding to $m\gamma=1$, the real part of the $\langle x^2 \rangle$ observable lies significantly away from the true solution. This points seems to be most affected by the convergence problem of the CLE.

In the lower panel, the simulation in the presence of the modified free theory kernel is presented. The outcome of the kernelled complex Langevin evolution is very close to the correct solution and shows only small statistical uncertainties. Note however that especially the observable $\langle x^2 \rangle$ still shows some deviation from the true result beyond the statistical error bars indicating that exact correct convergence has not yet been achieved\footnote{This behavior may be understood in terms of boundary terms. The kernel manages to significantly reduce the magnitude of boundary terms for $\langle x^2 \rangle$ where it differs from the true solution the boundary terms, while small, are not exactly zero.}

The above results are promising, as they indicate that in principle the convergence problem of real-time complex Langevin can be attacked by use of a kernel. At the same time explicitly constructed kernels, such as the modified free theory kernel are limited in the range of real-time extent in which they are effective. The question at hand is how to systematically construct kernels that will restore convergence at even larger real-time extent.

\section{Learning optimal kernels}
\label{sec:optimal_kernels}

In this section, we introduce our novel strategy to systematically construct kernels to improve the convergence of real-time complex Langevin. Our goal is to overcome the limitations of explicitly parametrized kernels, such as the one of \cref{eq:MassMatrixMod}. While optimal parameter values $g$ and $m_g$ were found for this kernel, they only achieved correct convergence for a limited $mt_{\rm max}\leq1$. Most importantly it is not clear how to systematically modify that kernel for realizing convergence at larger real-time extent. 

Instead we set out to use a generic parametrization of the kernel. We propose to use an expansion in a set of complete basis functions of the dynamical d.o.f. In this study, as a proof of principle, we will restrict ourselves to a field-independent kernel, which can be understood as the first term in an expansion in powers of the field. This field-independent kernel for the quantum anharmonic oscillator on the Schwinger-Keldysh contour will take the form of a $\tau_L$ independent matrix $K$ with $(2N_t+N_\tau)^2$ entries, multiplying the $2N_t$ d.o.f. on the forward and backward contour and the $N_\tau$ ones on the imaginary time branch. It is the values of these matrix entries that we set out to tune in order to achieve optimal convergence. 

And even though simple model systems indicate that a field-dependent kernel is needed to achieve correct convergence in case of strong complex drift terms, we find that an optimal field-independent kernel can already extend the range of convergence of the anharmonic oscillator out to  $mt_{\rm max} = 1.5$, three times larger than the previous record set for CL in ref.~\cite{BergesSexty2007}.

In order to obtain kernel values that restore correct convergence, we formulate an optimization problem based on a cost functional, which incorporates prior knowledge about the system of interest. Taking advantage of modern programming techniques that allow us to compute the dependence of a full complex-Langevin simulation on the entries of the kernel we propose to iteratively learn the optimal kernel. The fact that we incorporate prior information into the simulation opens a novel path to beat the notorious sign problem, i.e. for the first time complex Langevin can be amended by system-specific information in order to restore correct convergence.

\subsection{The optimization functional}
\label{sec:outlineOfStrategy}

In order to guarantee that a complex Langevin simulation converges to the true solution we must fulfill the correctness criteria of \cite{Aarts:2011ax}. First we must ensure the absence of boundary terms and second that the late-time distribution of the complex Fokker-Planck equation is indeed ${\rm exp[i S_M]}$. Constructing a loss function for both criteria however is only feasible for very low dimensional models, as it entails calculating the eigenvalues and eigenvectors of the complex Fokker-Planck operator, which is prohibitively expensive already for the anharmonic oscillator discussed here.

Instead we will retain only the first ingredient of the correctness criteria, the absence of boundary terms and use other prior information in order to guide the kernelled complex FP equation to the correct stationary solution. The boundary terms can be calculated via the expectation value $\langle L_c \mathcal{O} \rangle_Y$ where $L_c$ is the Langevin operator and $\mathcal O$ refers to any observable (for a detailed discussion see e.g. \cite{Scherzer:2018hid}). In \cref{sec:correctnessCriterionWithAKernel} we demonstrate that the correctness criterion still holds with a kernel and how to calculate these boundary terms. 

Besides the boundary terms, we often possess additional relevant prior information about the system at hand. We can e.g. compute correlation functions in Euclidean time using conventional Monte-Carlo methods. In addition, we know that in thermal equilibrium the correlation functions on the forward and backward branch are related due to the KMS relation. In order to exploit this prior information it is vital for the CL equations to be formulated on the canonical SK contour, whose real-time branches lie parallel to each other and connect to the Euclidean branch at the origin. In a tilted contour setup, access to the Euclidean branch is limited and the comparison of the values on the forward and backward branch is much more involved. In addition, symmetries provide powerful constraints to the simulation, as e.g. time-translation invariance in a thermal system renders local observables such as $\langle x^n(\gamma) \rangle$ constant along the full contour.

We quantify the distance of the simulated result from the behavior dictated by prior knowledge via a loss function $L^{\rm prior}$. The comparison is carried out on the level of expectation values of observables, where apriori known values from conventional Euclidean simulations are referred to as $\langle {\cal O}\rangle_{\rm MC}$ and those from the complex Langevin simulation in the presence of a kernel by $\langle {\cal O}\rangle_K$. 

In principle one can distinguish between four categories of prior knowledge: 
\begin{itemize}
    \item Euclidean correlators ($L^{\textrm{eucl}}$), which are accessible via conventional Monte-Carlo simulations: $$L^{\textrm{eucl}} = \sum_{\cal O} \int d\tau \Big| \left\langle \mathcal O(\tau) \right\rangle_K - \left\langle \mathcal O(\tau) \right\rangle_{\rm MC} \Big|^2/\sigma^2_{\langle {\cal O}(\tau)\rangle_K} \quad \tau \in \textrm{imaginary time}$$
    \item Model symmetries ($L^{\textrm{sym}}$), which exploit that the expectation values of observables ${\cal O}$ must remain invariant under a symmetry transformation $T_\xi$ governed by a continuous (or discrete) parameter $\xi$: $$L^{\textrm{sym}} = \sum_{{\cal O}}\int d\xi |\langle T_\xi{\cal O} \rangle_K - \langle {\cal O}\rangle_K |^2/\sigma^2_{\langle{\cal O}(\tau)\rangle_K}$$
    \item Contour symmetries ($L^{\textrm{rt}}$), which arise predominantly in systems in thermal equilibrium:
    $$L^{{\cal C}} = \sum_{\cal O} \int d\gamma \Big| \langle \mathcal O^{\cal C^+}(\gamma) \rangle_K -F[ \langle \mathcal O^{\cal C^-}(\gamma) \rangle_K] \Big|^2/\sigma^2_{\langle{\cal O}(\tau)\rangle_K} \quad {\rm with\,F\,analytically\,known}\label{eq:lossfunctions_rt_general}$$
    \item Boundary terms ($L^{\textrm{BT}}$), which can be explicitly computed from the outcome of the kernelled Langevin simulation: $$L^{\textrm{BT}} = \sum_{\cal O} \Big|\langle L_c(K) \mathcal O \rangle_K\Big|^2/\sigma^2_{\langle{\cal O}(\tau)\rangle_K} \label{eq:lossfunctions_bv}$$
\end{itemize}

In practice one wishes to combine as many of these different contributions as possible. To this end they must be added as dimensionless quantities. This is why each of the terms above is normalized by the variance of the complex Langevin simulation. In order for the combined functional to provide a meaningful distinction of the success of convergence (also in the case of e.g. the free theory as shown in \cref{fig:HO_freeKernel}) we propose to introduce an overall normalization for the combined prior functional 
\begin{align}
    L^{\rm prior}={\cal N}_{\rm tot} \Big( L^{\textrm{eucl}} +  L^{\textrm{sym}} + L^{{\cal C}} +L^{\textrm{BT}} \Big).
\end{align}
There is an element of arbitrariness in what overall normalization to choose, and we find that the best distinction between wrong and correct convergence is achieved if one uses the relative error of the most difficult observable to reproduce. In case of the systems studied here this amounts to the relative error of the equal time correlator obtained in the complex Langevin simulation with respect to the correct known value from Euclidean simulations ${\cal N}_{\rm tot}={\rm max}_\gamma \{\sigma_{\langle x^2\rangle_K}(\gamma)/ \langle x^2\rangle_{\rm MC}(\gamma) \} $. 

We carry the subscript $K$ in the expectation values above, in order to emphasize that the loss functional depends implicitly on the choice of kernel used in the underlying complex Langevin simulation. The number of observables ${\cal O}$ contained in the cost functional is not specified here and depends on the problem at hand. In practice, we find that often including the apriori known Euclidean one- and two-point functions already allow us to reliably distinguish between correct and incorrect convergence.

In the next section, we will discuss both fully general numerical strategies to locate the minimum of the optimization functional, as well as an approximate low-cost approach, which we have deployed in the present study.

\subsection{Optimization strategies\label{sec:updateKernel}}

\subsubsection{General approach}

The task at hand is to find the critical point of a cost functional that is comprised of a subset of the contributions listed in the previous section, i.e. of $L^{\rm eucl}$,$L^{\rm sym}$,$L^{\cal C}$ or $L^{\rm BT}$. Generically each contribution can be written as the expectation value of a known function $G$, depending on the dynamical degrees of freedom $x$ and the kernel $K$, i.e. $L^{\rm prior}[K] = \left| \left\langle G[x,K] \right\rangle_K \right|^2$. In order to make the dependence of the expectation value on the kernel explicit we consider the d.o.f. within the simulation to explicitly depend on $K$ as $x(K)$. This allows us to remove the subscript $K$ from the expectation value so that $L^{\rm prior}[K] = \left| \left\langle G[x(K),K] \right\rangle \right|^2$.

Let us characterize the kernel via a set of variables $\kappa$. We emphasize that this does not limit the general nature of the approach, as $\kappa$ may refer to the prefactors of a general expansion of the kernel in a complete set of basis functions.

To efficiently locate its critical point we deploy standard numerical optimization algorithms, which utilize the information of the gradient of the functional with respect to the parameters of the kernel. The computational challenge lies in determining the gradient robustly. In the continuum, the gradient of the loss reads
\begin{align}
    \nabla_\kappa L^{\rm prior}[K] =& 2\frac{\langle G[x(K),K] \rangle}{\left|\langle G[x(K),K] \rangle \right|}  \left\langle \nabla_\kappa G[x(K),K] \right\rangle \\ 
     =& 2\frac{\langle G[x(K),K] \rangle}{\left|\langle G[x(K),K] \rangle \right|} \left\{ \left\langle \nabla_x G[x(K),K] \cdot \nabla_\kappa {\bf x} \right\rangle + \left\langle \nabla_\kappa G[x,K]\right\rangle \right\}
    \label{eq:gradientLoss_b}
\end{align}

In order to evaluate \cref{eq:gradientLoss_b} we need to compute the change in the field $x(K)$, which depends on the kernel. This requires taking the gradient of the CL simulation itself. While a demanding task, dedicated methods to evaluate such gradients have been developed, which underpin the recent progress in the machine learning community. They are known as differential programming techniques (for an in-depth review see e.g. ref.~\cite{baydin2018automatic}).

As a first option, we considered using direct auto-differentiation\footnote{Auto-differentiation is a method to compute derivatives to machine precision on digital computers based on an efficient use of the chain rule, exploiting elementary arithmetic operations in the form of dual variables (see e.g.\cite{harrison2021brief}). We have used the Julia library \textit{Zygote.jl}\cite{Zygote.jl-2018} and \textit{ForwardDiff.jl} for computing gradients.} on the full loss function, as we are dealing with the standard setting of estimating the gradient of a highly dimensional functional whose output is a single number. For small systems with a number of degrees of freedom ${\cal O}(10)$, forward-auto-differentiation is feasible as it requires multiple runs of the full CL simulation. As the number of independent d.o.f. grows, backward-auto-differentiation offers us to reduce the number of necessary simulation runs, trading computational cost for increased memory demands to store intermediate results of the chain rule it computes internally. We find that already for the quantum anharmonic oscillator this direct computation of the gradient is too costly and thus not practical.

A more advanced approach, which promises to avoid the cost and memory limitations of direct auto-differentiation are so-called sensitivity analysis methods, such as e.g. adjoint methods for stochastic differential equations. A detailed discussion of these methods is beyond the scope of this paper and the interested reader is referred to refs.~\cite{cao2003adjoint,schafer2020differentiable,rackauckas2020universal} for further details.

We find that for the specific case of real-time complex Langevin, these methods in their standard implementation, as provided e.g. in \cite{rackauckas2020universal} are challenged in estimating the gradient robustly. We believe that the difficulty here lies in the stiffness of the underlying stochastic differential equation. One possible way out is to deploy sensitivity analysis methods specifically developed for chaotic systems, such as \textit{Least Square Shadowing algorithms}, discussed e.g. in Refs.~\cite{wang2014least,Ni_2017}. While these methods at this point are still too slow to be deployed in CL simulations, the rapid development in this field over the past years is promising.

Our survey of differential programming techniques indicates that while possible in principle, the optimization of the loss functional $L(K)$ is currently plagued by issues of computational efficiency. We believe that implementing by hand the adjoint method for the real-time complex Langevin systems considered here will offer a significant improvement in speed and robustness compared to the generic implementations on the market. This line of work goes beyond the scope of this manuscript and will be considered in an upcoming study.

To make headway in spite of these methods limitations we in the following propose an approach to compute an approximate low-cost gradient, which in practice allows us to significantly reduce the values of the optimization functional.

\subsubsection{A low cost update from an heuristic gradient}

Our goal is to compute a gradient, which allows us to approximately minimize the cost functional $L^{\rm prior}[K]$ without the need to take derivatives of the CL simulation. The approach we propose here relies on using a different optimization functional, whose form is motivated by the need to avoid boundary terms. While updating the values of the kernel according to a heuristic gradient obtained from this alternative functional, we will monitor the values of the true optimization functional, selecting the kernel which achieves the lowest value of $L^{\rm prior}[K]$.

We saw that the optimal kernel for the free theory reduces the drift term to a term in the direction of $-x$. This drift term points towards the origin. In this spirit we construct a functional that penalizes drift away from the origin. 

The starting point is the following expression, where we define $D=-iK\partial S_M/\partial x$ as the drift term modified by the kernel
\begin{equation}
    D(x,K)\cdot(-x) = ||D(x,K)||||x||\cos{\theta}.
\end{equation}
Here ${\rm cos}\theta$ denotes the angle between the drift and the optimal direction. As we wish to align the drift and $-x$, our optimization problem becomes finding a kernel $K$ such that
\begin{equation}
    \min_K \left\{ D(x,K)\cdot(-x) - ||D(x,K)|| \; ||x|| \right\}.
\end{equation}
We can write down different loss functionals which encode this minimum 
\begin{equation}\label{eq:driftLoss}
    L_{D} =   \left\langle \Big| D(x) \cdot (-x) - ||D(x)||\;||x|| \Big|^\xi\right\rangle = \frac1T \int \Big| D(x(\lt)) \cdot (-x(\lt)) - ||D(x)||\;||x|| \Big|^\xi
\end{equation}
The choice of $\xi$ determines how steep the gradients on the functional are and we find that in practice a value between $1<\xi<2$ leads to most efficient minimization, when $L_D$ is used to construct the heuristic gradient we describe below.

Note that turning the drift towards the origin differs from the strategy employed by dynamic stabilization. The scalar counterpart to minimizing the unitarity norm is driving the values of the complexified $x$ towards the real axis. In addition, in dynamical stabilization a non-holomorphic term is added to the action. Here the CL equation is modified only by a kernel, which still leads to a holomorphic complex Langevin equation that leaves the correctness criteria intact.

The exact gradient of the functional $L_D$ of \cref{eq:driftLoss} also contains the costly derivatives over the whole CL simulation. However we find that in practice for values $1\leq\xi\leq2$ in $L_D$ these contributions can be neglected. We believe the reason to lie in the fact that $L_{D}$ consists of the difference between two terms that contain the same powers of $x$. I.e. we find that carrying out the optimization using only the explicit dependence of $L_{D}$ on the kernel $K$, which is computed using standard auto-differentiation. The approximate gradient allows us to locate kernel values, which significantly reduce the values of the true optimization functional $L^{\rm prior}[K]$. The kernels identified in this way in turn achieves correct convergence on contours with larger real-time extent than previously possible.\footnote{Note that disregarding the costly terms in the gradient of the true cost functional $L^{\rm prior}[K]$ introduced in \cref{sec:outlineOfStrategy} did not lead to a viable minimization of its values.}.

\noindent The full optimization scheme can be summarized as follows:
\begin{enumerate}
    \item Initialize the kernel parameters yielding the initial kernel $K_1$
    \item Carry out the CL simulation with $K_1$ and save the configurations $\{ x_j \}_1$, where the subscript indicates that this is the first iteration
    \item Compute the values of the loss functions $L_D$ and $L^{\rm prior}[K_1]$
    \item Compute the gradients of the loss function $L_D(\{ x_j \}_1,K_1)$ with respect to the kernel parameters using auto-differentiation
    \item Update the kernel parameters using one step of the ADAM optimization scheme
    \item Rerun the CL simulation with the new kernel $K_{i+1}$ and save a new set of configurations $\{ x_j \}_{i+1}$
    \item Loop over step 3 - 6 for $N$ steps, or until $L_D$ have reached a minimum and then select the kernel parameters with the smallest $L^{\rm prior}[K_i]$
\end{enumerate}

We will demonstrate the efficiency of the proposed optimization based on the heuristic gradient in the next section, where we learn optimal kernels for the quantum harmonic and anharmonic oscillator on the thermal Schwinger-Keldysh real-time contour.

\subsection{Learning optimal kernels for the thermal harmonic oscillator}
\label{sec:num_res}

To put the strategy laid out in the previous section to a test we set out here to learn a field-independent kernel for the quantum harmonic oscillator on the canonical Schwinger-Keldysh contour at finite temperature. In \cref{sec:freeTheoryKernel_FreeTheory} we had identified one kernel by hand, which actually minimizes the low-cost functional \cref{eq:driftLoss}. We will compare it to the learned kernel at the end of this section.

We simulate on the canonical Schwinger-Keldysh contour with real-time extent $mt_{\rm max}=10$ and an imaginary time branch of length $m\beta=1$. The contour will be discretized with steps of equal magnitude $|a_i|=|a|$ such that $N_t=25$ points are assigned to the forward and backward branch each and $N_\tau=5$ to the imaginary time axis.

 The field-independent kernel therefore is a complex $55\times55$ matrix, which we parametrize via two real matrices $A$ and $B$ such that $K = e^{A + iB}$. This choice is arbitrary and is based on the observation that the minimization procedure is more robust for the exponentiated matrices than when using $A+iB$ directly. The kernel is initialized to unity before the start of the optimization by setting all elements of $A$ and $B$ to zero. The optimization itself, as discussed in the previous section, is carried out using the approximate gradient following from $L_D$ with a choice of $\xi=2$.

One needs to choose the actual cost functional $L^{\rm prior}$ based on prior knowledge through which to monitor the optimization success. We decide to include the known values of the Euclidean two-point correlator $\langle x(0)x(-i\tau)\rangle$ and exploit the knowledge about the symmetries of the system, which require that $\langle x\rangle = \langle x^3\rangle=0$, as well as $\langle x^2(\gamma)\rangle=\langle x^2(0)\rangle$. This leads us to the following functional
\begin{align}
    &L^{\rm prior} = {\cal N}_{\rm tot} \Big(\label{eq:lossfunctions_sym}\\
    \nonumber&\sum_{i \in SK} \Big\{ |\langle x(\gamma_i) \rangle_K|^2/\sigma^2_{x_{\rm K}} + |\langle x^3(\gamma_i) \rangle_K|^2/\sigma^2_{x^3_{\rm K}}+ | \langle x^2(0)\rangle_{\rm MC}- \langle x^2(\gamma_i) \rangle_K|^2/\sigma^2_{x^2_{\rm K}} \Big\}\\
    \nonumber+&\sum_{i\in{\rm Eucl.}} | \langle x(0) x(\tau_i) \rangle_{\rm MC} - \langle x(0) x(\tau_i) \rangle_K|^2/\sigma^2_{xx_{\rm K}}\Big)
\end{align}
where the first sum runs over all points of the discretized Schwinger-Keldysh contour, while the second sum only contains the correlator on the Euclidean branch. As discussed before, the overall normalization is based on the uncertainty of the equal-time $x^2$ correlator.

Since we start from a trivial kernel, we must make sure that our simulation algorithm provides a regularization and remains stable even for stiff dynamics. Therefore we solve the complex Langevin stochastic differential equation using the \textit{Implicit Euler-Maruyama scheme} with implicitness parameter $\theta = 1.0$ and adaptive step-size. For every update of the CL configurations we simulate $30$ different trajectories up to a Langevin time of $m\tau_L=30$, with a thermalization regime of $m\tau_L=5$ in Langevin time before we start collecting the configurations at every $m\Delta_{\tau_L}=0.05$ in Langevin time. To calculate the expectation values, we compute sub-averages from the saved configurations in each trajectory separately. The final mean and variance are then estimated from the results of the different trajectories.

The iterative optimization of the kernel values, based on the low-cost functional and its approximate gradient, is performed using the ADAM (Adaptive Moment) optimizer with a learning rate of $0.001$. This is an improved \textit{gradient descent} optimizer, which combines gradient descent momentum and an adaptive learning rate.

Since we know that the complex Langevin simulation will be the slowest part of the optimization scheme we will only run the full CL simulation for every five optimization steps. For this simple model it would not be a computation time problem to update the expectation values in $L_D$ after every kernel update, but for realistic models in higher dimensions this might be too expensive. As the distributions of the observables should be similar for a small change in the kernel we indeed find that not updating the CL configurations at every update steps still allows us to obtain a good estimate of the heuristic gradient.

Starting with the unit kernel, the functionals $L_D=6.58\times 10^{11}$ and $L^{\rm prior}=107$ show appreciable deviation from zero. After 32 steps of the ADAM optimizer we manage to find values of $K$ which reduce the value of $L^{\rm prior}=26.5$ indicating that the apriori known information has been well recovered.

\begin{figure}\centering
    \includegraphics[scale=0.4]{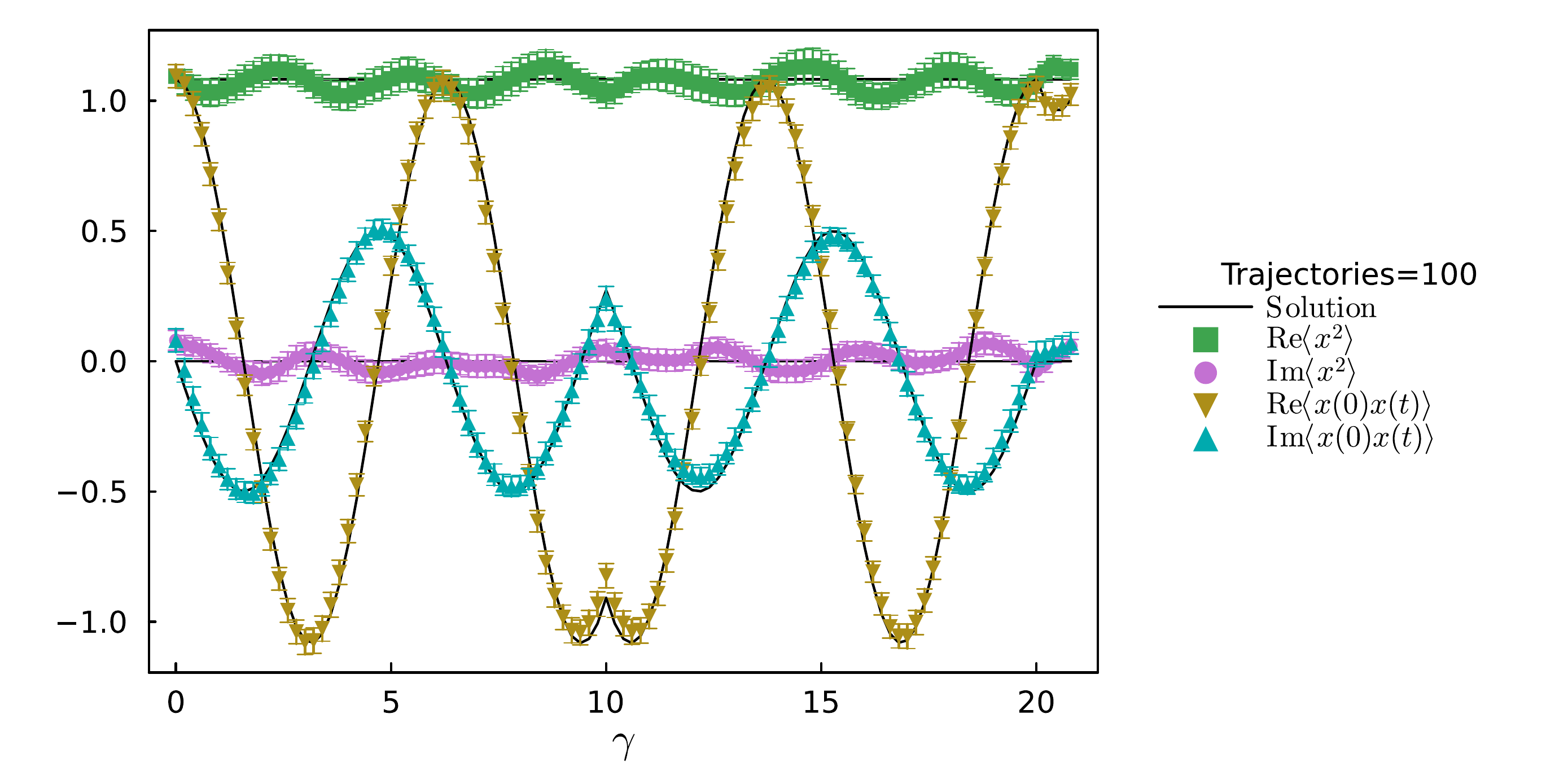}
    \includegraphics[scale=0.4]{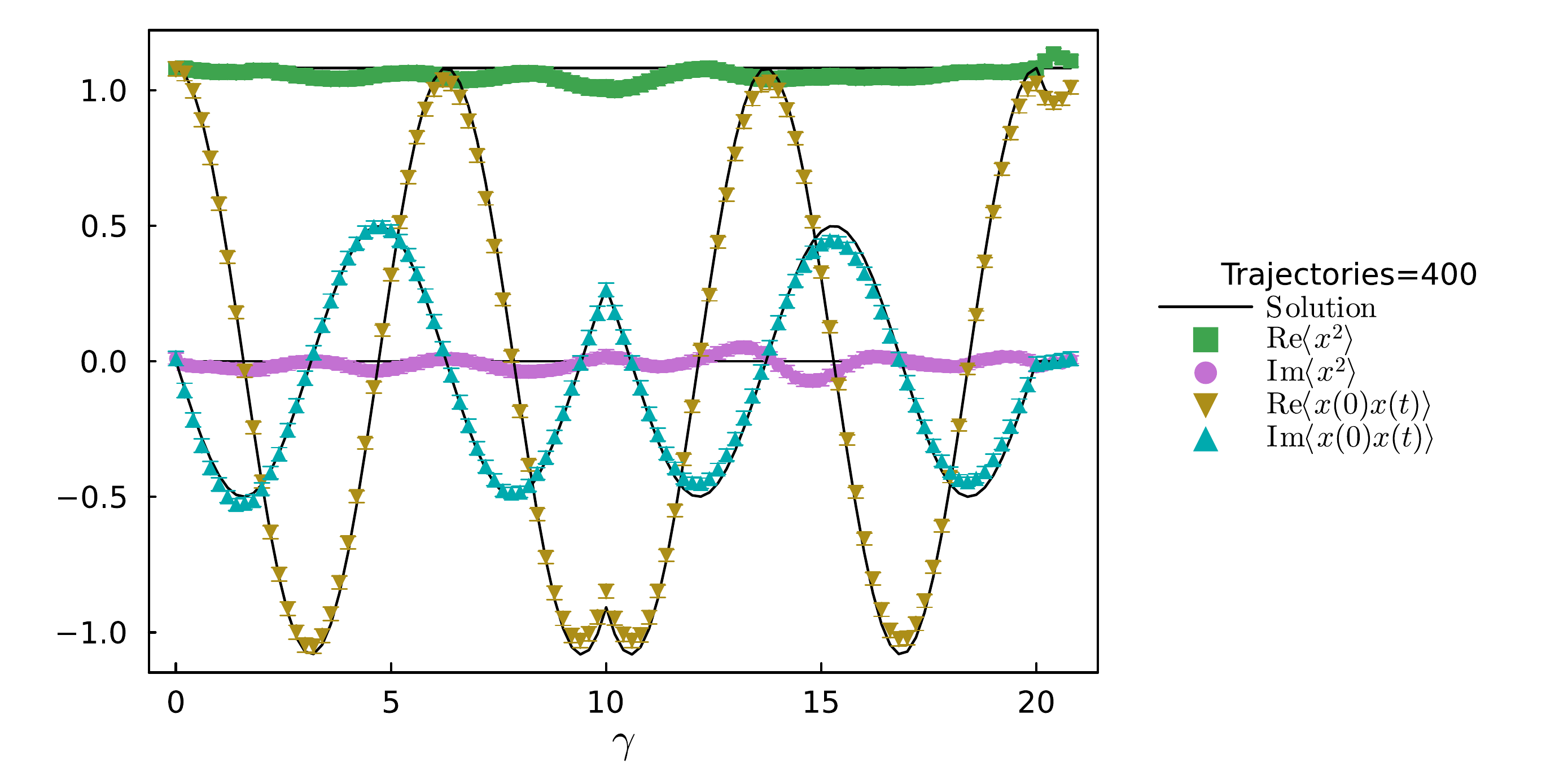}
    \caption{Harmonic oscillator in the presence of our learned optimal kernel, based on the heuristic gradient from the loss function in \cref{eq:driftLoss}. We simulate on the SK contour with $mt_{\rm max}=10$ in real-time and choose $m=1$. Simulating up to $m\tau_L=100$ in Langevin time, we combine samples from 100 (top) and 400 (bottom) different trajectories. Note that improved statistics diminishes the residual oscillatory artifacts in $\langle x^2\rangle$. Values of the correlators from the solution of the Schr\"odinger equation are given as solid black lines. }
    \label{fig:ResultsForFreeTheoryLearnedKernel}
\end{figure}

The results for the simulation with the optimal learned kernel are plotted in \cref{fig:ResultsForFreeTheoryLearnedKernel}, based on $100$ trajectories (top) and $400$ trajectories (bottom) each of which progresses up to $m\tau_L=100$ in Langevin time . The x-axis refers to the contour parameter $\gamma$, such that at $m \gamma=10$ we are at the turning point of the real-time branch of the Schwinger-Keldysh contour and at $m \gamma=20$ the contour has returned to the origin, before extending along the imaginary axis to $mt=-i$. We plot the real- and imaginary part of the unequal time correlator $\langle x(0)x(\gamma)\rangle$ as orange and blue data points, while the real- and imaginary part of the equal time expectation value $\langle x^2(\gamma)\rangle$ are given in green and pink respectively. The analytically known values from solving the Schr\"odinger equation are underlaid as black solid lines.

How has the learned kernel improved the outcome? When comparing to a simulation without kernel in the top panel of \cref{fig:HO_freeKernel} we see that using the same amount of numerical resources (i.e. 100 trajectories at $m\tau_L=100$) the learned kernel has reduced the resulting errorbars significantly. On the other hand in the top panel of \cref{fig:ResultsForFreeTheoryLearnedKernel} residual oscillations in $\langle x^2 \rangle$ seem to persist. One may ask whether these indicate incorrect convergence, which is why we provide in the lower panel the result after including 400 trajectories at the same Langevin time extent. One can see that not only the errorbars further reduce but also that the oscillatory artifacts have diminished. The improvement amounts to another factor of two in terms of $L^{\rm prior}$ from the value $L^{\rm prior}=26.5$ in the top panel to $L^{\rm prior}=13.4$ in the lower panel. We emphasize that we \textit{did not} use the analytically known solution of the system for the optimization procedure. 

\begin{figure}\centering
     \includegraphics[scale=0.45, trim=0 0.3cm 0 0]{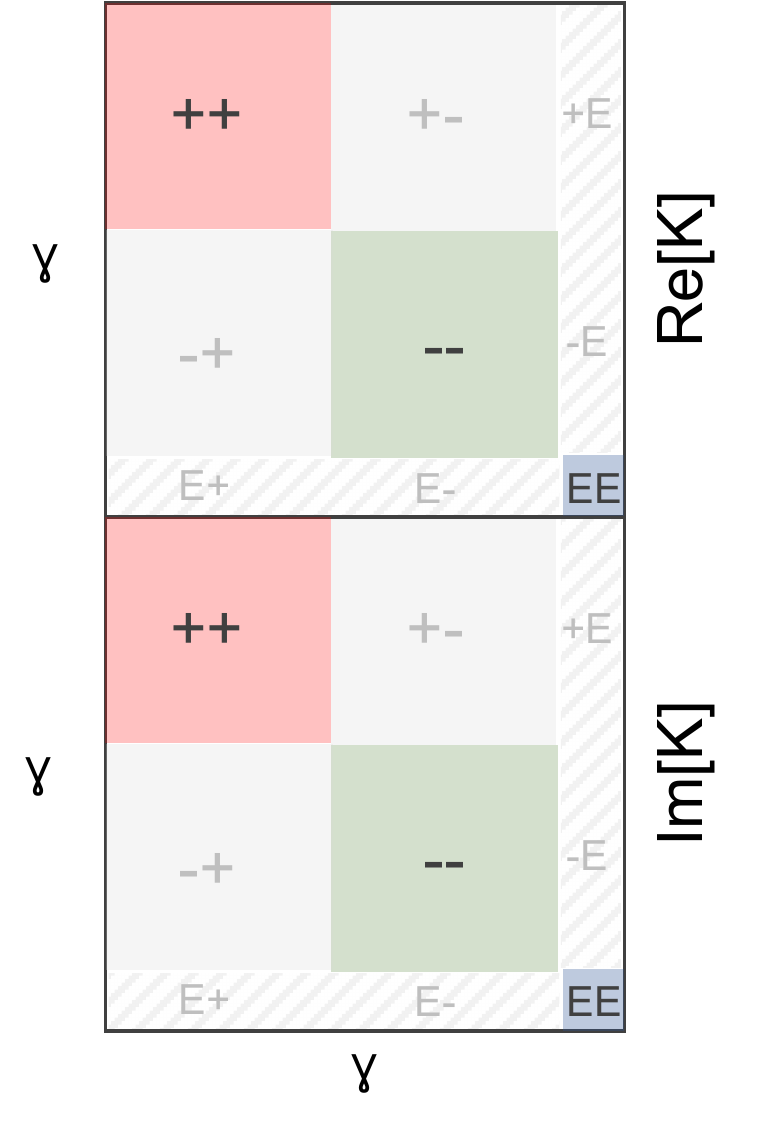}    
     \includegraphics[scale=0.065]{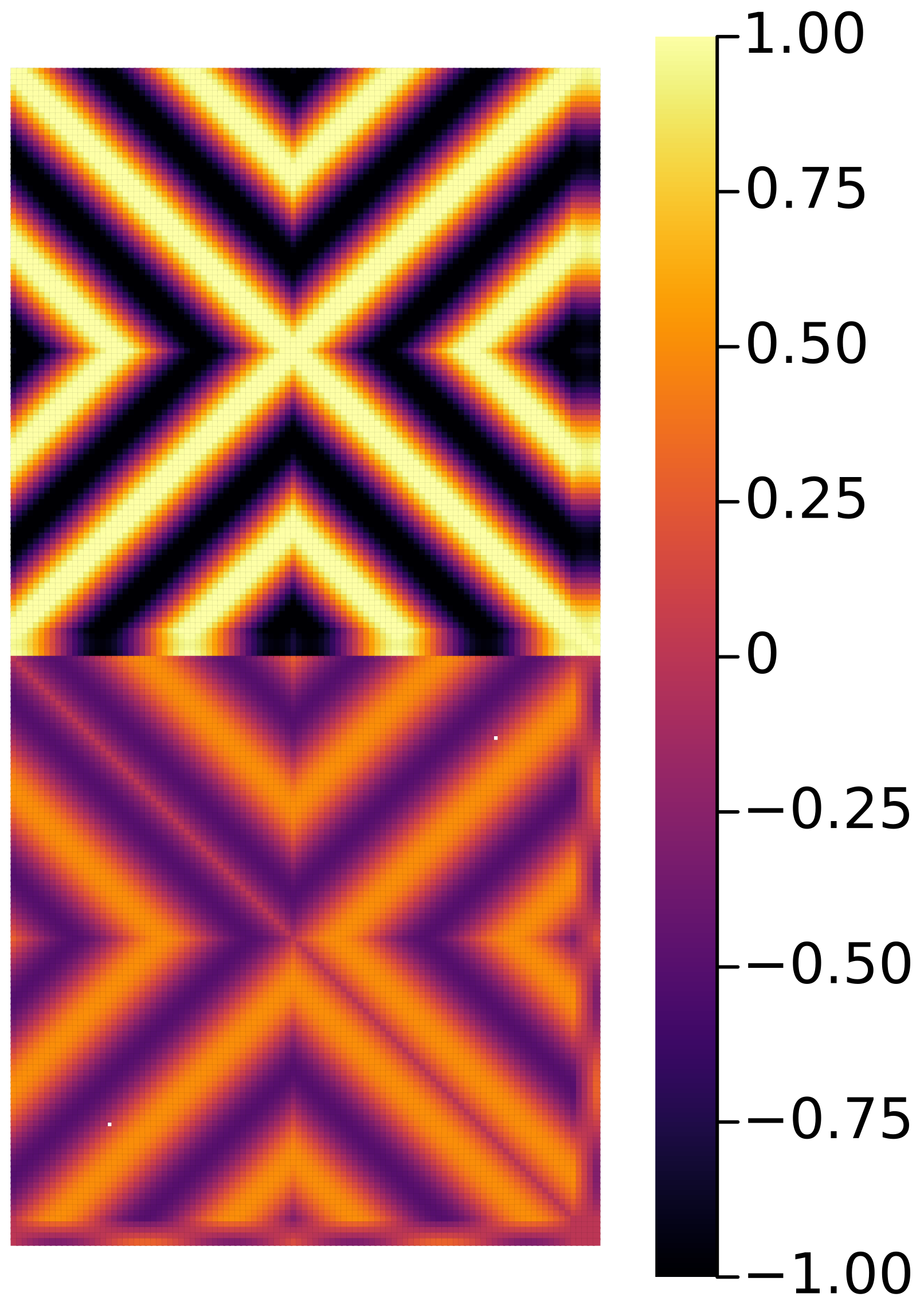}
    \includegraphics[scale=0.065]{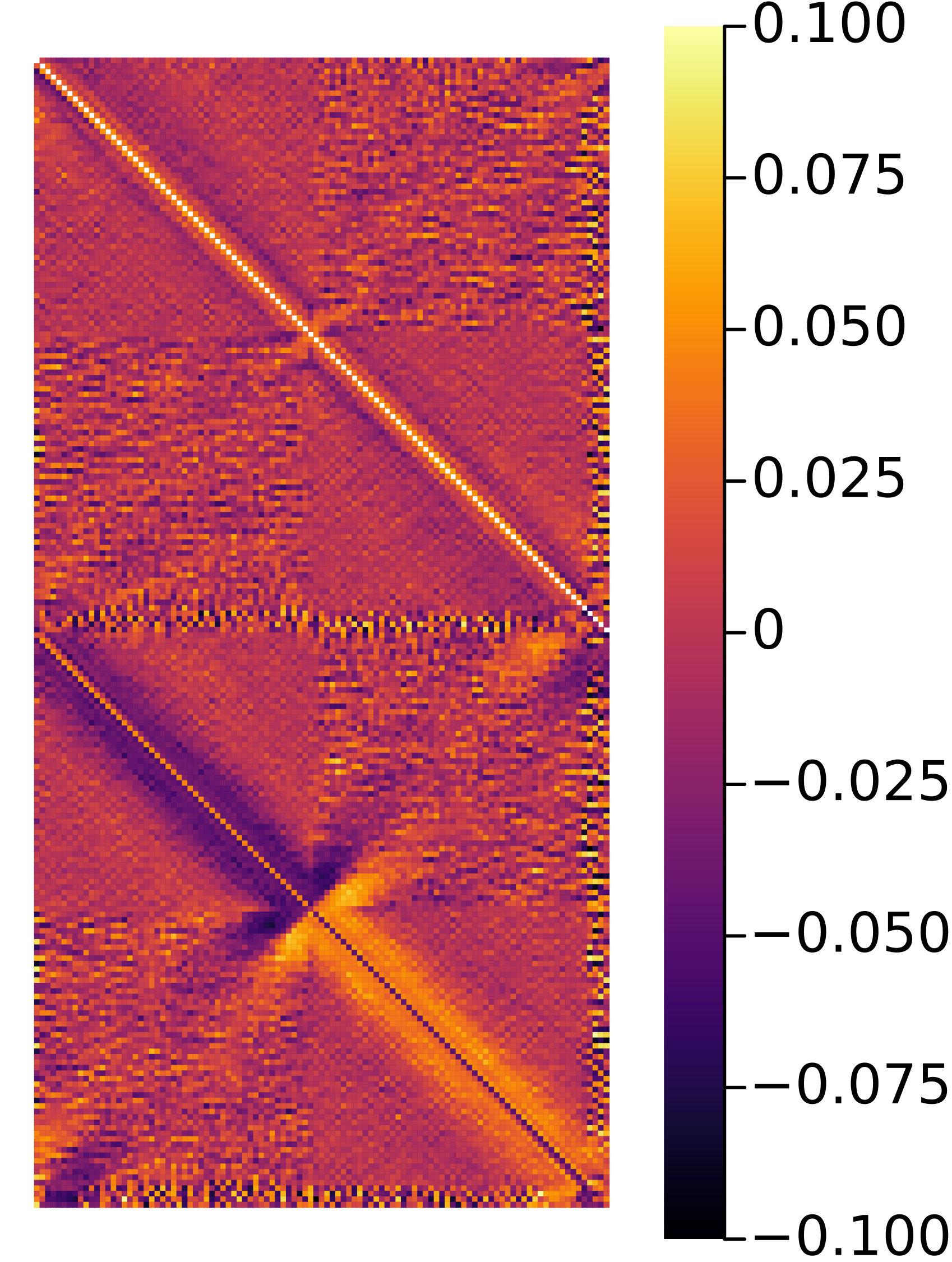}
    \caption{(left) An explanatory sketch of our visualization of the complex kernel used in the simulations. The top major panel denotes the values of the real part and the lower major panel those of the imaginary part of the kernel. Inside each panel the values of the kernel are ordered along the contour parameter, indicating which parts of the kernel couple which range on the contour. 
    (center) The free theory kernel in \cref{eq:freeTeoryPropKernel_discrete}, constructed explicitly in the previous section for $m t_{\textrm{max}}=10$. The repeating pattern indicates an oscillatory behavior in coordinate space arising from the fact that this kernel is just the propagator of the free theory. (right) In the optimal learned kernel based on the low-cost update we have subtracted the unit matrix from the real-part to avoid it dominating the other structures. We find that the learned kernel exhibits some of the structure of the manually constructed kernel but in general has a more simple form, which nevertheless manages to achieve correct convergence of the complex Langevin dynamics.}
    \label{fig:ResultKernelForFreeTheoryLearnedKernel}
\end{figure}

Let us inspect the learned kernel and compare it to the free theory propagator kernel of \cref{eq:freeTeoryPropKernel_discrete}. In \cref{fig:ResultKernelForFreeTheoryLearnedKernel} we visualize the structures of the kernel by plotting a heat-map of the matrix entries of the complex matrix kernel. The right sketch shows how the matrix is structured, where the top panel refers to the real part and the lower panel to the imaginary part. The entries of the matrices are laid out corresponding to the contour parameter $\gamma$. The smaller regions inside the two panels indicate how the kernel mixes points along the time contour. The $++$ corresponds to the mixing of the forward branch of the contour, while $+-$ mixes the forward and backward branch time points. There exists also a small strip involving the Euclidean points, mixing with the real-time points ($E+$ and $E-$), as well as a small corner ($EE$) mixing within the Euclidean points. 

The different regions shown in the sketch can easily be recognized in the two kernel structure plots. Note that we have subtracted the unit matrix from the real-part of the optimized kernel to more clearly expose off-diagonal structures, if present. The manually constructed free theory propagator kernel (middle), as expected from being the inverse free propagator, exhibits an oscillatory pattern. It leads to a significant coupling between the forward and backward time points, due to an anti-diagonal structure in the real and imaginary parts, forming an oscillatory cross pattern. This anti-diagonal behavior is much less pronounced in the optimized kernel (left). In its real-part it mainly exhibits a diagonal which is not as wide as in the manually constructed kernel. There is however a small negative structure present, an off-diagonal band, similar to the black part in the middle panel.

For the imaginary part, the patterns close to the diagonal are similar between the manually constructed kernel and the optimal learned kernel. Both possess a diagonal close to zero and a broad sub/super-diagonal that switches sign at the turning points between the $++$ and $--$ part of the time contour. We also see that the anti-diagonal structure is similar for a very short part in the $+-$ and $-+$ quadrants in the imaginary panel. The rest of the $+-$ and $-+$ quadrant seems to contain noise.

While some similarities exist between the explicit kernel and the optimal learned kernel, it appears that correct convergence requires some  non-trivial structure in the imaginary part of K. The learned kernel achieves correct convergence with much less structure than the manually constructed one.

\subsection{Learning optimal kernels for the strongly coupled anharmonic oscillator}

After successfully testing the learning strategy for a field-independent kernel in the free theory in the previous section, we are now ready to attack the central task of this study: learning an optimal kernel for a strongly coupled quantum system in order to extend the correct convergence of the corresponding real-time complex Langevin simulation.

We deploy the same parameter set as before with $m=1$ and $\lambda=24$. In \cref{sec:FreeTheoryKernelInteractive} we showed that for a real-time extent of $mt_{\textrm{max}} = 1$ an explicit kernel based on insight from the free theory can be constructed, which allows us to restore correct convergence within statistical uncertainties (see \cref{fig:AHO_freeKernel}).

Here we set out to learn an optimal kernel based only on the combination of our low-cost functional and prior knowledge of the Euclidean two-point functions and time-translation invariance of the thermal system. Since we restrict ourselves to a field-independent kernel we expect that our approach will be able to improve on the manually constructed kernel but will itself be limited in the maximum real-time extent up to which correct convergence can be achieved.

As testing ground we selected three different real-time extents, $mt_{\textrm{max}}=1$, $mt_{\textrm{max}}=1.5$ and $mt_{\textrm{max}}=2$, all of which show convergence to the wrong solution when performing naive complex Langevin evolution.

We discretize the real-time contour with a common magnitude of the lattice spacing $|a_i|=|a|$. I.e. depending on the maximum real-time extent the number of grid points changes. E.g. in case of $m t^{\textrm{max}}=2$ we use $N_t=20$ on the forward and backward part of the real-time contour each, and $N_\tau=10$ for the imaginary part of the contour. Due to the stiffness of the complex Langevin equations in the interacting case, all CL simulations are performed with the \textit{Euler-Maruyama} scheme with $\theta=0.6$ and adaptive step-size. We simulate $40$ different trajectories up to $m\tau_L=40$ in Langevin time, computing observables at every $m\Delta\tau_L=0.02$ step.

The setup for learning the optimal kernel is very similar to that in the previous section. The kernel parametrization is given by $K=e^{A + iB}$, where $A$ and $B$ are real matrices. We search for the critical point of the true loss function $L^{\rm prior}$ (see \cref{eq:lossfunctions_sym}) via the heuristic gradient obtained from the loss function $L_D$ of \cref{eq:driftLoss}. We find that minimization proceeds efficiently, when choosing the parameter $\xi=1$ in $L_D$. The optimal kernel is chosen according to the lowest values observed in $L^{\rm prior}$. 

\begin{figure}\centering
    \includegraphics[scale=0.28]{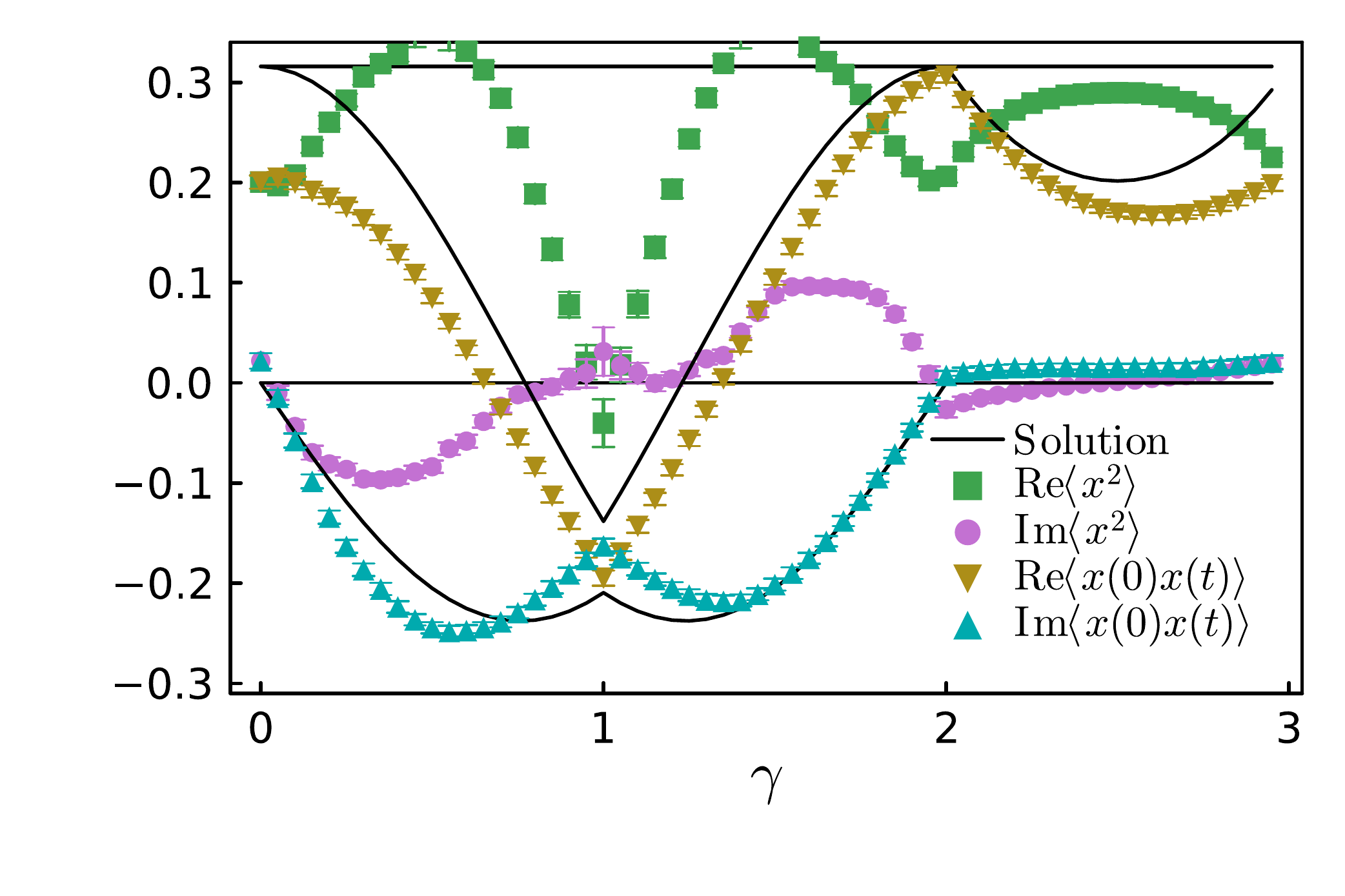}
    \includegraphics[scale=0.28]{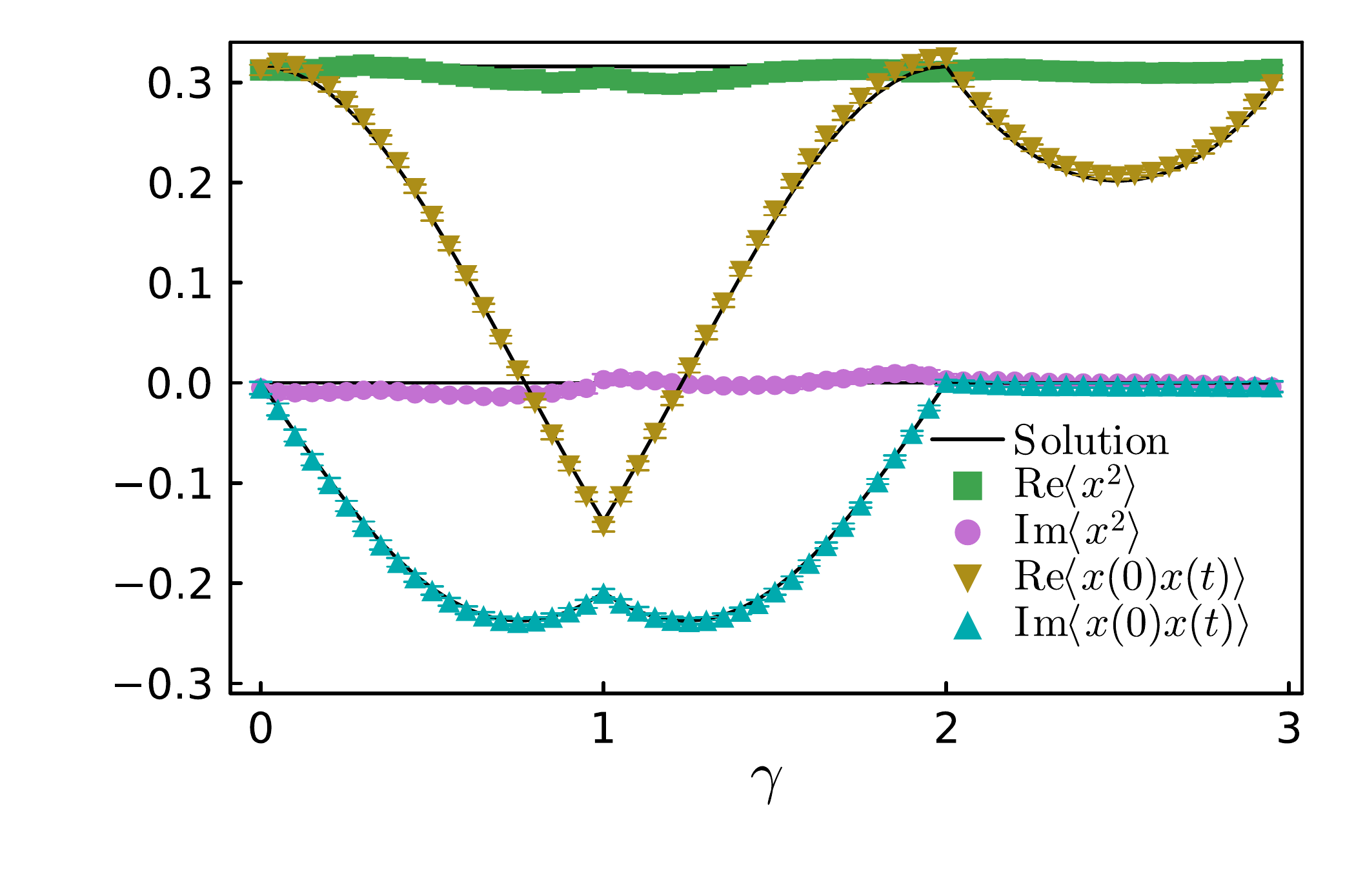}
    \includegraphics[scale=0.28]{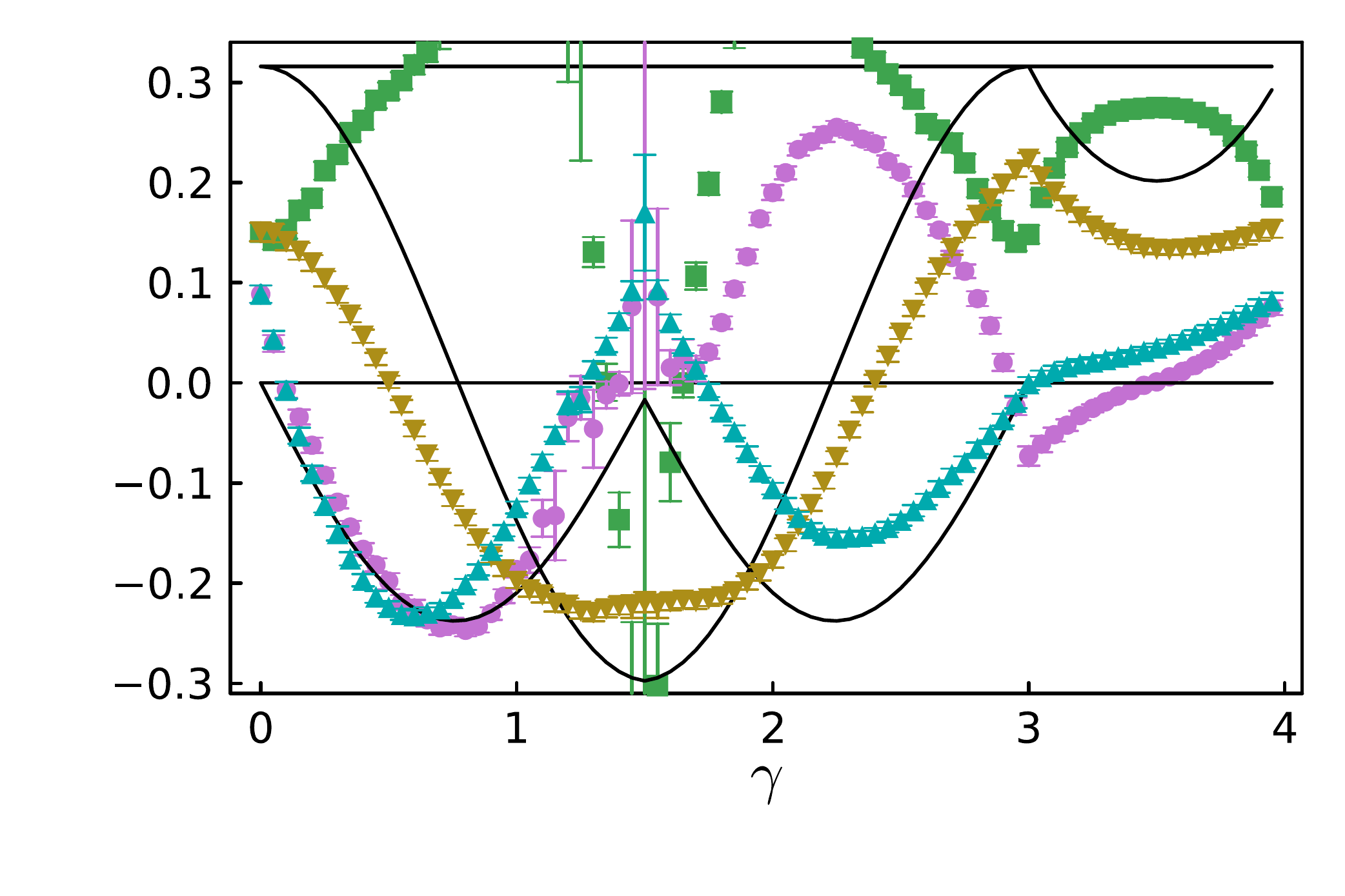}
    \includegraphics[scale=0.28]{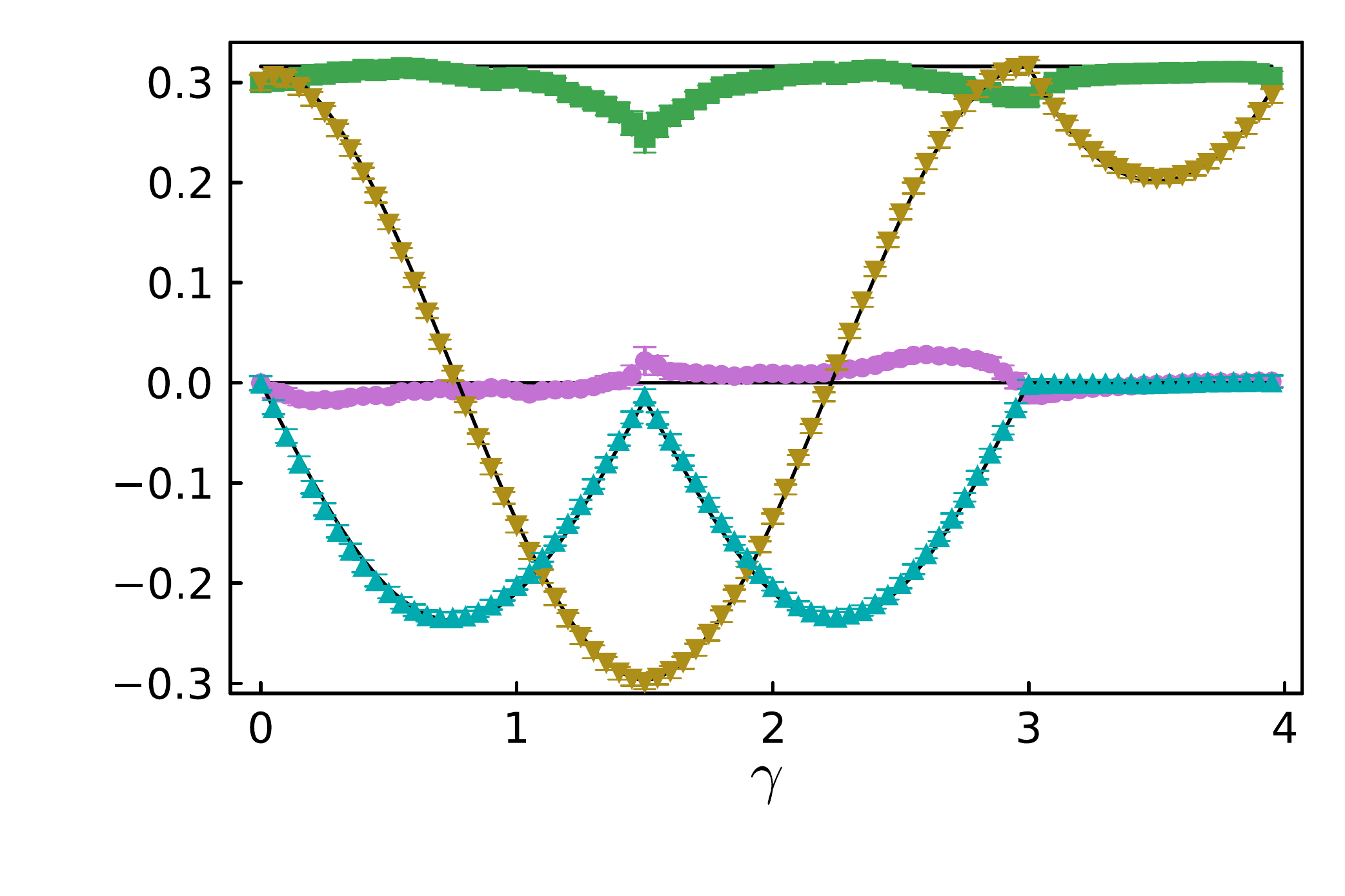}
    \includegraphics[scale=0.28]{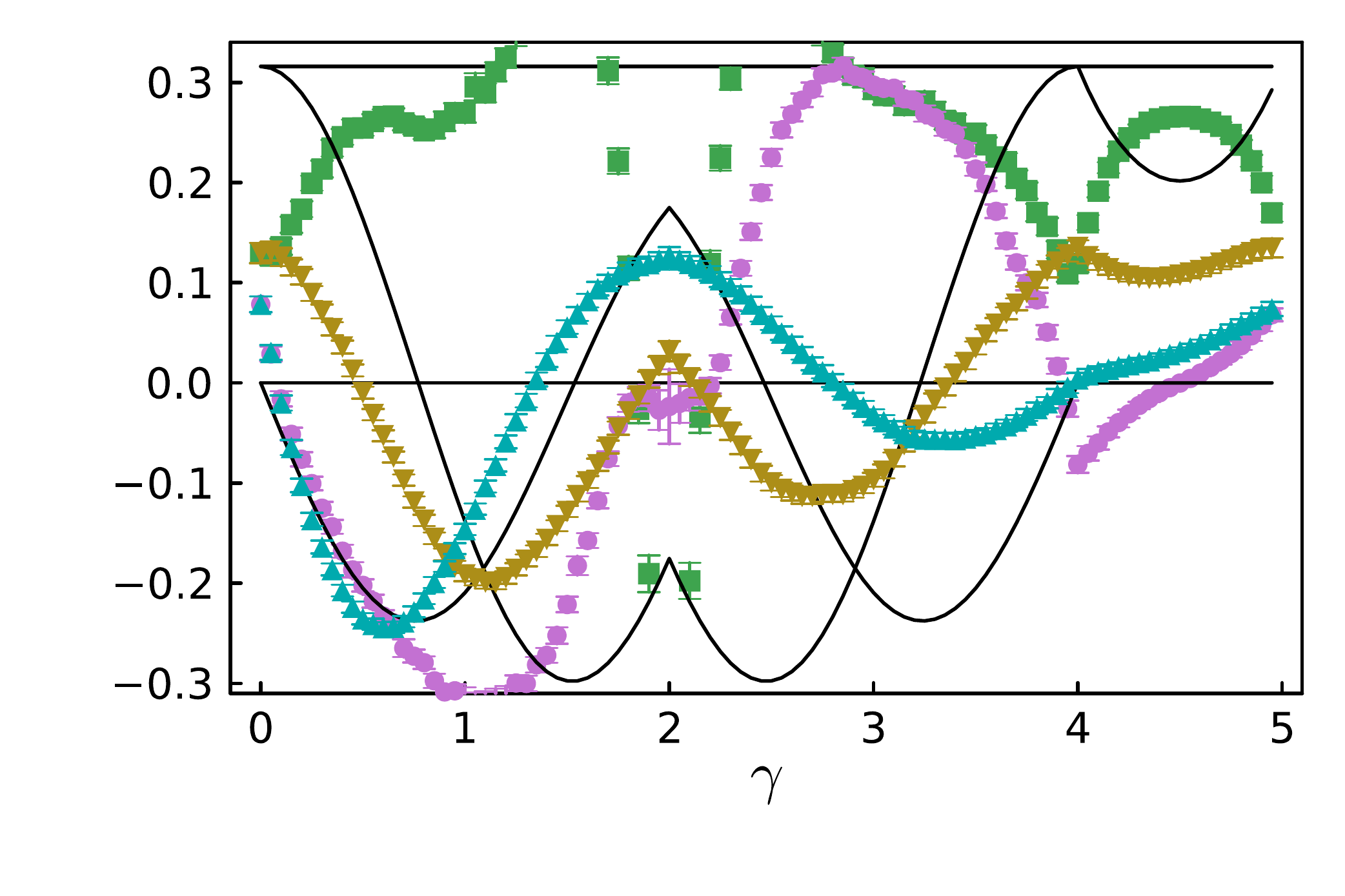}
    \includegraphics[scale=0.28]{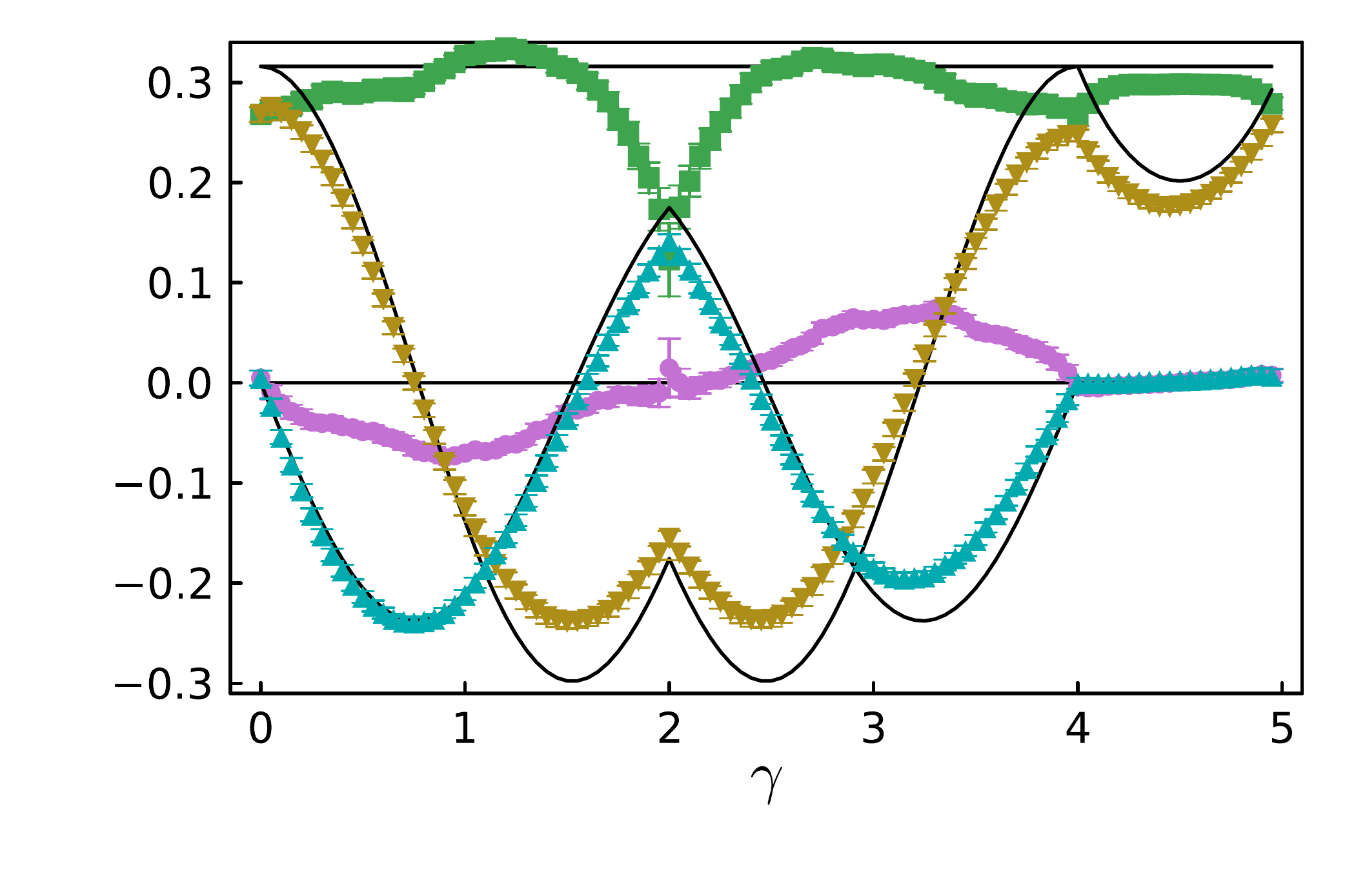}
    \caption{Complex Langevin simulation of the strongly coupled anharmonic oscillator on the thermal Schwinger-Keldysh contour in the absence (left) of a kernel and in the presence of the optimal learned field-independent kernel (right). The top row corresponds to results from a contour with real-time extent $mt_{\textrm{max}}=1$, while the center row shows results for $mt_{\textrm{max}}=1.5$ and the bottom row for $mt_{\textrm{max}}=2$. Values of the correlators from the solution of the Schr\"odinger equation are given as solid black lines.}
    \label{fig:ResultObsIntPlots}
\end{figure}

We find that for a trivial unit kernel where complex Langevin fails, the cost functional $L^{\rm prior}$ based on prior information indicates values of $L^{\rm prior}_{mt_{\rm max}=1}=942$, $L^{\rm prior}_{mt_{\rm max}=1.5}=597320$ and $L^{\rm prior}_{mt_{\rm max}=2}=12923$. The left column of \cref{fig:ResultObsIntPlots} shows from top to bottom the rows correspond to results of the naive CL simulation for $mt_{\textrm{max}}=1$, $mt_{\textrm{max}}=1.5$ and $mt_{\textrm{max}}=2$ respectively. As in previous comparison plots the real- and imaginary part of the unequal time correlation function $\langle x(0)x(\gamma)\rangle$ is given by orange and blue data points, while the real- and imaginary part of the equal time expectation value $\langle x^2(\gamma)\rangle$ is represented by the green and pink symbols respectively. The analytically known values from solving the Schr\"odinger equation are underlaid as black solid lines.

The results of real-time CL in the presence of the optimal learned kernel for the anharmonic oscillator are shown in the right column of \cref{fig:ResultObsIntPlots}. For $mt_{\textrm{max}}=1$ we achieve to lower the value of $L^{\rm prior}_{mt_{\rm max}=1}=14.3$. At this low value all the correlation functions plotted, agree with the true solution within uncertainties. Note that we manage to restore correct convergence for the unequal time correlation function on the real-time axis, even though no prior information about these points was provided in $L^{\rm prior}$ nor $L_{D}$. In contrast to the use of the modified free theory kernel, we see here that $\langle x^2\rangle$ does not show a systematic shift on the real-time branches anymore.

We continue to the second row, where, via an optimal learned kernel, we achieve extending the correctness of CL into a region inaccessible to the modified free theory kernel at $mt^{\textrm{max}}=1.5$. The value of the functional encoding our prior knowledge has reduced to $L^{\rm prior}_{mt_{\rm max}=1.5}=48.1$. We find that the unequal time correlation function values are reproduced excellently, while the real- and imaginary part of $\langle x^2\rangle$ show residual deviations from the correct solution around those points along the SK contour, where the path exhibits sharp turns, i.e. at the end point $\gamma=t_{\rm max}$ and the point where the real-time and Euclidean branch meet $\gamma=2t_{\rm max}$.

The results shown in the third row clearly spell out the limitation of the field-independent kernel we deploy in this study. At $mt^{\textrm{max}}=2$ we do not manage to reduce the value of the cost functional below $L^{\rm prior}_{mt_{\rm max}=2}=759$. Correspondingly in the bottom row of \cref{fig:ResultObsIntPlots} it is clear that CL even in the presence of the field-independent kernel fails to converge to the correct solution. Interestingly the imaginary part of the unequal-time two-point correlator still agrees very well with the true solution on the forward branch while its real part already shows significant deviations from the correct solution. This deviation of the unequal time correlation function affects also the values of the equal-time correlation function which is far from constant and thus leads to a penalty in $L^{\rm prior}$, correctly indicating failure of correct convergence.

There are two possible reasons behind the failure of convergence at $mt^{\textrm{max}}=2$. One is that the low-cost gradient obtained from $L_D$ is unable to bring the kernel close to those values required for restoring correct convergence. The other is that the field-independent kernel is not expressive enough to encode the change in CL dynamics needed to restore correct convergence. In simple models it is e.g. known from ref.~\cite{Okano:1991tz} that field-independent kernels may fail to restore correct convergence for large imaginary drift. We believe that, as a next step, the investigation of field dependent kernels is most promising.

\begin{figure}\centering
    \includegraphics[scale=0.4]{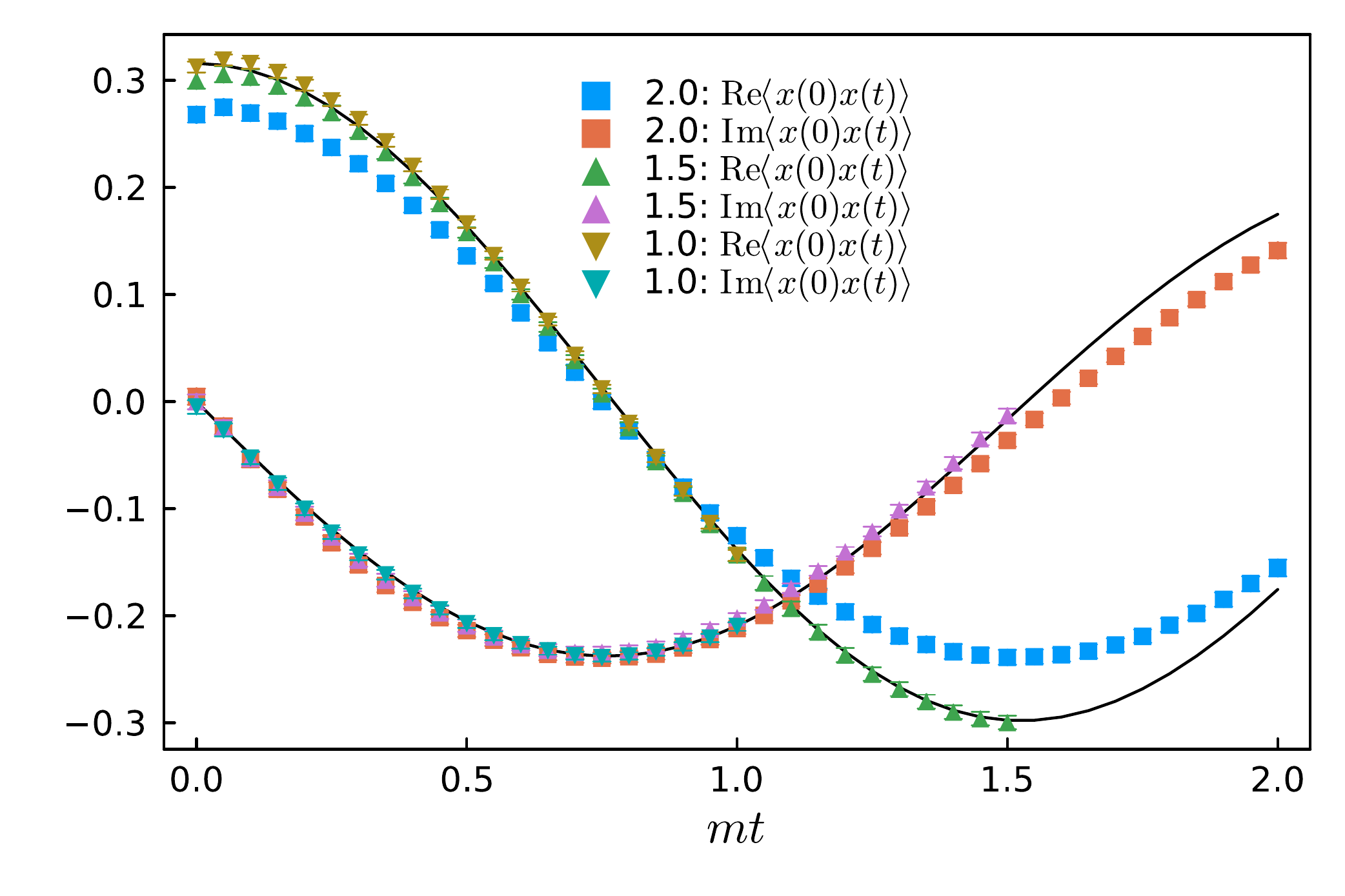}
    \caption{A detailed comparison of the unequal time correlation functions $\langle x(0)x(t)\rangle$ from \cref{fig:ResultObsIntPlots} evaluated in the presence of the optimal learned field-independent kernel on contours with $mt_{\textrm{max}}=1,1.5$ and $2$ respectively.The different colored circles correspond to the real-part while the squares to the imaginary part of the correlator. Values of the correlators from the solution of the Schr\"odinger equation are given as solid black lines.}
    \label{fig:ComparingingCorr0t}
\end{figure}

The unequal time correlation function is most relevant phenomenologically, as it encodes the particle content and occupation numbers in the system. We thus compare in \cref{fig:ComparingingCorr0t} the values of $\langle x(0) x(t) \rangle$ along the forward real-time extent of the contour for $mt^{\textrm{max}}=1.0,1.5$ and $2.0$ to the correct solution given as black solid line. Here we can see in more detail that for a real-time extent of $1$ and $1.5$ CL with the optimal learned kernel converges to the true solution within uncertainties. At $2$ the real part of the correlator begins to deviate from the correct solution. Note that the most difficult points to achieve convergence at are $t=0$ and at $t=t_{\rm max}$. Similarly we find that these points are also the ones, where the equal time correlator deviates the most from the correct solution, an important fact as this allows this deviation to contribute to the penalty in $L^{\rm prior}$.

\begin{figure}\centering
    \includegraphics[scale=0.066, trim= 0 0.4cm 0 0]{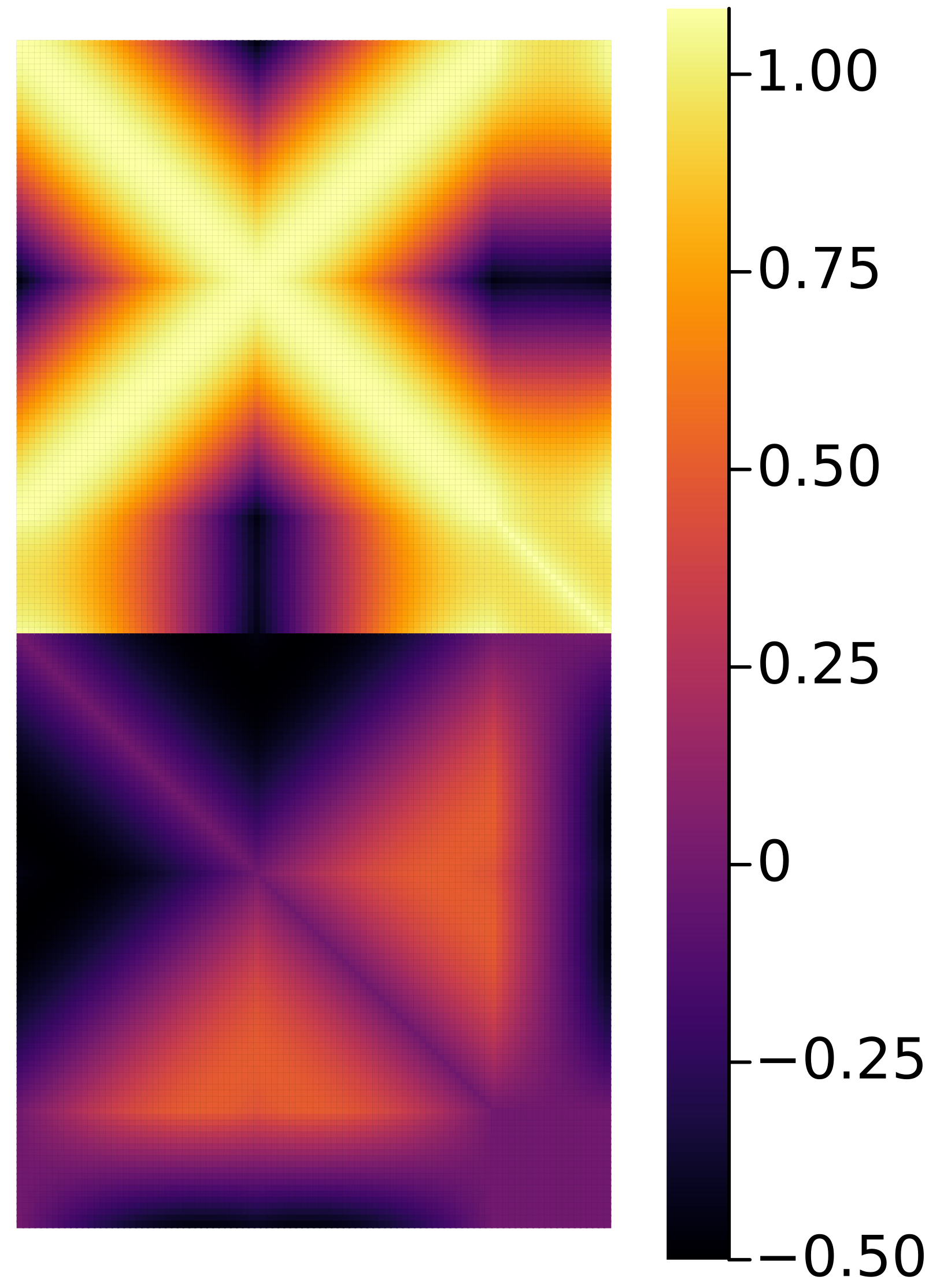}
    \includegraphics[scale=0.067]{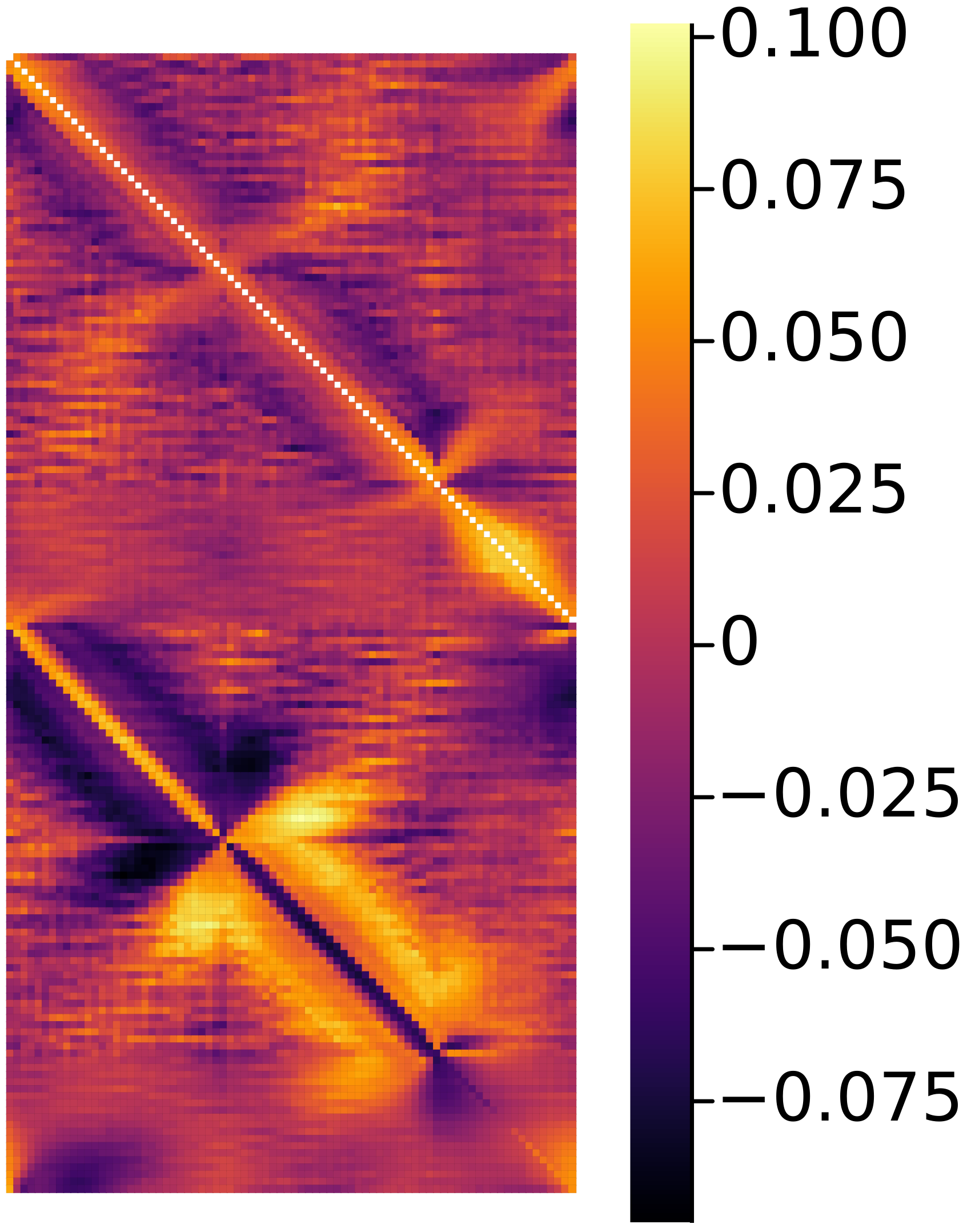}
    \includegraphics[scale=0.067]{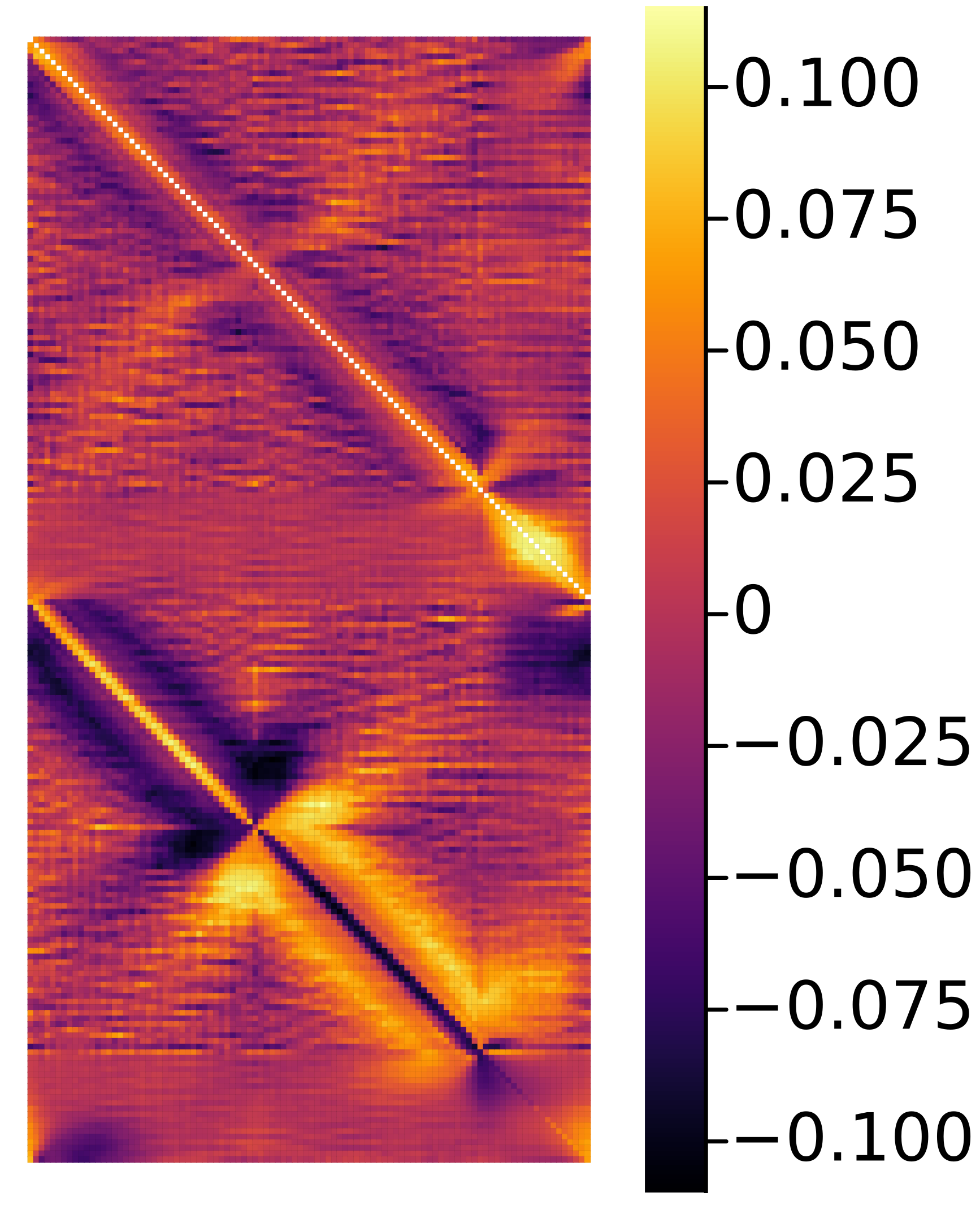}
    \caption{(left) Free theory kernel for the SK contour with $mt_{\rm max}=1.5$. (center) The optimal learned kernel in the interacting theory for $mt_{\rm max}=1.5$, which achieves correct convergence of CL. The diagonal entries with values close to unity are subtracted from the kernel. (right) The kernel obtained as a result of the optimization procedure in the case of $mt_{\rm max}=2.0$, which does not achieve correct convergence. At the turning point at $t_{\rm max}$ and when connecting to the Euclidean domain the kernel for the interacting theory shows nontrivial structure not present in the free theory.}
    \label{fig:ResultKernelForInteractingTheoryLearnedKernel}
\end{figure}

In \cref{fig:ResultKernelForInteractingTheoryLearnedKernel} we plot a heat map of the values of the kernels with $mt_{\rm max}=1.5$ (center) and $mt_{\rm max}=2$ (right) compared to the free theory propagator kernel from \cref{eq:freeTeoryPropKernel_discrete} (left) for $mt_{\rm max}=1.5$. (for a sketch of the structure of the heat map see the left panel of \cref{fig:ResultKernelForFreeTheoryLearnedKernel}). We have subtracted the unit matrix from the real-part of the two optimized kernels. They both exhibit a diagonal band in the real part, which is thinner than the one in the free theory kernel. It is interesting to see that both show non-trivial structures passing through the $t^{\textrm{max}}$ point and when connecting to the Euclidean branch. In the imaginary part the structures have more similarity with the free theory propagator kernel, where a sign change occurs as one moves away from the diagonal. The difference in the optimal kernels between $mt_{\rm max}=1.5$ and $mt_{\rm max}=2$ is small overall. 

\section{Summary and Conclusion}

In this paper we proposed a novel strategy to recover correct convergence of real-time complex Langevin simulations by incorporating prior information into the simulation via a learned kernel. The effectiveness of the strategy was demonstrated for the strongly coupled anharmonic oscillator on the Schwinger Keldysh contour by extending correct convergence in this benchmark system up to $mt_{\rm max}=1.5$, three times the previously accessible range of $mt_{\rm max}=0.5$.

After discussing the concept of neutral and non-neutral modifications of Langevin dynamics by use of real and complex kernels, we demonstrated that an explicitly constructed complex kernel can be used to improve the convergence behavior of real-time complex Langevin on the Schwinger-Keldysh contour. Taking insight from a single d.o.f. model and the harmonic oscillator, approximately correct convergence in the strongly coupled anharmonic oscillator was achieved up to $mt_{\rm max}=1$. As no systematic extension to the explicit construction of that kernel exists, we instead proposed to learn optimal kernels using prior information.

The ingredients to learning an optimal kernel are prior information and an efficient prescription for computing gradients. Prior information comes in the form of apriori known Euclidean correlation functions, known symmetries of the theory and the Schwinger-Keldysh contour, as well as information on the boundary terms. Here we included only the first two types of information, which sufficed to achieve improvements in convergence. We surveyed different modern differential programming techniques that in principle allow a direct optimization of the kernel based on the full prior information, but found that in their standard implementations they are of limited use in practice due to runtime or memory limitations. Instead we constructed an approximate gradient based on an alternative optimization functional, inspired by the need to avoid the presence of boundary terms. This optimization functional possesses a gradient, which can be approximated with much lower cost than that of the original optimization functional. The low-cost gradient in practice is computed using standard auto-differentiation. By minimizing with this gradient and monitoring success via the full prior information cost functional we proposed, we were able to locate optimal kernels.

Our strategy was successfully applied first to the harmonic oscillator on the thermal SK contour. We managed to restore correct convergence with an optimal learned field-independent kernel that shows a simpler structure compared to the manually constructed kernel. This result bodes well for future studies, where we will investigate in detail the structure of the optimal learned kernel to draw conclusions about the optimal analytic structure for extending the approach to a field-dependent kernel.

The central result of our study is the restoration of correct convergence in the strongly correlated anharmonic oscillator on the thermal SK contour up to a real-time extent of $mt_{\rm max}=1.5$, which is beyond the reach of any manually constructed kernel proposed so far. We find some remnant deviations of the equal-time correlation function $\langle x^2\rangle$ from the true solution at the turning points of the SK contour. The phenomenologically relevant unequal-time correlation function $\langle x(0)x(t)\rangle$ on the real-time branch on the other hand reproduces the correct solution within statistical uncertainty. 

While our strategy based on a field-independent kernel is successful in a range three times the previous state-of-the-art, we find that the restricted choice of kernel limits its success at larger real-time extent.

We conclude that our study provides a proof-of-principle for the restoration of correct convergence in complex Langevin based on the inclusion of prior information via kernels. Future work will  focus on extending the approach to field-dependent kernels, carefully reassess the discretization prescription of the SK at the turning points and improve the efficiency of the differential programming techniques necessary to carry out a minimization directly on the full prior knowledge cost functional.

\section*{Acknowledgements}
The team of authors gladly acknowledges support by the Research Council of Norway under the FRIPRO Young Research Talent grant 286883. The numerical simulations have been partially carried out on computing resources provided by  
UNINETT Sigma2 - the National Infrastructure for High Performance Computing and Data Storage in Norway under project NN9578K-QCDrtX "Real-time dynamics of nuclear matter under extreme conditions"

\appendix

\section{Correctness criterion in the presence of a kernel}
\label{sec:correctnessCriterionWithAKernel}

In this section we discuss the correctness criterion in the presence of a kernel in the CL evolution. As mentioned in \cref{sec:complexKernelLangevin} there are two parts to the correctness criterion that need to be fulfilled in order for complex Langevin to converge to the correct solution. We must avoid boundary terms for the real-valued distribution $\Phi(x^R,x^I, \lt)$ and the complex Fokker-Planck \cref{eq:CFP} must have the correct equilibrium distribution. If both conditions are fulfilled the equal sign of \cref{eq:corrconv} holds. 

To check if the equilibrium distribution of $\rho(x,\tau_L)$ is ${\rm exp[i S_M]}$ we need to either solve the Fokker-Planck equation explicitly, or inspect the eigenvalue spectrum of the Fokker-Planck equation \cite{Klauder:1985kq}. To make inference about correct convergence based on the eigenspectrum, the eigenvectors of the Fokker-Planck operator must form a complete set, as otherwise there exist non-orthogonal zero modes competing with the $e^{iS_M}$ stationary distribution. For a non-self-adjoint operator this is not always the case. 

To show the connection between the eigenvalues of the Fokker-Planck equation and the equilibrium distribution we use a similarity transform to define the operator $G$ from the Fokker-Planck operator $L$ including the kernel
\begin{equation}\label{eq:SimilaryTranform}
\begin{aligned}
    G(x) =& U L(x) U^{-1} = e^{-\frac12 iS_M(x)}L(x)e^{\frac12 iS_M(x)} \\
         =& \left( \frac{\partial}{\partial x} + \frac12 i\frac{\partial S_M}{\partial x} \right) K[x] \left(  \frac{\partial}{\partial x} - \frac12 i\frac{\partial S_M}{\partial x} \right),
\end{aligned}
\end{equation}
which by definition has the same eigenvalues as $L$. The transformation is carried out here to follow closely the conventional way of proving the correct convergence for a real action $S$. I.e., when $S$ is real, $G$ becomes a self-adjoint and hence negative semi-definite operator. For complex actions, $iS_M$, this transformation is not necessary for the following arguments. It is however useful in practice as a pre-conditioner for calculating the eigenvalues of the Fokker Planck operator. The complex distribution $\rho(x,\tau_L)$ is also transformed based on the same transformation, such that
\begin{equation}\label{eq:FP_after_similarity_transform}
    \tilde{\rho}(x,\tau_L) = e^{-i\frac12 S_M}\rho(x,\tau_L), \quad \textrm{where} \quad \dot{\tilde{\rho}}(x,\tau_L) = G(x)\tilde\rho(x,\tau_L) 
\end{equation}
is the Fokker-Planck equation for the transformed operator. Since we are interested in the stationary distribution, we construct the eigenvalue equation
\begin{equation}\label{eq:EigenvalueProblem_G}
    G(x)\psi_n(x) = \lambda_n \psi_n(x).
\end{equation}
Due to the form of the operator we know that it must have at least one zero eigenvalue, $\lambda=0$, associated with the eigenvector $e^{i\frac12 S_M}$. 

The formal solution of the Fokker-Planck equation after the similarity transform of \cref{eq:FP_after_similarity_transform} is given by
\begin{equation}
    \tilde \rho(x;\tau_L) = e^{\tau_LG(x)} \tilde \rho(x,0)
\end{equation}
and by expanding $\tilde \rho(x;0)$ in the eigenbasis $\psi_n$, and using $\psi_0e^{\lambda_0 t} = e^{\frac12 iS_M}$ we get
\begin{align}
    \rho(x;\tau_L) =& e^{\frac12 iS_M(x)}\tilde \rho(x;\tau_L) \\
              =& e^{\frac12 iS_M(x)}\sum_{n=0}^\infty a_n \psi_n(x)e^{\lambda_n \tau_L} = ce^{iS_M(x)} + \sum_{n=1}^\infty a_n \psi_n(x)e^{\lambda_n \tau_L}
\end{align}
such that when $\tau_L\rightarrow \infty$ only the first term is left, namely the equilibrium distribution ${\rm exp[i S_M]}$. This is however only true if $\textrm{Re}\; \lambda_n \leq 0$, in which case the spectrum of $G$ provides information of the equilibrium distribution of the Fokker-Planck equation. 

The second condition, which needs to be satisfied is that the sampling of CL gives the same distribution as the complex Fokker-Planck equation. To establish that it does, we follow the correctness criterion of ref.~\cite{Aarts:2011ax}. Let us show that the criterion also holds in the presence of a kernel by revisiting some central steps of the original proof. We start with the Fokker-Planck equation for complex Langevin \cref{eq:FP_KCLE}, which operates on a real distribution $\Phi(x^R,x^I;t)$ for the complexified degrees of freedom $x^R$ and $x^I$. Let us take a look at the Fokker-Planck equation, which evolves the distribution of an observable ${\cal O}$
\begin{equation}
\begin{aligned}
        \partial_{\tau_L} {\cal O}&(x^R,x^I) = \left[( H_R \partial_{x^R} +  H_I \partial_{x^I})^2 +  {\rm Re}\left\{iK[x^R+ix^I]\nabla S_M + \frac{\partial K[x^R+ix^I]}{\partial x^R}\right\} \partial_{x^R} \right. \\
        & \left. +  {\rm Im}\left\{iK[x^R+ix^I]\nabla S_M + \frac{\partial K[x^R+ix^I]}{\partial x^R}\right\} \partial_{x^I} \right]{\cal O}(x^R,x^I) = L_K^T{\cal O}(x^R,x^I),
\end{aligned}
\end{equation}
where we can identify the operator $L_K$ to be the bilinear adjoint of the Fokker-Planck operator $L_K^T$\cite{Aarts:2011ax}. If we assume that ${\cal O}$ is holomorphic, we know that $\partial_{x^I} {\cal O} = i\partial_{x^R} {\cal O} \rightarrow i\partial_z {\cal O}$, where for the last equality we have used the following relation between derivatives $\partial_{x^R} {\cal O}(x^R+ix^I) \rightarrow \partial_z f(z)$ with $z=x^R+ix^I$. Replacing derivatives yields the following Langevin equation for the holomorphic observable ${\cal O}$ expressed in the complex variable $z$
\begin{align}
    \partial_t {\cal O} &= \left[K[z] \partial_z^2 + iK[z] \nabla S_M \partial_z  +  \frac{\partial K[z]}{\partial z} \partial_z \right] f = \left[\partial_z + i\nabla S_M \right] K[z] \partial_z  {\cal O} = \tilde L_K^T {\cal O},
\end{align}
where in the last equality we have used that $K[z]\partial_z^2 + (\partial_z K[z])\partial_z =\partial_z K[z] \partial_z $ based on \textit{integration by parts}. We have now shown that $(\tilde L_K^T - L_K^T){\cal O} = 0$ for a Fokker-Planck equation with a field dependent kernel. In turn, we conclude that the correctness criterion also holds for a kernelled complex Langevin equation. 

For the above derivation to hold there may not arise any boundary terms given by\cite{Scherzer:2018hid}
\begin{equation}\label{eq:BoundaryTerms}
    B_n = \int dx^R dx^I \Phi(x^R,x^I) \left ( \tilde L_K ^T \right)^n \mathcal O(x^R+ix^I)
\end{equation}
where $\tilde L_K ^T$ is the Langevin operator given by
\begin{equation}\label{eq:LangevinOperator}
    \tilde L_K ^T = (\partial_z + i\nabla S_M) K[z] \partial_z.
\end{equation}
The formal criterion is then that the observable $\langle \tilde L_K ^T \mathcal O \rangle$ should be zero. This expression for $B$ includes contributions from the full range of values of the d.o.f. between $-\infty$ to $\infty$. Including all of these will introduce significant amounts of noise in the expectation value. This can be avoided by introducing a cut-off $\Omega$ for the values for $x^R$ and $x^I$ in the calculation of the observable. The boundary terms of \cref{eq:BoundaryTerms} are thus calculated using
\begin{equation}\label{eq:BoundaryTermsOperator}
    B_n^{\Omega} = \left\langle \left( \tilde L_K ^T \right)^n \mathcal O(x^R+ix^I) \right\rangle_\Omega = \left\langle
\begin{cases}
     \left( \tilde L_K^T \right)^n \mathcal O(x^R+ix^I) ,& \text{if } x^R \leq \Omega_{x^R} \text{ and } x^I \leq \Omega_{x^I}\\
    0,              & \text{otherwise}
\end{cases}
\right\rangle
\end{equation}
where $\Omega_{x^R}$ and $\Omega_{x^I}$ denote the individual cutoffs for the real- and imaginary part respectively. In the case of scalar fields (which in contrast to gauge fields do not feature a compact dimension), we need to cut off in both $x^R$ and $x^I$ direction. We will in this paper stick to considering the cut-off to be a square. For all the values outside the square we set the contributions to the expectation value to zero. 

Since the observable of interest in the simple models is $z^2$ (i.e. it is the most difficult to capture accurately), we find the boundary terms observable from \cref{eq:BoundaryTermsOperator} to be
\begin{align}
   \nonumber \tilde L _K ^T \; z^2 =& (\nabla_z + i\nabla S_M)K(z)\nabla_z z^2 = (\nabla_z + i \nabla S_M)K(z) 2z \\
   \nonumber          =& 2( (\nabla_z K(z)) z + K(z) + i \nabla S_M K(z) z) \\
             =& 2K(z)(1 + i \nabla S_M z) + 2(\nabla_z K(z)) z,
\end{align}
which for a field-independent kernel reduces to $\langle \tilde L _K ^T \; z^2 \rangle_\Omega = \langle  2K + i z \nabla S_M \rangle_\Omega$. 

We have discussed both ingredients necessary to establish correct convergence of our simulation in the presence of a kernel, i.e. the behavior of the Fokker-Planck spectrum and boundary terms. The boundary terms can be calculated in practice without problems, while the eigenvalues of the Fokker-Planck operator of \cref{eq:EigenvalueProblem_G} so far remain out of reach for realistic systems, due to computational cost.

\section{Constant kernels and correct convergence in simple models}
\label{sec:limitations}

In this appendix we investigate concrete examples of our optimization procedure and the corresponding learned kernels in one-degree of freedom models, for which in the literature (see e.g. \cite{Okamoto:1988ru,Okano:1991tz}) kernels have been constructed by hand. The motivation behind this appendix is to understand how the kernels affect the behavior of the complex Langevin simulation, in particular how  they are connected to the idea of minimizing the drift loss in \cref{eq:driftLoss}. To this end we connect complex Langevin to the Lefschetz thimbles and the correctness criterion \cite{Aarts:2011ax}. 

We investigate the one-degree of freedom model with the action
\begin{equation}\label{eq:simpleModelAction}
    S=\frac{1}{2}\sigma x^2 + \frac{\lambda}{4}x^4,
\end{equation}
which leads to the following partition function
\begin{equation}
    Z = \int dx e^{-S},
\end{equation}
i.e. we use the same convention as in the  literature \cite{Okamoto:1988ru,Okano:1991tz,Aarts:2013fpa}. Note that this is a different convention from the main text as $S$ can now have a imaginary part. This model is interesting as it exhibits similar properties as the interacting real-time model: the convergence problem appears, breaking both the boundary term condition and the equilibrium distribution of the Fokker-Planck equation for various parameters. 

We will therefor take a closer look at two specific sets of parameters. The first one is $\sigma=4i$ and $\lambda = 2$ where we can find an optimal kernel, and as second parameter we choose $\sigma=-1 + 4i$ with the same $\lambda=2$, where for correct convergence we have to go beyond a constant, field-independent kernel.

In \cref{sec:simplestRTModel} we looked at a variant of this model corresponding to $\sigma = i$ and $\lambda=0$ in \cref{eq:simpleModelAction}. The optimal field independent kernel $K=-i$ transforms the complex Langevin equation such that it samples exactly on the Lefschetz thimble. In contrast, the models considered here have more than one critical point, and hence the relation to the Lefschetz thimbles is not as simple. The critical points for \cref{eq:simpleModelAction}, can be found via
\begin{equation}\label{eq:criticalPoints}
    \frac{\partial S(x)}{\partial x} = 0.
\end{equation}
which are located at $x=0,\pm\sqrt{\sigma/\lambda}$ \cite{Aarts:2011ax}. We see that the smaller the real-part of the $\sigma$ parameter becomes, the further out into the complex plane the two critical points away from the origin are located.

\subsection{Non-uniqueness of the optimization}

In this study we used the optimization functional 
\begin{equation}
    L_{D} =   \left\langle \Big| D(x) \cdot (-x) - ||D(x)||\;||x|| \Big|^\xi\right\rangle 
\end{equation}
with $D=K\delta S/\delta x$, to compute an approximate gradient for the minimization of the true cost functional $L^{\rm prior}$. $L_D$ was constructed with the idea in mind that in order to remove boundary terms we wish to penalize drift away from the origin. In this appendix we discuss the fact that there exist multiple critical points to $L_D$, which may or may not correspond to a kernel that restores correct convergence. In practice we distinguish between these solutions by testing the success of the corresponding kernel in restoring correct convergence via the value of $L^{\rm prior}$.

Let us start with the parameter set $\sigma = 4i$ and $\lambda=2$ in \cref{eq:simpleModelAction}. For this choice ref.~\cite{Okamoto:1988ru} showed that a constant kernel can be constructed that restores correct convergence.

In a one-degree of freedom model, where the constant kernel is nothing but a complex number, we can optimize by brute force. Using the parametrization
\begin{equation}
    K = e^{i\theta}, \quad H = \sqrt{K} = e^{i\frac{\theta}{2}}
\end{equation}
we only have to consider a single compact parameter: $\theta \in [0,2\pi)$. A scan of the $\theta$ values reveals two minima of the $L_D$ loss function. One at $\theta_1 = \frac{\pi}{3}$ and one at $\theta_2 = \frac{2\pi}{3}$, where the first one corresponds to the kernel found manually in ref~\cite{Okamoto:1988ru}. When deriving the optimal kernel, the authors also obtained two solutions, which correspond to these two kernels. They selected the correct one by requiring the kernel to belong to the first Riemann sheet when taking a square root. In our case, we too need to select the correct one and in this simple model can use the correctness criteria directly to do so. 

\begin{figure}\centering
    \includegraphics[scale=0.04]{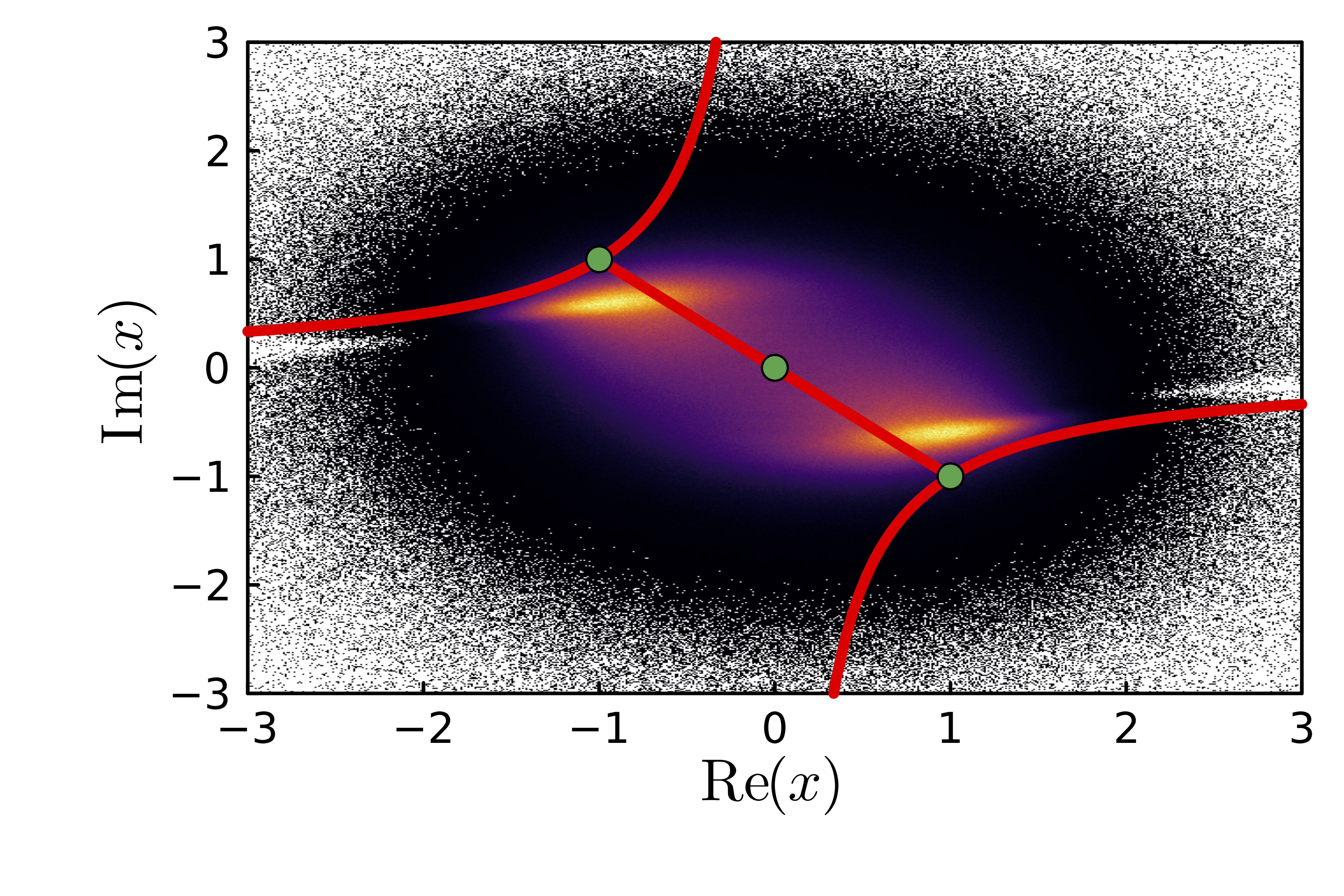}
    \includegraphics[scale=0.04]{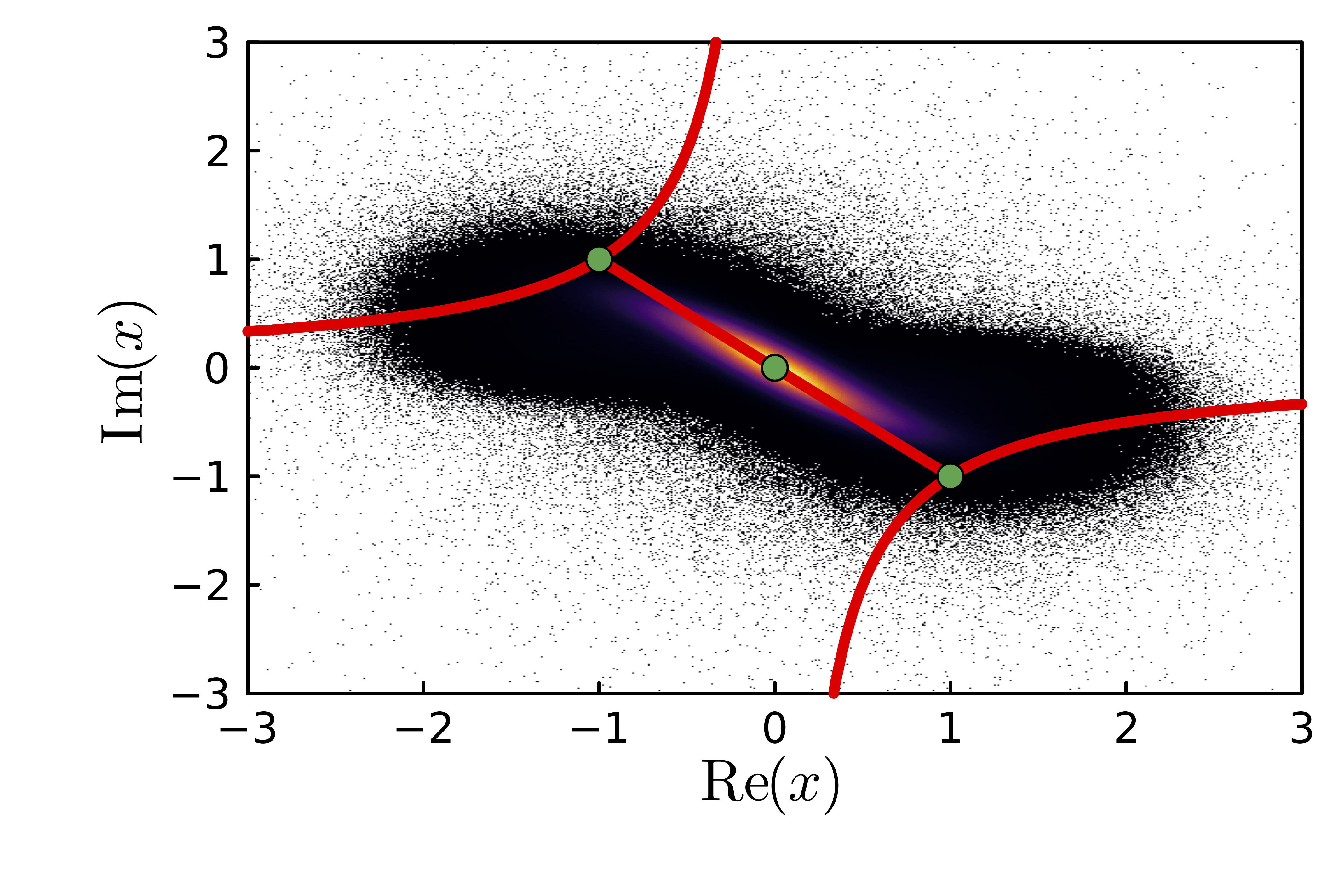}
    \includegraphics[scale=0.04]{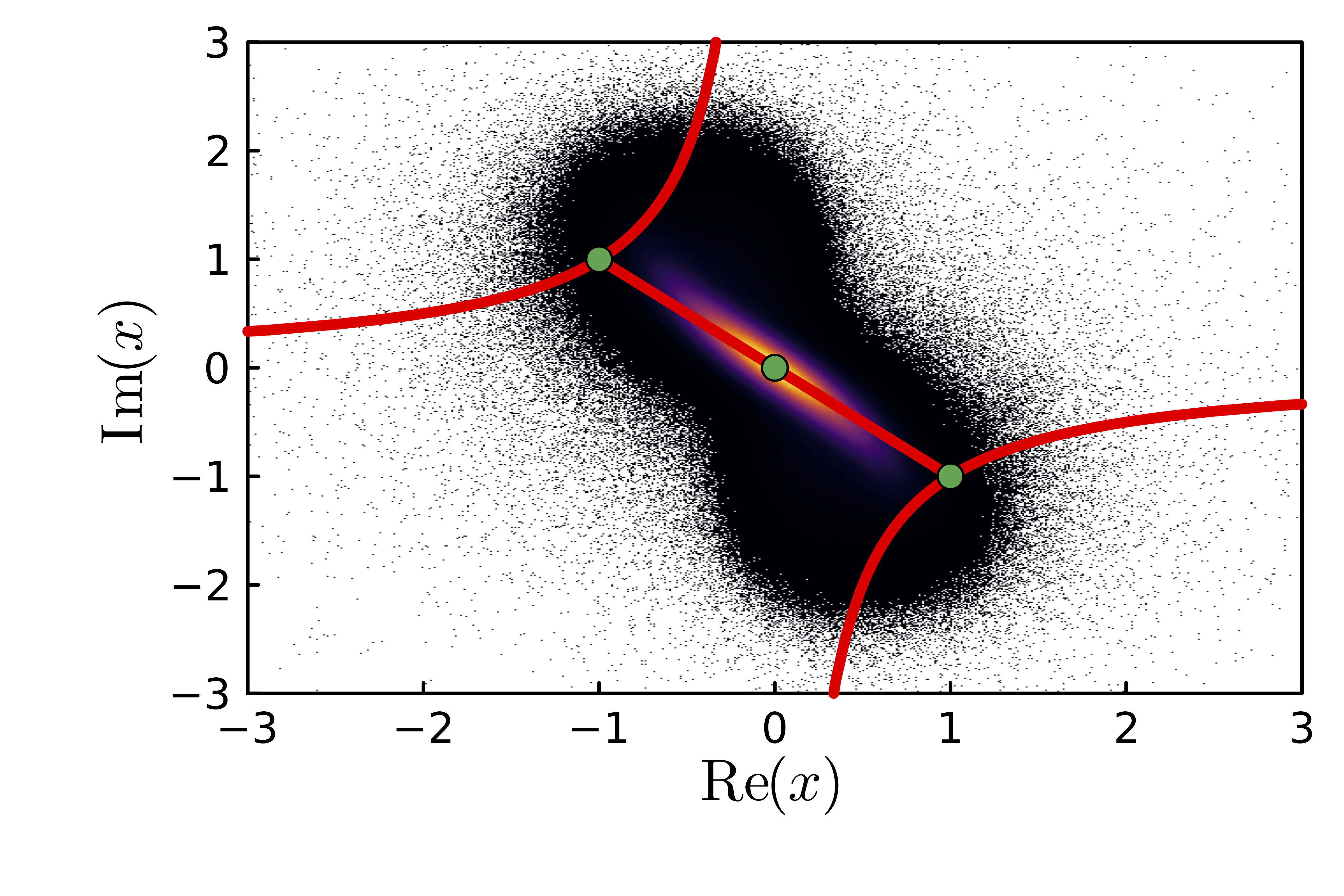}
    \caption{Distribution of the complex Langevin simulation and the Lefschetz thimble (red line) for the model of \cref{eq:simpleModelAction} with $\sigma =4i$ and $\lambda=2$, using different kernels; $K_0=1$ (top left), $K_1={\rm exp}[-i\pi/3]$ (top right) and$K_2={\rm exp}[-i2\pi/3]$ (bottom). The green points denote the critical points given by the solution to \cref{eq:criticalPoints}. The color in the distribution heat map corresponds to the number of samples at the corresponding position (a lighter color refers to a higher value).}
    \label{fig:Dist042}
\end{figure}

To proceed in this direction, let us take a look at the complex Langevin distribution according to the two kernels found in the optimization process and compare them to the Lefschetz thimble structure of the model. The thimble here consist of three different parts as shown by red lines in \cref{fig:Dist042}, together with the critical points (green points). Note that the thimbles always cross through the critical points. The distribution of the complex Langevin evolution is shown as a point cloud. The three different distributions shown in each panel correspond to the case of (top left) $K_0=1$, (top right) $K_1={\rm exp}[-i\pi/3]$ and (bottom) $K_2={\rm exp}[-i2\pi/3]$. One can clearly see that for the trivial kernel complex Langevin tries to sample parallel to the real axis. As we saw in \cref{sec:const_kernel} the angle parametrizing the kernel translates into a preferred sampling direction. 

In the top right and bottom of \cref{fig:Dist042}, we have plotted the complex Langevin distribution obtained after introducing one of the two kernels that minimize $L_D$. Again we find that the angle of the noise term decides where CL samples. We see that the highest density of the CL distribution lies along the direction in which the thimble passes through the critical point at the origin. Further out from the origin, the distribution follows closely the angle of the noise term, which is $H_1 = \sqrt{e^{-i\pi/3}} = e^{-i\pi/6}$ for the first kernel (top left) and $H_2=\sqrt{e^{-i2\pi/3}} = e^{-i\pi/3}$ for the second (bottom). I.e. we can distinguish that sampling with the first kernel leads to samples slightly closer to the thimbles going out along the real-axis, compared to the other kernel which favors sampling more closely along the parts of the thimble that eventually run off to infinity. We will give a formal explanation for this behavior in the next paragraphs. 

\begin{figure}\centering
    \includegraphics[scale=0.28]{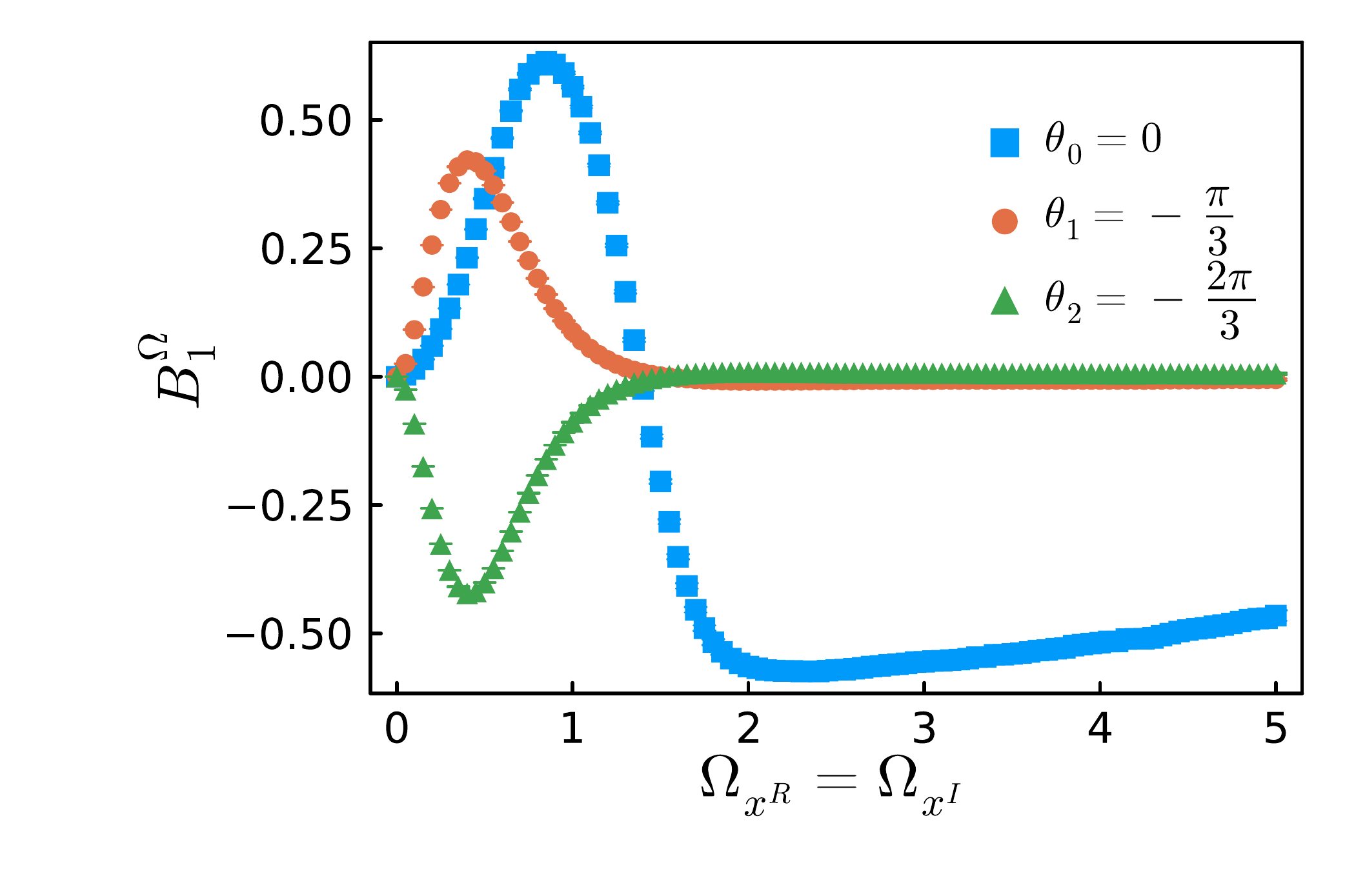}
    \includegraphics[scale=0.28]{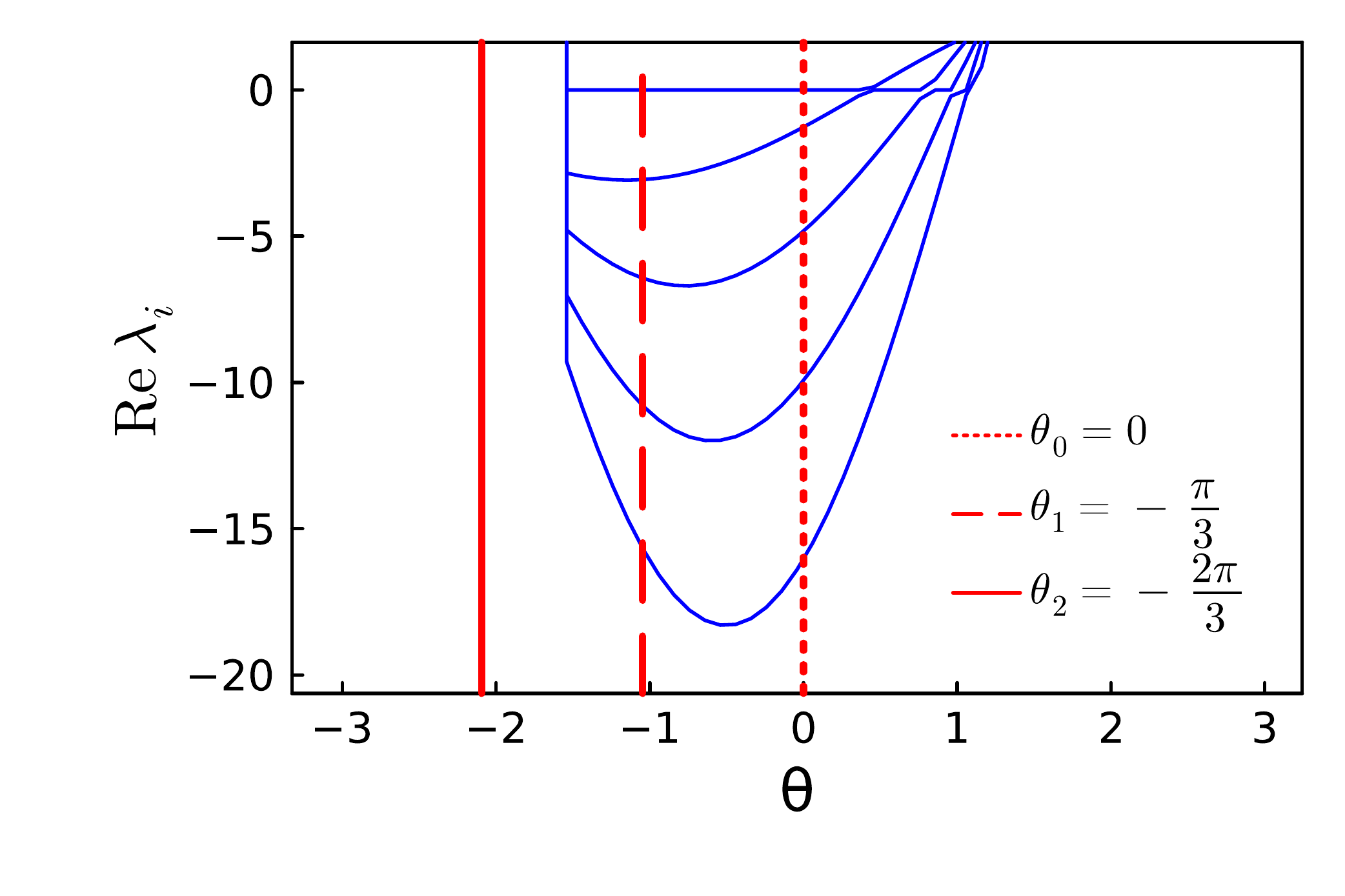}
    \caption{(left) Boundary term according to the $x^2$ observable for the model of \cref{eq:simpleModelAction} with $\sigma =4i$ and $\lambda=2$, evaluated for the three different kernels $K_i$ discussed in the main text. (right) the five eigenvalues of the Fokker-Planck operator with the largest real part (blue lines) plotted against the kernel parameter $\theta$. The position of the two kernels that optimize $L_D$ are indicated by red lines.}
    \label{fig:BT_Eigen_042}
\end{figure}

As shown in \cref{sec:correctnessCriterionWithAKernel} the correctness criterion consist of two parts, the first one states that no boundary terms may appear and the second requires that the eigenvalues of the complex Fokker-Planck equation need to have a negative real part. For this simple model we can compute both of these criteria, which is illustrated in \cref{fig:BT_Eigen_042}. 

The left plot contains the boundary terms for the real-part of the observable $\langle x^2 \rangle$. Each of the curves corresponds to one of the three kernels $K_i$. They are computed using the boundary term expectation value of \cref{eq:BoundaryTermsOperator}. We see that both of the kernels lead to very small values of the boundary terms for this observable, while the complex Langevin process without kernel exhibits a clear boundary term. However at this point we cannot yet say which of the two kernels produces the correct solution, if any.

In order to see which of them is correct, we need to look at the right plot in \cref{fig:BT_Eigen_042} where the five eigenvalues of the Fokker-Planck operator are plotted, which have the largest real-part (blue lines). They are plotted against different kernel parameters $\theta$ and the red lines indicate the position of the two kernels that optimize $L_D$. The eigenvalue calculation is carried out using a restarted Arnoldi method solver, which internally uses a Krylov-Schur method. We see that there is a region of $\theta$ where the eigenvalues are all satisfying $\textrm{Re} (\lambda ) \leq 0$, which includes the kernel $\theta_1=-\frac{\pi}{3}$. It is exactly this kernel, which, when incorporated into the complex Langevin evolution gives the right solution for the model. For smaller $\theta$s, the eigenvalues will eventually cross the zero. This is the region where one finds the second kernel $\theta_2=-\frac{2\pi}{3}$. We can therefore attribute the failure to restore correct convergence with the second kernel to a violation of the correctness criterion pertaining to the spectrum of the complex Fokker-Planck equation.

The interesting point here is that the boundary terms do not seem to distinguish between the two kernels as both lead to quickly diminishing distributions.

\subsection{Limitation of constant kernels and boundary terms}
\label{sec:kernel_and_boundaryTerms}

Let us now go to the set of parameters $\sigma=-1+4i$ and $\lambda=2$, for which there does not exist a constant kernel, which restores correct convergence. It is however possible to construct a field-dependent kernel that solves the problem \cite{Okano:1991tz}. 

We can understand this behavior, as the constant kernel that is optimal in the sense of removing boundary terms, does not achieve correct convergence of the complex Fokker-Planck equation to the correct $e^{-S}$.

This can be seen again by plotting the CL distribution for some of the local minima of the $L_D$ loss function. For this parameter space there are more than two, but we have picked out two of the solutions which have the interesting property that they both have no boundary terms, and still do not converge to the true solution. The kernels that are picked have the parameters $\theta_3 = -\frac{3\pi}{4}$ and $\theta_4 = \frac{\pi}{2}$

\begin{figure}\centering
    \includegraphics[scale=0.04]{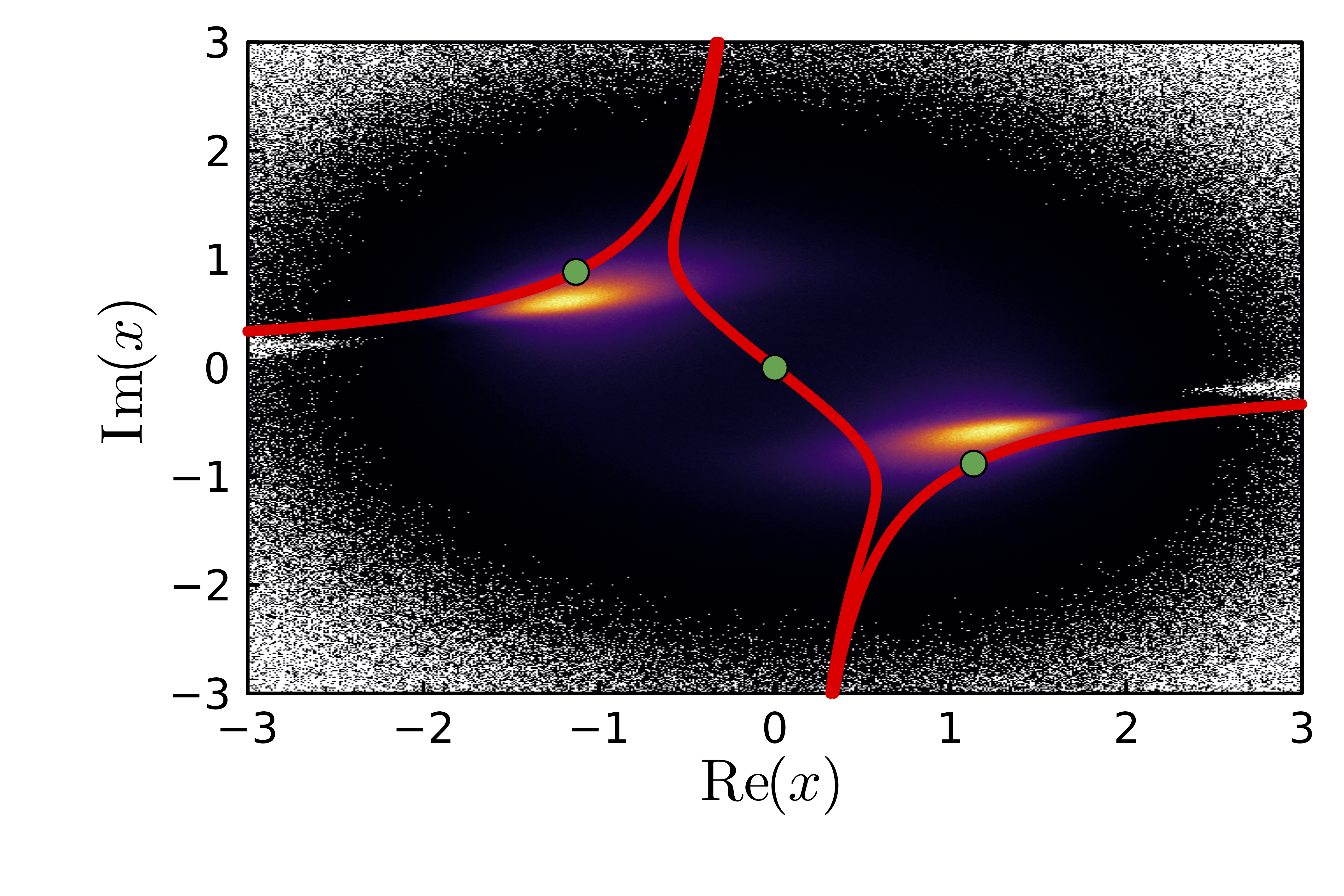}
    \includegraphics[scale=0.04]{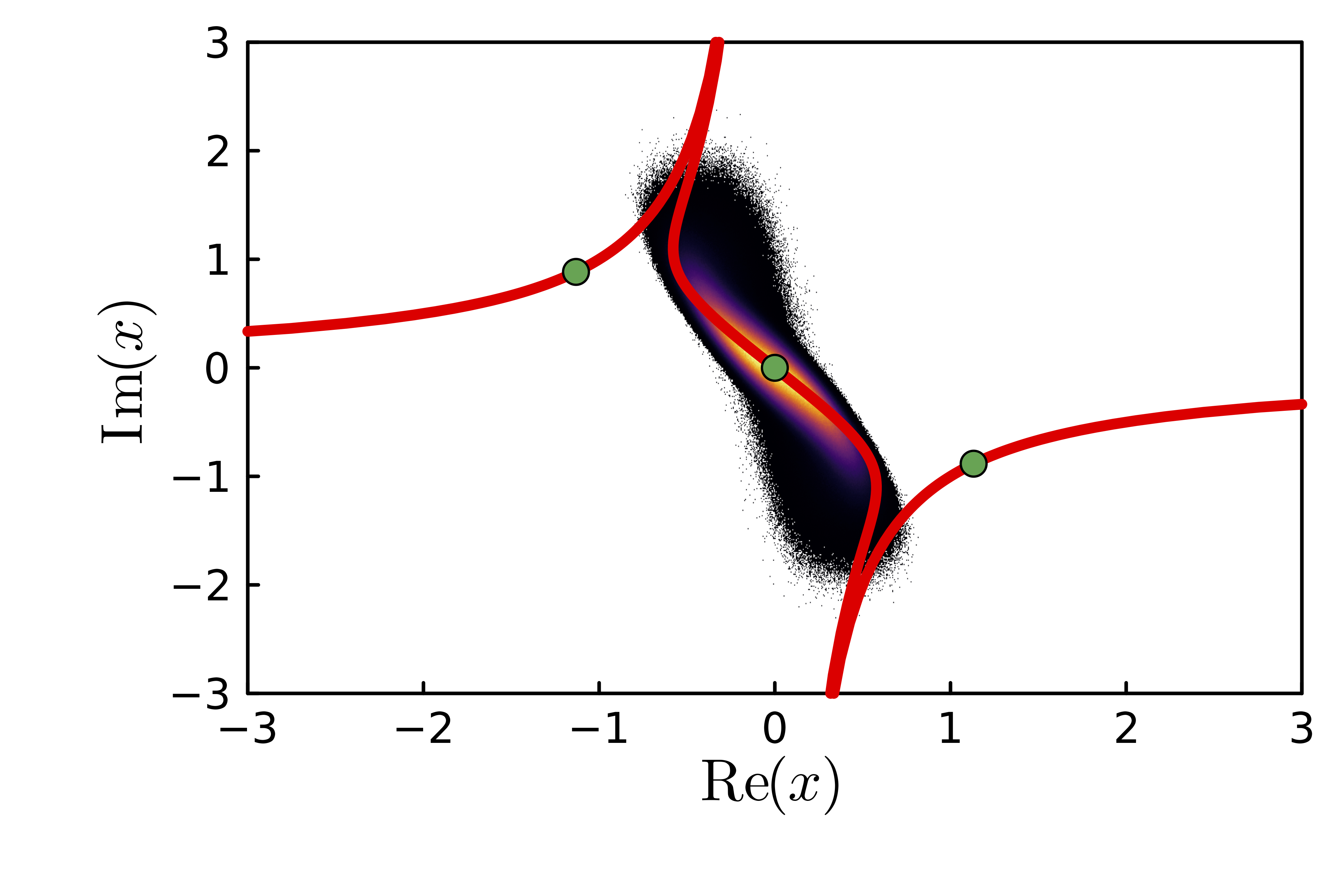}
    \includegraphics[scale=0.04]{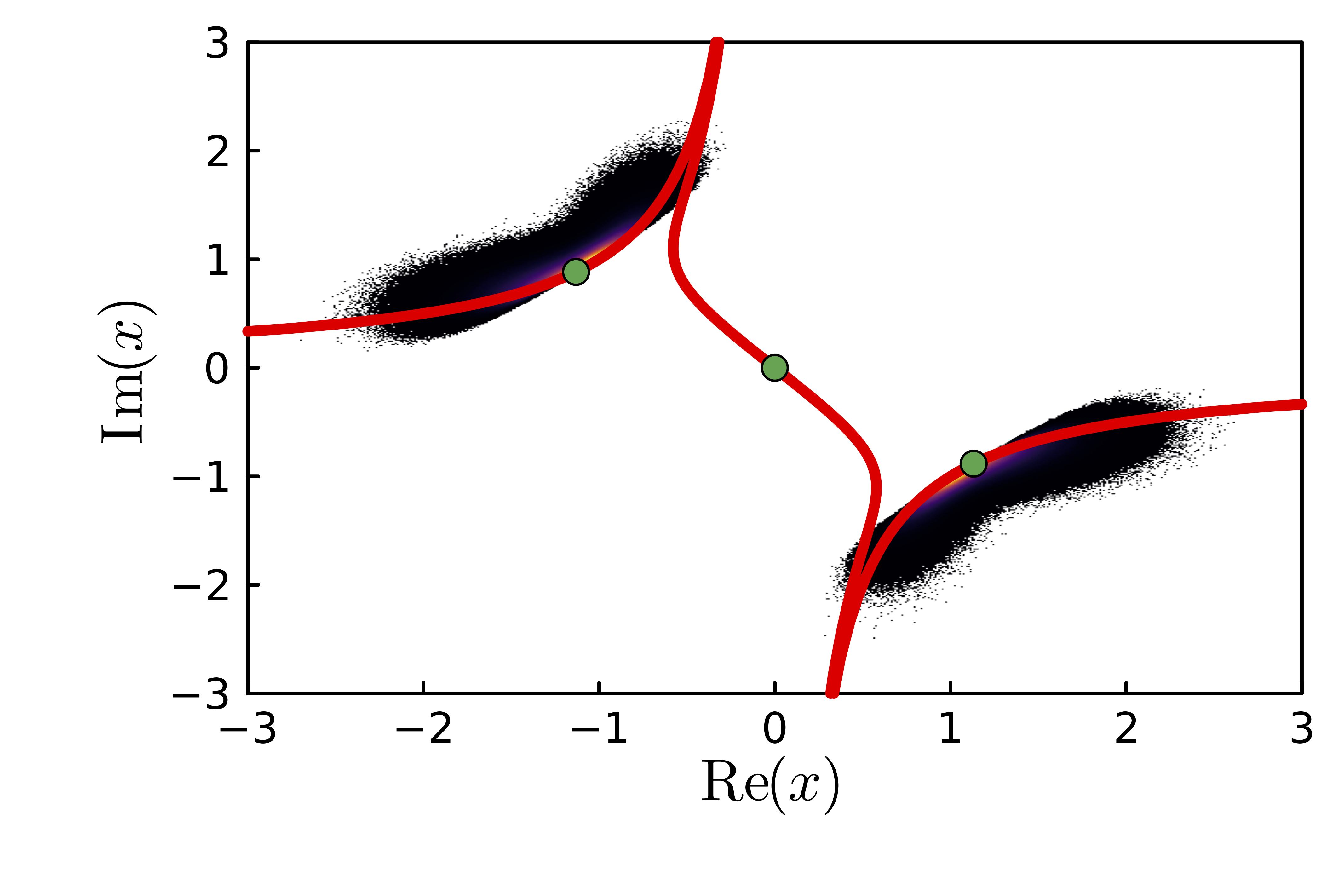}
    \caption{Distribution of the complex Langevin simulation and the Lefschetz thimbles (red line) for the model of \cref{eq:simpleModelAction} with $\sigma = -1 + 4i$ and $\lambda=2$, using different kernels; $K_0=1$ (top left), $K_3=e^{-i\frac{3\pi}{4}}$ (top right) and $K_4=e^{i\frac{\pi}{2}}$ (bottom). The green points are the critical points given by the solution to \cref{eq:criticalPoints}.}
    \label{fig:Dist-142}
\end{figure}

The CL distribution together with the thimble is plotted in \cref{fig:Dist-142} for the three different kernels $K_0=1$ (top left), $K_3=\exp(i\theta_3)$ (top right) and $K_4=\exp(i\theta_4)$ (bottom). We see that the thimbles show three distinct structures, connecting at infinity. To obtain the thimbles we evolve the gradient flow equation starting from a small offset from the critical points (which all are saddle points) and then combine the six part of the thimbles. The CL distribution without a kernel (top left plot in \cref{fig:Dist-142}) again favors sampling parallel to the real-axis, while the two other kernels sample  completely different parts of the thimbles. The distribution for $K_3$ is located along the thimble crossing the origin. The other kernel ($K_4$), follows the other two thimbles crossing the critical points away from the origin. We can explain this behavior with the angle of the noise coefficient. For $K_3$ we have an angle of $-\frac{3\pi}{8}$ against the real axis and for $K_4$ we have an angle of $\frac{\pi}{4}$ against the real axis. 

\begin{figure}\centering
    \includegraphics[scale=0.28]{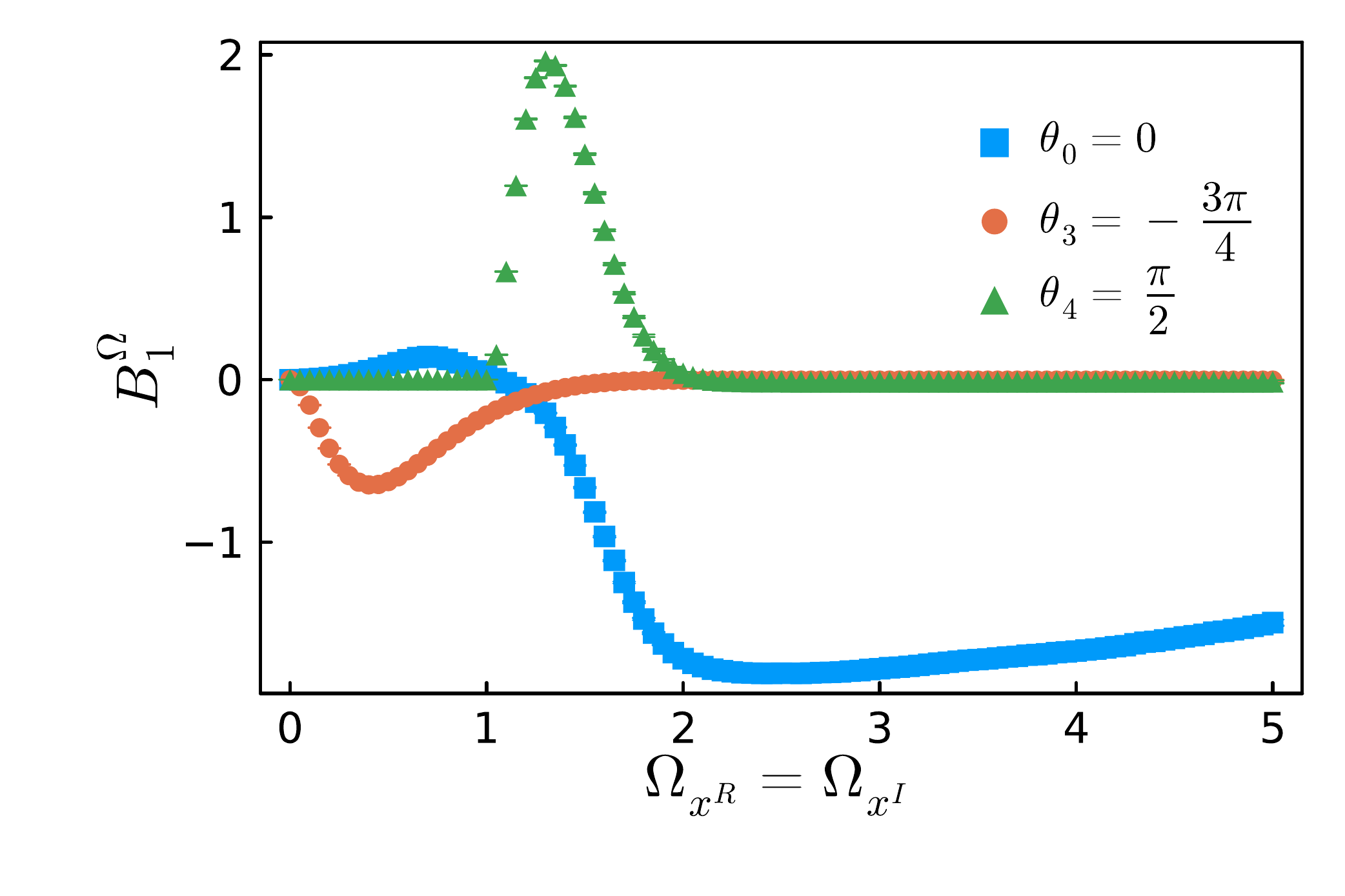}
    \includegraphics[scale=0.28]{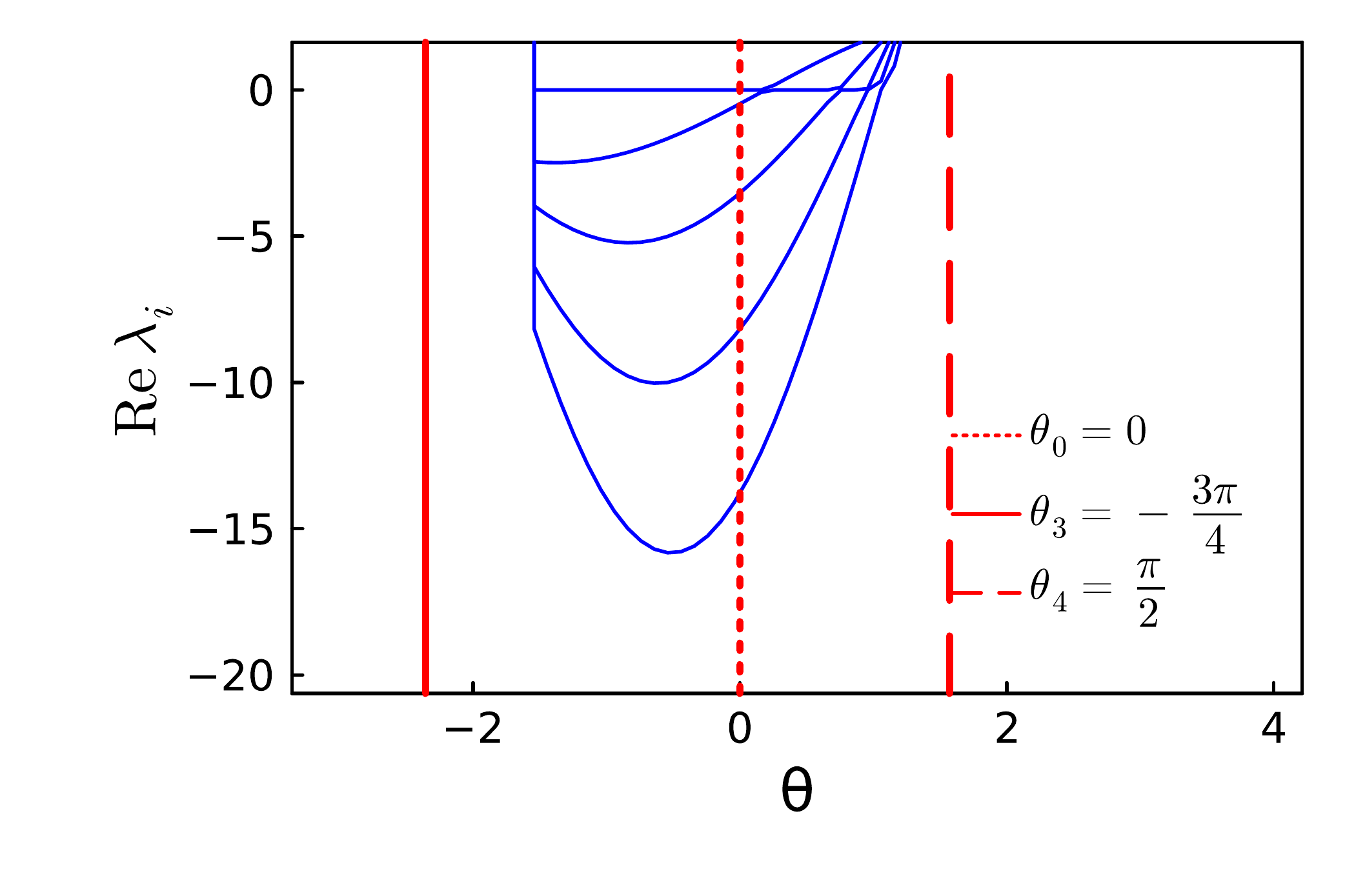}
    \caption{(left) Boundary term according to the $x^2$ observable for the model of \cref{eq:simpleModelAction} with $\sigma = -1 + 4i$ and $\lambda=2$, evaluated for the three different kernels $K_i$ discussed in the main text. (right) the five eigenvalues of the Fokker-Planck operator with the largest real part (blue lines) plotted against the kernel parameter $\theta$. The position of the two kernels that optimize $L_D$ are indicated by red lines.}
    \label{fig:BTEigen-142}
\end{figure}

In \cref{fig:BTEigen-142} (left) the boundary terms for this set of model parameters is calculated for the observable $\textrm{Re}\; \langle x^2 \rangle$ and plotted for increasing square box cutoff. We see that without a kernel, there are boundary terms present, as the blue datapoints do not go to zero for large cutoff. This can also been seen directly from the distribution in \cref{fig:Dist-142} which exhibits a large spread and hence the falloff of the distribution is not fast enough. For the two kernels, $K_3$ and $K_4$, that correspond to a local minimum in $L_D$, the system does not show any boundary terms. This is an important point as even though we have avoided boundary terms, the CL dynamics under the kernels $K_3$ and $K_4$ still does not converge to the correct solution. In turn it appears that it is in general not enough to remove the boundary terms to achieve correct convergence. In fact one also needs to be sure that the complex Fokker-Planck equation converges to the desired equilibrium distribution. 

In \cref{fig:BTEigen-142} (right) we show the five eigenvalues of the Fokker-Planck equation with the largest real-part plotted against the parameter $\theta$ which determines the kernel $K=e^{i\theta}$. For the parameters chosen here, we find that both kernels lie outside of the admissible region\footnote{An interesting observation was made in \cite{Okano:1991tz}, that combining kernels, which sample different parts of the thimble into a field-dependent kernel seems to work well. The motivation was to find a kernel that would reduce the drift term to $-x$ when either the $x^2$ or  the $x^4$ term in the action dominates. A similar argument for constructing a field-dependent kernel can now be made via the minima of the $L_D$ loss function, which favor sampling different parts of the thimbles.}, where $\Re \lambda \leq 0$. Interestingly at $\theta=0$ the eigenvalues actually all lie in the lower half complex plane but there the boundary criterion is not fulfilled.
But as one increases the imaginary part of $\sigma$, e.g. at $\sigma=-1+5i$, one finds that the eigenvalues  for the identity kernel $K=1$ already take on positive real-parts.

Including the calculation of eigenvalues in the cost functional would be possible for simple models such as the one of \cref{eq:simpleModelAction}, but for larger, more realistic systems the dimension of the Fokker-Planck operator scales as $N^d$, where $N$ is the number of points along in each dimensions $d$. Even for the anharmonic oscillator on the SK contour, the calculation of the Fokker-Planck eigenvalues is too costly in practice. We therefore need a different way of distinguishing which kernel leads to correct convergence. As discussed in detail in the main text of this manuscript we thus propose to collect as much prior information about the system as possible in the cost functional $L^{\rm prior}$, based on which the success of the optimal kernel according to $L_D$ is judged.


\FloatBarrier

\begin{backmatter}

\section*{Competing interests}
  The authors declare that they have no competing interests.

\section*{Author's contributions}
    \begin{itemize}
        \item D. Alvestad: concept development, algorithmic development, code implementation, data analysis, writing
        \item R. Larsen: supervising code development
        \item A. Rothkopf: project conceptualization, funding acquisition, supervision, writing
    \end{itemize}


\bibliographystyle{stavanger-mathphys}


\bibliography{references}


\end{backmatter}


\end{document}